\pdfoutput=1

\documentclass[12pt,a4paper]{article}

\usepackage{ifthen}
\newboolean{pdflatex}
\setboolean{pdflatex}{true}

\newboolean{articletitles}
\setboolean{articletitles}{true}

\newboolean{uprightparticles}
\setboolean{uprightparticles}{false}

\newboolean{inbibliography}
\setboolean{inbibliography}{false}

\def\phani{\phantom{1}}
\def\phanm{\phantom{-}}
\def\phanii{\phantom{11}}
\def\phaniii{\phantom{111}}

\newcommand{\mpr}        {\mbox{$m^\prime$}}
\newcommand{\thpr}       {\mbox{$\theta^\prime$}}

\usepackage{microtype}
\usepackage{lineno}
\usepackage{xspace}

\usepackage{caption}

\usepackage{graphicx}
\usepackage{rotating}
\usepackage{color}
\usepackage{colortbl}

\usepackage{amsmath}
\usepackage{amssymb}
\usepackage{amsfonts}
\usepackage{upgreek}

\newcommand*\patchAmsMathEnvironmentForLineno[1]{
\expandafter\let\csname old#1\expandafter\endcsname\csname #1\endcsname
\expandafter\let\csname oldend#1\expandafter\endcsname\csname
end#1\endcsname
 \renewenvironment{#1}
   {\linenomath\csname old#1\endcsname}
   {\csname oldend#1\endcsname\endlinenomath}
}
\newcommand*\patchBothAmsMathEnvironmentsForLineno[1]{
  \patchAmsMathEnvironmentForLineno{#1}
  \patchAmsMathEnvironmentForLineno{#1*}
}
\AtBeginDocument{
\patchBothAmsMathEnvironmentsForLineno{equation}
\patchBothAmsMathEnvironmentsForLineno{align}
\patchBothAmsMathEnvironmentsForLineno{flalign}
\patchBothAmsMathEnvironmentsForLineno{alignat}
\patchBothAmsMathEnvironmentsForLineno{gather}
\patchBothAmsMathEnvironmentsForLineno{multline}
\patchBothAmsMathEnvironmentsForLineno{eqnarray}
}

\usepackage{hyperref}
\usepackage[all]{hypcap}

\def\lhcb {\mbox{LHCb}\xspace}

\def\babar  {\mbox{BaBar}\xspace}

\def\MagUp {\mbox{\em Mag\kern -0.05em Up}\xspace}

\ifthenelse{\boolean{uprightparticles}}
{

 \def\Peta        {\ensuremath{\upeta}\xspace}

 \def\Ppi         {\ensuremath{\uppi}\xspace}

 \def\Ppsi        {\ensuremath{\uppsi}\xspace}

 \def\PDelta      {\ensuremath{\Delta}\xspace}
 \def\PXi      {\ensuremath{\Xi}\xspace}
 \def\PLambda      {\ensuremath{\Lambda}\xspace}
 \def\PSigma      {\ensuremath{\Sigma}\xspace}
 \def\POmega      {\ensuremath{\Omega}\xspace}
 \def\PUpsilon      {\ensuremath{\Upsilon}\xspace}

 \def\PB      {\ensuremath{\mathrm{B}}\xspace}
 
 \def\PD      {\ensuremath{\mathrm{D}}\xspace}

 \def\PJ      {\ensuremath{\mathrm{J}}\xspace}
 \def\PK      {\ensuremath{\mathrm{K}}\xspace}

 \def\Pb      {\ensuremath{\mathrm{b}}\xspace}
 \def\Pc      {\ensuremath{\mathrm{c}}\xspace}

 \def\Pi      {\ensuremath{\mathrm{i}}\xspace}

 \def\Pp      {\ensuremath{\mathrm{p}}\xspace}

 \def\Ps      {\ensuremath{\mathrm{s}}\xspace}

}
{

 \def\Peta        {\ensuremath{\eta}\xspace}

 \def\Ppi         {\ensuremath{\pi}\xspace}

 \def\Ppsi        {\ensuremath{\psi}\xspace}
 
 \mathchardef\PDelta="7101
 \mathchardef\PXi="7104
 \mathchardef\PLambda="7103
 \mathchardef\PSigma="7106
 \mathchardef\POmega="710A
 \mathchardef\PUpsilon="7107
 
 \def\PB      {\ensuremath{B}\xspace}
 
 \def\PD      {\ensuremath{D}\xspace}

 \def\PJ      {\ensuremath{J}\xspace}
 \def\PK      {\ensuremath{K}\xspace}

 \def\Pb      {\ensuremath{b}\xspace}
 \def\Pc      {\ensuremath{c}\xspace}

 \def\Pi      {\ensuremath{i}\xspace}

 \def\Pp      {\ensuremath{p}\xspace}

 \def\Ps      {\ensuremath{s}\xspace}

}

\makeatletter
\ifcase \@ptsize \relax
  \newcommand{\miniscule}{\@setfontsize\miniscule{4}{5}}
\or
  \newcommand{\miniscule}{\@setfontsize\miniscule{5}{6}}
\or
  \newcommand{\miniscule}{\@setfontsize\miniscule{5}{6}}
\fi
\makeatother

\DeclareRobustCommand{\optbar}[1]{\shortstack{{\miniscule (\rule[.5ex]{1.25em}{.18mm})}
  \\ [-.7ex] $#1$}}

\def\squark    {{\ensuremath{\Ps}}\xspace}

\def\cquark    {{\ensuremath{\Pc}}\xspace}

\def\bquark    {{\ensuremath{\Pb}}\xspace}

\def\pion   {{\ensuremath{\Ppi}}\xspace}
\def\piz    {{\ensuremath{\pion^0}}\xspace}

\def\pip    {{\ensuremath{\pion^+}}\xspace}
\def\pim    {{\ensuremath{\pion^-}}\xspace}
\def\pipm   {{\ensuremath{\pion^\pm}}\xspace}

\def\kaon    {{\ensuremath{\PK}}\xspace}
  \def\Kbar    {{\kern 0.2em\overline{\kern -0.2em \PK}{}}\xspace}

\def\KorKbar    {\kern 0.18em\optbar{\kern -0.18em K}{}\xspace}

\def\Kp      {{\ensuremath{\kaon^+}}\xspace}
\def\Km      {{\ensuremath{\kaon^-}}\xspace}
\def\Kpm     {{\ensuremath{\kaon^\pm}}\xspace}

\def\KS      {{\ensuremath{\kaon^0_{\rm\scriptscriptstyle S}}}\xspace}

\def\Kstarz  {{\ensuremath{\kaon^{*0}}}\xspace}

\def\Kstar   {{\ensuremath{\kaon^*}}\xspace}

\def\Kstarbsubz  {{\ensuremath{\Kbar{}^*_0}}\xspace}

\newcommand{\etapr}{\ensuremath{\Peta^{\prime}}\xspace}

  \def\Dbar    {{\kern 0.2em\overline{\kern -0.2em \PD}{}}\xspace}
\def\D       {{\ensuremath{\PD}}\xspace}

\def\DorDbar    {\kern 0.18em\optbar{\kern -0.18em D}{}\xspace}
\def\Dz      {{\ensuremath{\D^0}}\xspace}
\def\Dzb     {{\ensuremath{\Dbar{}^0}}\xspace}
\def\Dp      {{\ensuremath{\D^+}}\xspace}
\def\Dm      {{\ensuremath{\D^-}}\xspace}

\def\Dstar   {{\ensuremath{\D^*}}\xspace}

\def\Dstarzb {{\ensuremath{\Dbar{}^{*0}}}\xspace}
\def\DorDstarzb{{\ensuremath{\Dbar{}^{(*)0}}}\xspace}
\def\Dstarp  {{\ensuremath{\D^{*+}}}\xspace}
\def\Dstarm  {{\ensuremath{\D^{*-}}}\xspace}

\def\B       {{\ensuremath{\PB}}\xspace}
\def\Bbar    {{\ensuremath{\kern 0.18em\overline{\kern -0.18em \PB}{}}}\xspace}

\def\BorBbar    {\kern 0.18em\optbar{\kern -0.18em B}{}\xspace}
\def\Bz      {{\ensuremath{\B^0}}\xspace}

\def\Bu      {{\ensuremath{\B^+}}\xspace}

\def\Bp      {{\ensuremath{\Bu}}\xspace}

\def\Bd      {{\ensuremath{\B^0}}\xspace}
\def\Bs      {{\ensuremath{\B^0_\squark}}\xspace}

\def\Bds     {\ensuremath{\B^0_{(\squark)}}\xspace}

\def\jpsi     {{\ensuremath{{\PJ\mskip -3mu/\mskip -2mu\Ppsi\mskip 2mu}}}\xspace}

  \def\Y#1S{\ensuremath{\PUpsilon{(#1S)}}\xspace}

\def\proton      {{\ensuremath{\Pp}}\xspace}
\def\antiproton  {{\ensuremath{\overline \proton}}\xspace}

\def\Lz          {{\ensuremath{\PLambda}}\xspace}
\def\Lbar        {{\ensuremath{\kern 0.1em\overline{\kern -0.1em\PLambda}}}\xspace}
\def\LorLbar    {\kern 0.18em\optbar{\kern -0.18em \PLambda}{}\xspace}

\def\Lbbar   {{\ensuremath{\Lbar{}^0_\bquark}}\xspace}

\def\to                 {\ensuremath{\rightarrow}\xspace}

\def\CP                {{\ensuremath{C\!P}}\xspace}

\def\AT#1     {\ensuremath{A_{\mathrm{T}}^{#1}}\xspace}

\def\C#1      {\ensuremath{\mathcal{C}_{#1}}\xspace}
\def\Cp#1     {\ensuremath{\mathcal{C}_{#1}^{'}}\xspace}
\def\Ceff#1   {\ensuremath{\mathcal{C}_{#1}^{\mathrm{(eff)}}}\xspace}
\def\Cpeff#1  {\ensuremath{\mathcal{C}_{#1}^{'\mathrm{(eff)}}}\xspace}
\def\Ope#1    {\ensuremath{\mathcal{O}_{#1}}\xspace}
\def\Opep#1   {\ensuremath{\mathcal{O}_{#1}^{'}}\xspace}

\newcommand{\tev}{\ifthenelse{\boolean{inbibliography}}{\ensuremath{~T\kern -0.05em eV}\xspace}{\ensuremath{\mathrm{\,Te\kern -0.1em V}}}\xspace}
\newcommand{\gev}{\ensuremath{\mathrm{\,Ge\kern -0.1em V}}\xspace}
\newcommand{\mev}{\ensuremath{\mathrm{\,Me\kern -0.1em V}}\xspace}
\newcommand{\kev}{\ensuremath{\mathrm{\,ke\kern -0.1em V}}\xspace}
\newcommand{\gevnsp}{\ensuremath{\mathrm{Ge\kern -0.1em V}}\xspace}
\newcommand{\mevnsp}{\ensuremath{\mathrm{Me\kern -0.1em V}}\xspace}
\newcommand{\kevnsp}{\ensuremath{\mathrm{ke\kern -0.1em V}}\xspace}
\newcommand{\ev}{\ensuremath{\mathrm{\,e\kern -0.1em V}}\xspace}
\newcommand{\gevc}{\ensuremath{{\mathrm{\,Ge\kern -0.1em V\!/}c}}\xspace}
\newcommand{\mevc}{\ensuremath{{\mathrm{\,Me\kern -0.1em V\!/}c}}\xspace}
\newcommand{\gevcc}{\ensuremath{{\mathrm{\,Ge\kern -0.1em V\!/}c^2}}\xspace}
\newcommand{\gevgevcccc}{\ensuremath{{\mathrm{\,Ge\kern -0.1em V^2\!/}c^4}}\xspace}
\newcommand{\mevcc}{\ensuremath{{\mathrm{\,Me\kern -0.1em V\!/}c^2}}\xspace}

\def\mm   {\ensuremath{\rm \,mm}\xspace}

\def\mum  {\ensuremath{{\,\upmu\rm m}}\xspace}

\def\fm   {\ensuremath{\rm \,fm}\xspace}

\def\invfb   {\ensuremath{\mbox{\,fb}^{-1}}\xspace}

\newcommand{\chisq}{\ensuremath{\chi^2}\xspace}

\newcommand{\chisqip}{\ensuremath{\chi^2_{\rm IP}}\xspace}

\def\gsim{{~\raise.15em\hbox{$>$}\kern-.85em
          \lower.35em\hbox{$\sim$}~}\xspace}
\def\lsim{{~\raise.15em\hbox{$<$}\kern-.85em
          \lower.35em\hbox{$\sim$}~}\xspace}

\newcommand{\Real}{\ensuremath{\mathcal{R}e}\xspace}

\def\sPlot{\mbox{\em sPlot}\xspace}

\def\ptot       {\mbox{$p$}\xspace}
\def\pt         {\mbox{$p_{\rm T}$}\xspace}

\def\evtgen     {\mbox{\textsc{EvtGen}}\xspace}

\def\geant      {\mbox{\textsc{Geant4}}\xspace}

\def\photos     {\mbox{\textsc{Photos}}\xspace}

\def\pythia     {\mbox{\textsc{Pythia}}\xspace}

\def\tell1  {TELL1\xspace}
\def\ukl1   {UKL1\xspace}

\newcommand{\ie}{\mbox{\itshape i.e.}\xspace}

\usepackage{cite}
\usepackage{mciteplus}
\usepackage{multirow}

\begin{document}

\renewcommand{\thefootnote}{\fnsymbol{footnote}}
\setcounter{footnote}{1}

\begin{titlepage}
\pagenumbering{roman}

\vspace*{-1.5cm}
\centerline{\large EUROPEAN ORGANIZATION FOR NUCLEAR RESEARCH (CERN)}
\vspace*{1.5cm}
\hspace*{-0.5cm}
\begin{tabular*}{\linewidth}{lc@{\extracolsep{\fill}}r}
\ifthenelse{\boolean{pdflatex}}
{\vspace*{-2.7cm}\mbox{\!\!\!\includegraphics[width=.14\textwidth]{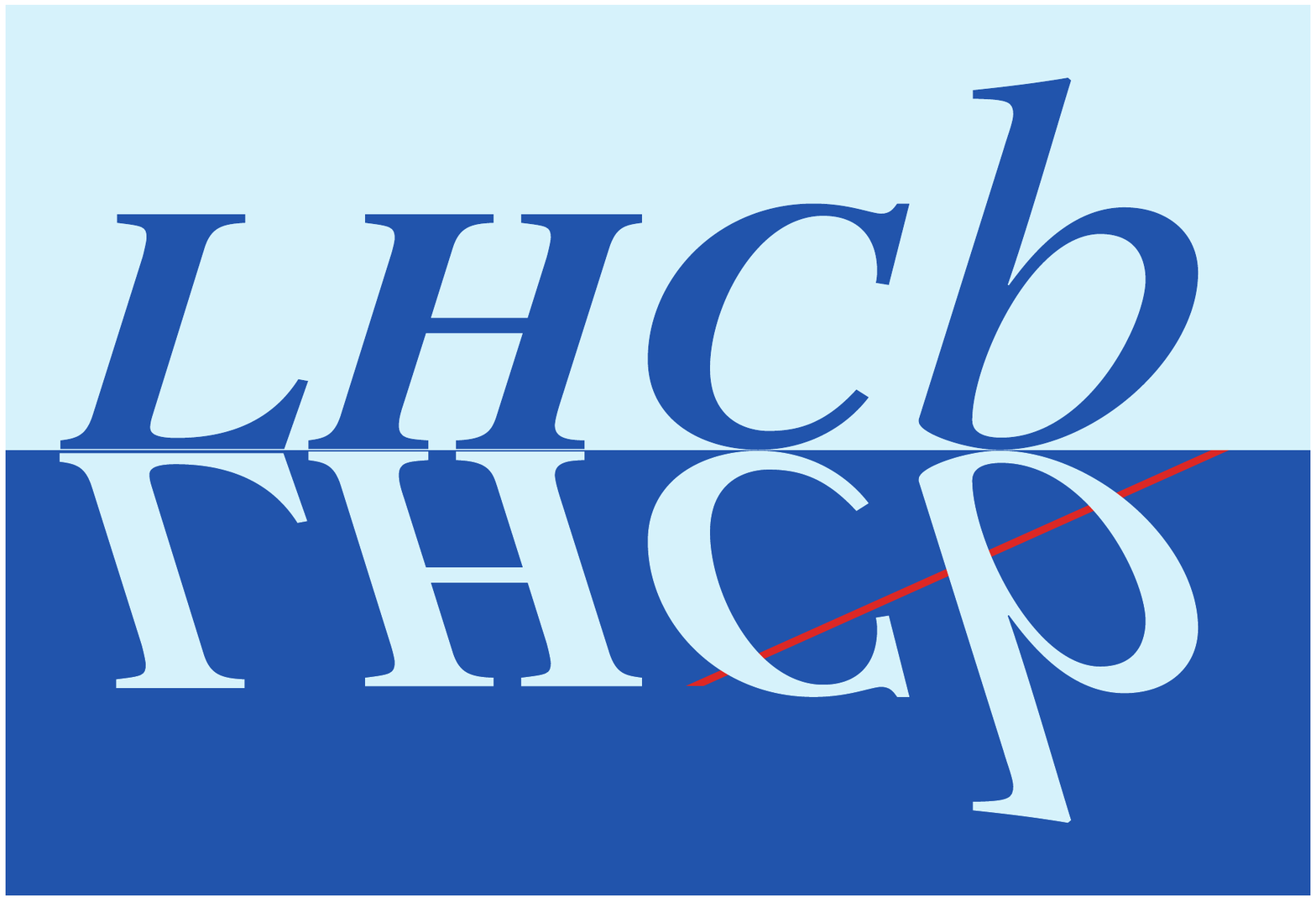}} & &}
{\vspace*{-1.2cm}\mbox{\!\!\!\includegraphics[width=.12\textwidth]{lhcb-logo.eps}} & &}
\\
 & & CERN-PH-EP-2015-107 \\
 & & LHCb-PAPER-2015-017 \\
 & & 3 August 2015 \\
 & & \\
\end{tabular*}

\vspace*{2.5cm}

{\bf\boldmath\huge
\begin{center}
  Amplitude analysis of $\Bd\to\Dzb\Kp\pim$ decays
\end{center}
}

\vspace*{1.5cm}

\begin{center}
The LHCb collaboration\footnote{Authors are listed at the end of this paper.}
\end{center}

\vspace{\fill}

\begin{abstract}
  \noindent
  The Dalitz plot distribution of $\Bd\to\Dzb\Kp\pim$ decays is studied using a data sample corresponding to $3.0\invfb$ of $pp$ collision data recorded by the LHCb experiment during 2011 and 2012.
  The data are described by an amplitude model that contains contributions
  from intermediate $\Kstar(892)^{0}$, $\Kstar(1410)^{0}$, $K^*_2(1430)^{0}$
  and $D^{*}_{2}(2460)^{-}$ resonances. The model also contains components to describe broad structures,
  including the $K^*_0(1430)^0$ and $D^*_0(2400)^-$ resonances, in the $K\pi$ S-wave and the $D\pi$ S- and P-waves.
  The masses and widths of the $D^{*}_{0}(2400)^{-}$ and $D^{*}_{2}(2460)^{-}$
  resonances are measured, as are the complex amplitudes and fit fractions for
  all components included in the amplitude model.
  The model obtained will be an integral part of a future
  determination of the angle $\gamma$ of the CKM quark mixing matrix using $\Bd\to\D\Kp\pim$ decays.
\end{abstract}

\vspace*{1.5cm}

\begin{center}
  Published in Phys.~Rev.~D
\end{center}

\vspace{\fill}

{\footnotesize
\centerline{\copyright~CERN on behalf of the \lhcb collaboration, licence \href{http://creativecommons.org/licenses/by/4.0/}{CC-BY-4.0}.}}
\vspace*{2mm}

\end{titlepage}

\newpage
\setcounter{page}{2}
\mbox{~}

\cleardoublepage

\renewcommand{\thefootnote}{\arabic{footnote}}
\setcounter{footnote}{0}
\pagestyle{plain}
\setcounter{page}{1}
\pagenumbering{arabic}

\section{Introduction}
\label{sec:introduction}

\begin{figure}[!b]
\centering
\includegraphics[width=0.49\textwidth]{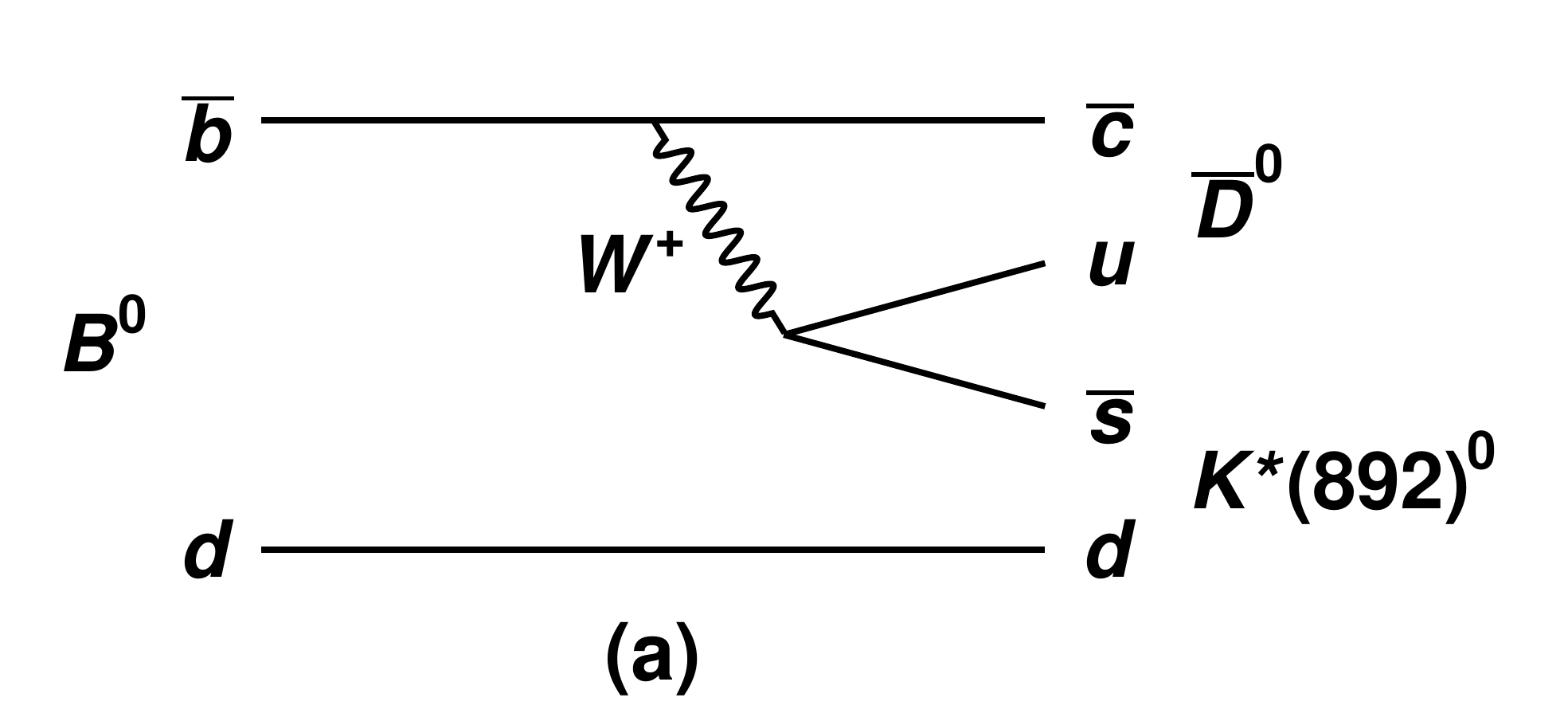}
\includegraphics[width=0.49\textwidth]{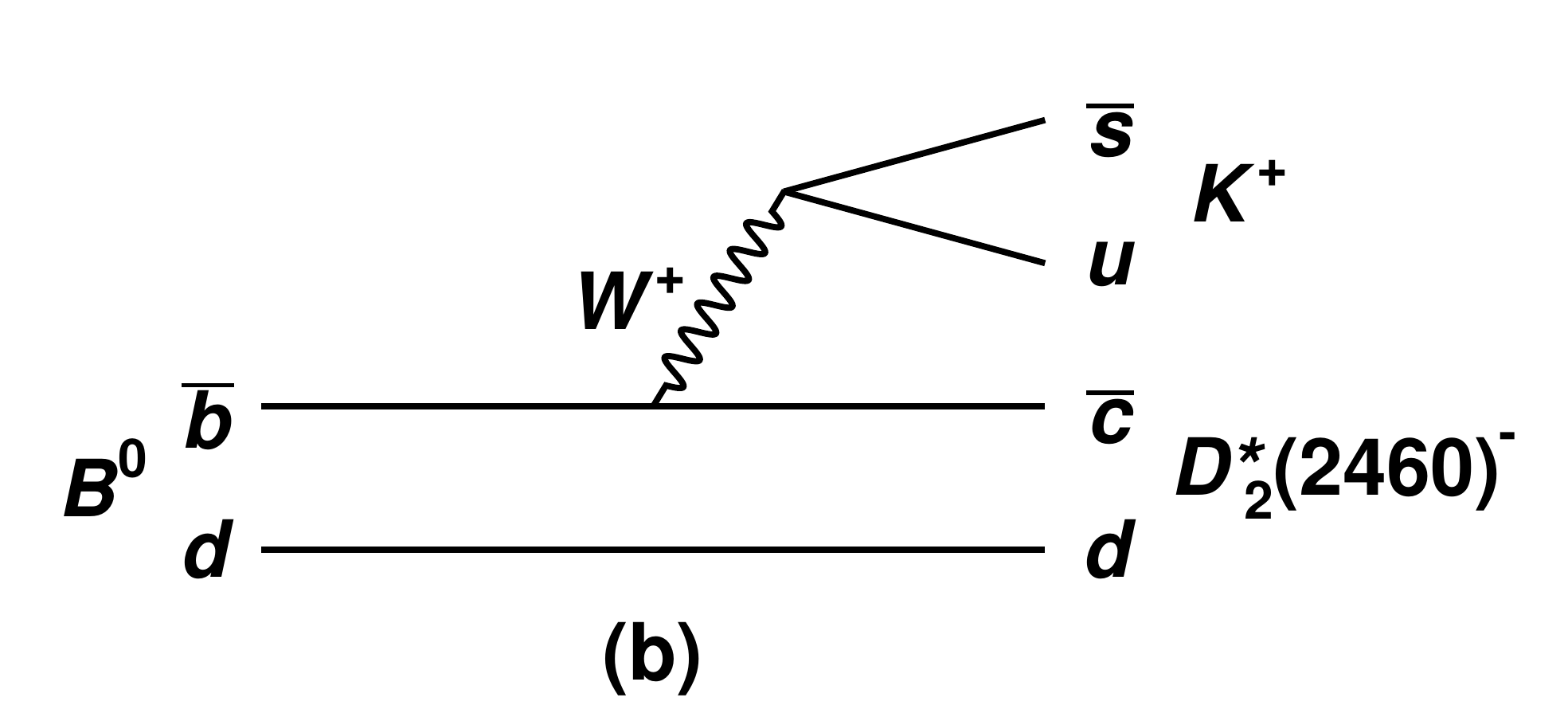}
\caption{\small
  Decay diagrams for the quasi-two-body contributions to $\Bd \to \D\Kp\pim$ from (a) $\Bd \to D\Kstar(892)^0$ and (b) $\Bd \to D^{*}_{2}(2460)^{-}\Kp$ decays.
}
\label{fig:decay}
\end{figure}

Dalitz plot (DP) analysis of $\Bd \to \D\Kp\pim$ decays has been proposed as a way to measure the unitarity triangle angle $\gamma$~\cite{Gershon:2008pe,Gershon:2009qc}.
The sensitivity to $\gamma \equiv \arg\left[ - V^{}_{ud}V_{ub}^*/(V^{}_{cd}V_{cb}^*) \right]$, where $V^{}_{xy}$ are elements of the Cabibbo-Kobayashi-Maskawa (CKM) quark mixing matrix~\cite{PhysRevLett.10.531,PTP.49.652}, originates from the interference of $\bar{b}\to \bar{c}u\bar{s}$ and $\bar{b}\to \bar{u}c\bar{s}$ amplitudes.
Such interference occurs when the neutral $D$ meson is reconstructed in a final state that is accessible to both \Dzb and \Dz decays and therefore corresponds to an admixture of the two states~\cite{Gronau:1990ra,Gronau:1991dp}.
One of the largest components of the $\Bd \to \D\Kp\pim$ final state is $\Bd \to D\Kstar(892)^0$, for which both $\bar{b}\to \bar{c}u\bar{s}$ and $\bar{b}\to \bar{u}c\bar{s}$ amplitudes are colour suppressed, making them comparable in magnitude and potentially enhancing $\CP$ violation effects~\cite{Dunietz:1991yd}.
Decay diagrams for the quasi-two-body contributions from $\Bd \to D\Kstar(892)^0$ and $\Bd \to D^{*}_{2}(2460)^{-}\Kp$ decays are shown in Fig.~\ref{fig:decay}.
Observables sensitive to $\gamma$ have been measured by LHCb with a quasi-two-body approach~\cite{LHCb-PAPER-2014-028}, but a DP analysis is expected to be more sensitive because interference between resonances provides the possibility to resolve ambiguities in the determination of $\gamma$.

In order to determine $\gamma$ with this method, it is necessary to have an amplitude model of the $\Bd \to \Dzb\Kp\pim$ decay that proceeds through the favoured $\bar{b}\to \bar{c}u\bar{s}$ transition.
This can be achieved by reconstructing the $\Dzb$ meson through the $\Kp\pim$ decay; in this way the contribution from the $\bar{b}\to \bar{c}u\bar{s}$ amplitude is dominant and effects due to $\bar{b}\to \bar{u}c\bar{s}$ amplitudes can be neglected.
While the analysis of this decay chain is not itself sensitive to $\gamma$, its outcome will be an integral part of a future analysis using, for example, $\D \to \Kp\Km$ decays where \CP violation effects are expected as the final state is common to both $\Dz$ and $\Dzb$ decays.

Dalitz plot analyses of $B$ meson decays to final states containing a charmed meson and two charged particles (either pions or kaons) also provide opportunities for studies of the spectroscopy of charmed mesons.
Results in this area have recently been obtained from DP analyses of $\Bs \to \Dzb \Km\pip$~\cite{LHCb-PAPER-2014-035,LHCb-PAPER-2014-036}, $\Bp \to \Dm\Kp\pip$~\cite{LHCb-PAPER-2015-007} and $\Bz \to \Dzb\pip\pim$~\cite{LHCb-PAPER-2014-070} decays, all from LHCb.
As the branching fraction for $\Bd \to \Dzb\Kp\pim$ decays is smaller than that for $\Bz \to \Dzb\pip\pim$ decays, the analysis presented in this paper is not as sensitive to the parameters of charm resonances as that of Ref.~\cite{LHCb-PAPER-2014-070}.
However, the much larger sample of \B mesons available at LHCb compared to that used in the only other published DP analysis of $\Bz \to \Dzb\pip\pim$ decays from the Belle collaboration~\cite{Kuzmin:2006mw} allows useful results on excited charm mesons to be obtained.
Moreover, results on charm meson spectroscopy obtained from DP analysis of $\Bd \to \Dzb\Kp\pim$ decays provide important independent cross-checks of results from studies of the $\Bz \to \Dzb\pip\pim$ DP, as possible biases due to other structures in the Dalitz plots are different between the two modes.

In this paper an amplitude analysis of the $\Bd \to \Dzb\Kp\pim$ decay is reported.
The inclusion of charge-conjugate processes is implied throughout the paper.
The analysis is based on a data sample corresponding to an integrated luminosity of $3.0  \,{\rm fb}^{-1}$ of $pp$ collision data collected with the LHCb detector in 2011 and 2012, when the collision centre-of-mass energy was $\sqrt{s} = 7 \tev$ ($1.0  \,{\rm fb}^{-1}$) and $\sqrt{s} = 8 \tev$ ($2.0  \,{\rm fb}^{-1}$), respectively.
Previously, the branching fraction for the three-body decay has been measured~\cite{Aubert:2005yt,LHCb-PAPER-2013-022}, and the $\Bd \to \Dzb\Kstar(892)^{0}$~\cite{Krokovny:2002ua,Aubert:2006qn} and $\Bd \to D^{*}_{2}(2460)^{-}\Kp$~\cite{Aubert:2005yt} contributions have been measured using quasi-two-body approaches; however this is the first DP analysis of the $\Bd \to \Dzb\Kp\pim$ decay.

The paper is organised as follows.
A description of the LHCb detector, reconstruction and simulation software is given in Sec.~\ref{sec:detector}.
The selection of signal candidates is described in Sec.~\ref{sec:selection},
and the determination of signal and background yields is presented in Sec.~\ref{sec:mass-fit}.
An overview of the Dalitz plot analysis formalism is given in Sec.~\ref{sec:dalitz-generalities}, and details of the implementation of the amplitude analysis are presented in Sec.~\ref{sec:dalitz}.
The evaluation of systematic uncertainties is described in Sec.~\ref{sec:systematics},
with results and a brief summary given in Sec.~\ref{sec:results}.

\section{LHCb detector and software}
\label{sec:detector}

The \lhcb detector~\cite{Alves:2008zz,LHCb-DP-2014-002} is a single-arm forward
spectrometer covering the \mbox{pseudorapidity} range $2<\eta <5$,
designed for the study of particles containing \bquark or \cquark
quarks. The detector includes a high-precision tracking system
consisting of a silicon-strip vertex detector~\cite{LHCb-DP-2014-001}
surrounding the $pp$ interaction region, a large-area silicon-strip detector
located upstream of a dipole magnet with a bending power of about
$4{\rm\,Tm}$, and three stations of silicon-strip detectors and straw
drift tubes~\cite{LHCb-DP-2013-003} placed downstream of the magnet.
The polarity of the dipole magnet is reversed periodically throughout data-taking.
The tracking system provides a measurement of momentum, \ptot, of charged particles with
a relative uncertainty that varies from 0.5\% at low momentum to 1.0\% at 200\gev (natural units with $c = \hbar = 1$ are used).
The minimum distance of a track to a primary vertex, the impact parameter (IP), is measured with a resolution of $(15+29/\pt)\mum$,
where \pt is the component of the momentum transverse to the beam, in \gev.
Different types of charged hadrons are distinguished using information
from two ring-imaging Cherenkov detectors~\cite{LHCb-DP-2012-003}.
Photon, electron and
hadron candidates are identified by a calorimeter system consisting of
scintillating-pad and preshower detectors, an electromagnetic
calorimeter and a hadronic calorimeter. Muons are identified by a
system composed of alternating layers of iron and multiwire
proportional chambers~\cite{LHCb-DP-2012-002}.

The trigger~\cite{LHCb-DP-2012-004} consists of a
hardware stage, based on information from the calorimeter and muon
systems, followed by a software stage, in which all tracks
with $\pt>500~(300)\mev$ are reconstructed for data collected in 2011 (2012).
The software trigger used in this analysis requires
a two-, three- or four-track secondary vertex with significant displacement
from any primary $pp$ interaction vertex~(PV). At least one charged particle
must have $\pt > 1.7\gev$ and be inconsistent with originating from a PV.
A multivariate algorithm~\cite{BBDT} is used for
the identification of secondary vertices consistent with the decay
of a \bquark hadron.

In the offline selection, the objects that prompted a positive trigger decision are associated with reconstructed particles.
Selection requirements can therefore be made not only on whether the hardware trigger decision was due to a signature in the calorimeters or in the muon system, but on whether the decision was due to the signal candidate, other particles produced in the $pp$ collision, or a combination of both.
Signal candidates are accepted offline if at least one of the final-state particles created a cluster in the hadronic calorimeter with sufficient transverse energy to fire the hardware trigger.
Events that are triggered at the hardware level by another particle in the event are also retained.
After all selection requirements are imposed, 67\,\% of events in the sample
were triggered by the decay products of the signal candidate, while the remainder were triggered only by another particle in the event.

Simulated events are used to characterise the detector response to signal and
certain types of background events.
In the simulation, $pp$ collisions are generated using
\pythia~\cite{Sjostrand:2006za,*Sjostrand:2007gs} with a specific \lhcb
configuration~\cite{LHCb-PROC-2010-056}.  Decays of hadronic particles
are described by \evtgen~\cite{Lange:2001uf}, in which final-state
radiation is generated using \photos~\cite{Golonka:2005pn}.
The interaction of the generated particles with the detector, and its response, are implemented using the \geant toolkit~\cite{Allison:2006ve, *Agostinelli:2002hh} as described in Ref.~\cite{LHCb-PROC-2011-006}.

\section{Selection requirements}
\label{sec:selection}

The selection requirements follow closely those used in Ref.~\cite{LHCb-PAPER-2014-036}.
The more copious $\Bd\to\Dzb\pip\pim$ decay is topologically and kinematically similar to the $\Bd\to\Dzb\Kp\pim$ channel, allowing it to be used as a control mode to optimise the selection requirements.
Loose initial requirements are used to obtain a visible signal peak of $\Dzb\pip\pim$ candidates.
The tracks are required to be of good quality and must be above thresholds in $p$, \pt and \chisqip, where \chisqip is defined as the difference in \chisq of a given PV reconstructed with and without the considered particle.
The $\Dzb\to\Kp\pim$ candidate must satisfy criteria on its vertex quality ($\chisq_{\rm vtx}$) and flight distance from any PV and from the $B$ candidate vertex, and must have invariant mass $m(\Kp\pim)$ in the range $1814$--$1914 \mev$.
A requirement on the output of a boosted decision tree that identifies $\Dzb \to \Kp\pim$ decays originating from \bquark hadron decays ($\Dzb$ BDT)~\cite{LHCb-PAPER-2012-025,LHCb-PAPER-2012-050} is also applied.
Candidate $B$ mesons are selected with requirements on invariant mass, $\chisqip$ and on the cosine of the angle between the $B$ momentum vector and the line from the PV under consideration to the $\B$ vertex ($\cos \theta_{\rm dir}$).
A requirement is placed on the $\chisq$ of a kinematic fit~\cite{Hulsbergen:2005pu} to the candidate's decay chain in which the $\Dzb$ mass is constrained to its nominal value.
The four final-state tracks must satisfy pion and kaon particle identification (PID) criteria.

A neural network~\cite{Feindt:2006pm} is used to discriminate between signal decays and combinatorial background.
The \sPlot\ technique~\cite{Pivk:2004ty}, with the \B candidate mass as the discriminating variable, is used to separate statistically $\Bd\to\Dzb\pip\pim$ decays from background.
The signal and background weights returned by this method are applied to the candidates, which are then used to train the network.
The network is trained using 16 variables.
They include the $\chisqip$ of the four final-state tracks and the following variables associated to the $\Dzb$ candidate:
$\chisqip$, $\chisq_{\rm vtx}$, the square of the flight distance from the PV divided by its uncertainty squared ($\chisq_{\rm flight}$), $\cos \theta_{\rm dir}$, and the output of the $\Dzb$ BDT.
In addition, the following variables associated to the $\B$ candidate are included:
$\pt$, $\chisqip$, $\chisq_{\rm vtx}$, $\chisq_{\rm flight}$, and $\cos \theta_{\rm dir}$.
The $\pt$ asymmetry~\cite{LHCb-PAPER-2014-036} and track multiplicity in a cone with half-angle of 1.5 units of the plane of pseudorapidity and azimuthal angle
(measured in radians) around the \B candidate flight direction, which contain information about the isolation of the \B candidate from the rest of the event, are also used.
The input quantities to the neural network depend only weakly on the position of the candidate in the $\B$ decay Dalitz plot and therefore any requirement on the network output cannot appreciably bias the DP distribution.
A requirement imposed on the network output reduces the combinatorial background remaining after the initial selection by a factor of five while retaining more than $90\,\%$ of the signal.

The $\Bd\to\Dzb\Kp\pim$ candidates must satisfy the same selection as the $\Bd\to\Dzb\pip\pim$ decays, except for the PID requirement on the positively charged track from the \B decay vertex, which is imposed to preferentially select kaons rather than pions.
The PID efficiency is evaluated from calibration samples of $\Dz \to \Km\pip$ decays from the $\Dstarp\to\Dz\pip$ decay chain.
The kinematics of this decay chain can be exploited to obtain clean samples without using the PID information~\cite{LHCb-DP-2012-003}.
The PID efficiency of the requirements on the four tracks in the
final state is around $50\,\%$ and varies depending on the kinematics of the
tracks, as described in detail in Sec.~\ref{sec:efficiency}.

Candidates are vetoed when the difference between $m(\Kp\pim\pim)$ and $m(\Kp\pim)$ lies within $\pm 2.5 \mev$ of the known $\Dstar(2010)^-$--$\Dzb$ mass difference~\cite{PDG2014} to remove background containing $\Dstarm \to \Dzb\pim$ decays.
Candidates are also rejected if a similar mass difference calculated with the pion mass hypothesis applied to the bachelor kaon satisfies the same criterion.
To reject backgrounds from $\Bd\to\Dm\Kp$, $\Dm\to\Kp\pim\pim$ decays, it is required that the combination of the pion from the $\Dzb$ candidate together with the two bachelor particles does not have an invariant mass in the range $1850$--$1890 \mev$.
Additionally, candidates are removed if the pion and kaon originating directly from the \Bd decay combine to give an invariant mass consistent with that of the $\Dz$ meson ($1835$--$1880 \mev$).
This removes candidates with the $\Dzb$ wrongly reconstructed as well as potential background from $\Bd\to\Dz\Dzb$ decays.
Other incorrectly reconstructed candidates are removed by vetoing candidates where the pion from the $\Dzb$ decay and the kaon originating directly from the \Bd decay give an invariant mass in the range $1850$--$1885\mev$.
At least one of the pion candidates is required to have no associated hits in the muon system to remove potential background from $\Bd\to\jpsi\Kstarz$ decays.
Charmless decays of $b$ hadrons are suppressed by the use of the $\Dzb$ BDT and further reduced to a negligible level by requiring that the $\Dzb$ candidate vertex is separated from the $\Bd$ decay vertex by at least $1\mm$.

Signal candidates are retained for further analysis if they have invariant mass in the range $5100$--$5900 \mev$.
After all selection requirements are applied, fewer than 1\,\% of selected events also contain a second candidate.
Such multiple candidates are retained and treated in the same manner as other candidates; the associated systematic uncertainty is negligible.

\section{Determination of signal and background yields}
\label{sec:mass-fit}

The signal and background yields are determined from an extended maximum likelihood fit to the $B$ candidate invariant mass distribution.
The fit allows for signal decays, combinatorial background and contributions from other $b$ hadron decays.
The decay chain $\Bd\to\Dstarzb\Kp\pim$ with $\Dstarzb\to\Dzb\gamma$ or $\Dzb\piz$ forms a partially reconstructed background that peaks at low $B$ candidate mass as the neutral particle is not included in the candidate.
Misidentified $\Bd \to \DorDstarzb\pip\pim$, $\Lbbar \to \DorDstarzb\Kp\antiproton$ and $\Bds \to \DorDstarzb\Kp\Km$ decays are found to contribute to the background.
A possible contribution from the highly suppressed $\Bs\to\Dzb\Kp\pim$ decay mode is also included in the fit.

The signal peak is modelled with the sum of two Crystal Ball~\cite{Skwarnicki:1986xj} functions, which have tails on opposite sides and which share a common mean.
The tail parameters are fixed to the values found in fits to simulated signal decays.
The relative normalisation and the ratio of widths of the two functions are constrained, within uncertainties, to the values determined in a fit to $\Bd\to\Dzb\pip\pim$ data.
The $\Bs\to\Dzb\Kp\pim$ shape is modelled identically to the signal peak, with the difference between the $\Bd$ and $\Bs$ mass peak positions fixed to its known value~\cite{PDG2014}.
An exponential shape is used to model the combinatorial background.

Misidentified and partially reconstructed backgrounds are modelled with smoothed non-parametric functions.
Simulated samples are used to obtain the shape for $\Bd\to\Dstarzb\Kp\pim$ decays, with the $\Dstarzb\to\Dzb\gamma$ and $\Dzb\piz$ contributions generated in the correct proportions~\cite{PDG2014}.
To account for differences between simulation and data, for example in the polarisation of the $\Dstarzb$ meson, the $B$ candidate invariant mass distribution for partially reconstructed $\Bd\to\Dstarzb\Kp\pim$ decays is allowed to be shifted by an offset that is a free parameter of the fit.
The misidentified background shapes are also obtained from simulated samples.
For $\Bd \to \DorDstarzb\pip\pim$ decays the $\Dzb$ and $\Dstarzb$ contributions are combined according to their relative branching fractions~\cite{PDG2014}.
The backgrounds from $\Lbbar \to \DorDstarzb\Kp\antiproton$ and $\Bds \to \DorDstarzb\Kp\Km$ decays are assumed to have equal branching fractions for the decays with $\Dzb$ and $\Dstarzb$ mesons in the final state, as the decays involving $\Dstarzb$ mesons have not yet been measured.
The simulated samples are reweighted according to the relevant particle identification and misidentification probabilities and to match known DP distributions~\cite{LHCb-PAPER-2014-070,LHCb-PAPER-2012-018,LHCb-PAPER-2013-056,Abe:2004cw}.
The (mis)identification probabilities take account of track kinematics and are calculated using calibration samples of $\Dstarp\to\Dz\pip$, $\Dz\to\Km\pip$ and $\Lz\to p\pim$ decays~\cite{LHCb-DP-2012-003,LHCb-DP-2014-002}.
The yields of the misidentified backgrounds relative to the signal are constrained, within uncertainties, to their expected values based on their known branching fractions~\cite{LHCb-PAPER-2014-070,LHCb-PAPER-2012-018,LHCb-PAPER-2013-056} and misidentification probabilities.

\begin{figure}[!tb]
\centering
\includegraphics[width=0.49\textwidth]{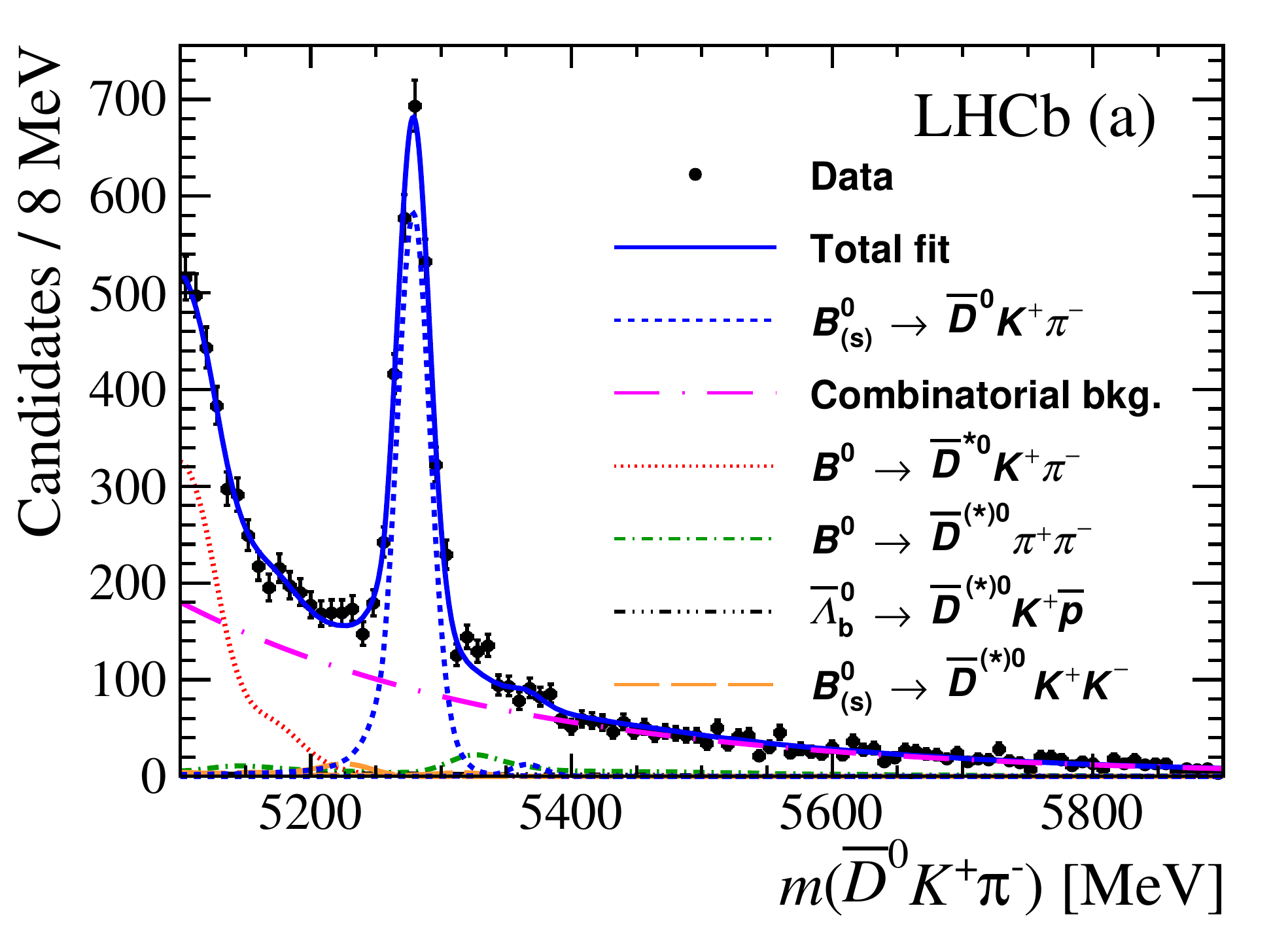}
\includegraphics[width=0.49\textwidth]{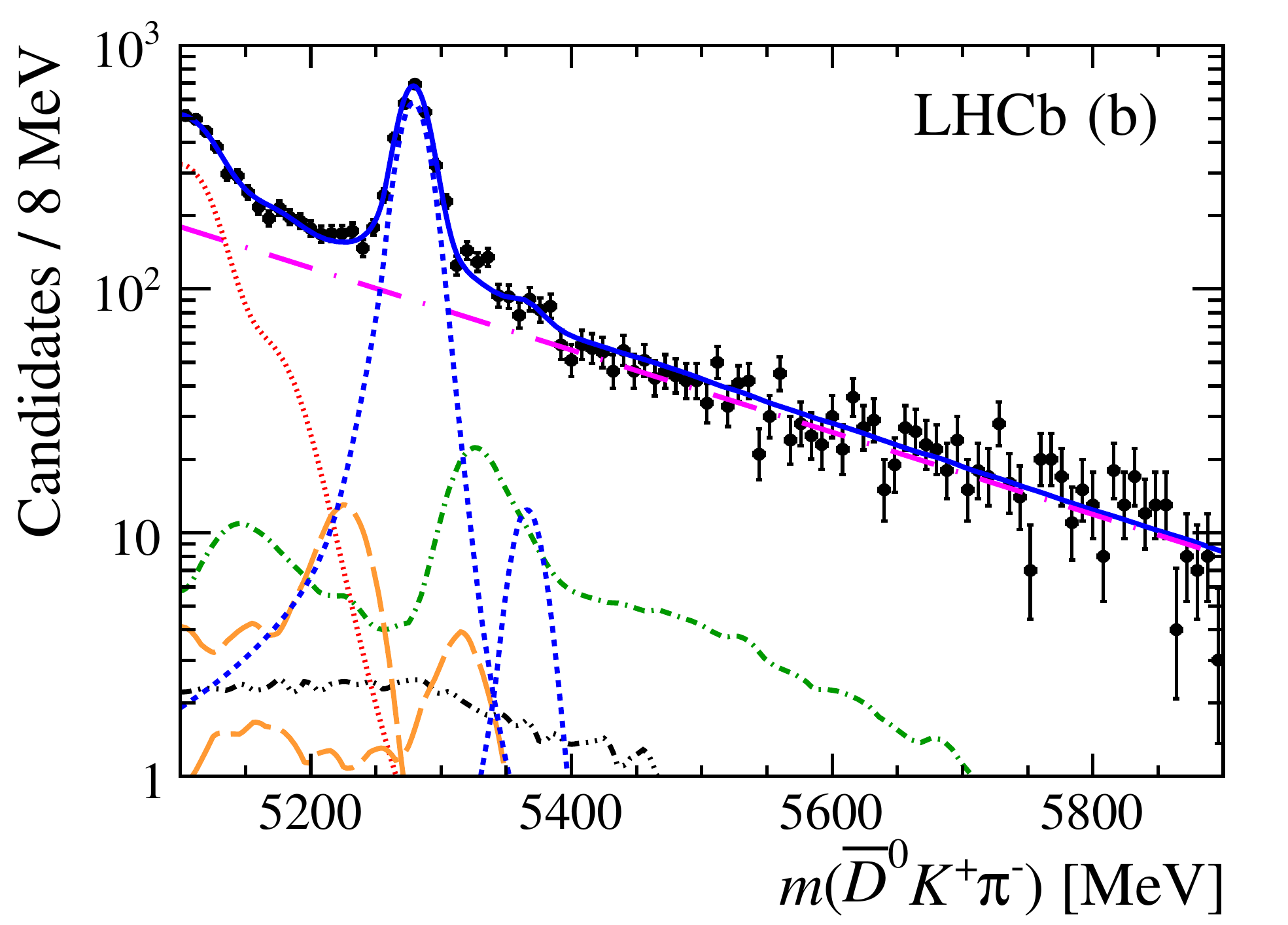}
\caption{\small
  Results of the fit to the \B candidate invariant mass distribution with (a) linear and (b) logarithmic $y$-axis scales. The components are as described in the legend.
}
\label{fig:massfit}
\end{figure}

\begin{figure}[!tb]
\centering
\includegraphics[width=0.49\textwidth]{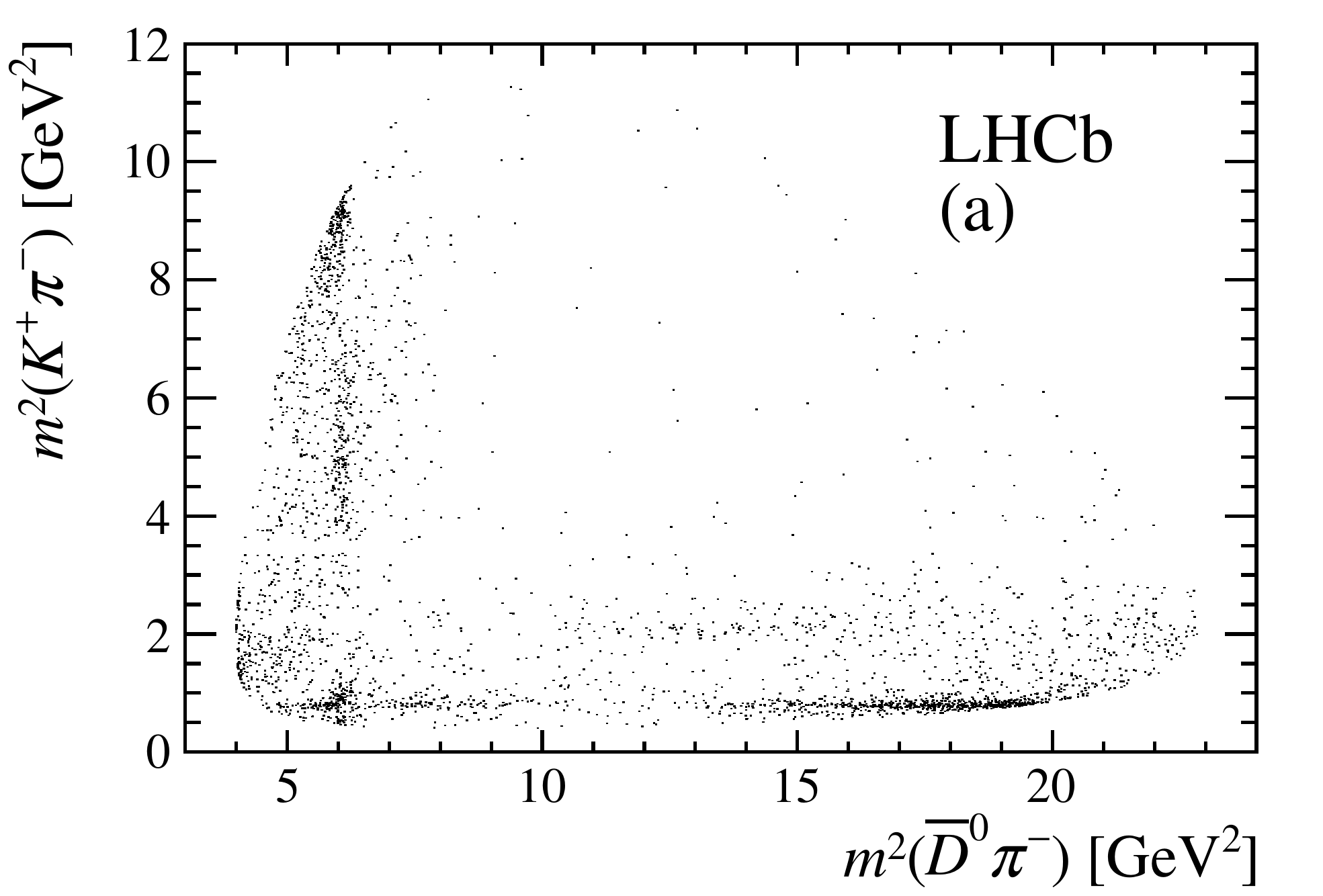}
\includegraphics[width=0.49\textwidth]{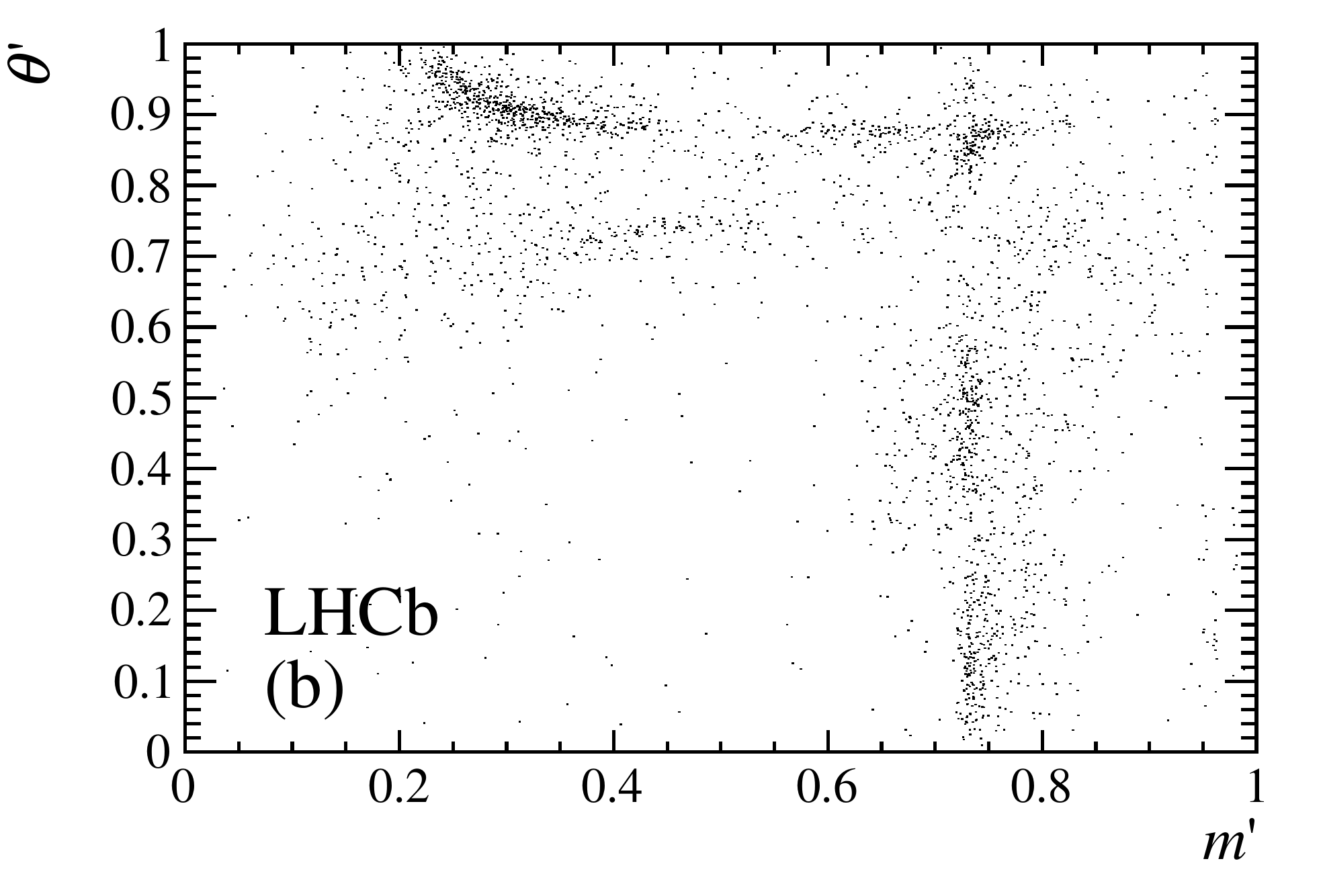}
\caption{\small
  Distribution of $\Bd \to \Dzb\Kp\pim$ candidates in the signal region over (a) the Dalitz plot and (b) the square Dalitz plot.
  The definition of the square Dalitz plot is given in Sec.~\ref{sec:efficiency}.
}
\label{fig:sigcands}
\end{figure}

There are 14 free parameters in the fit model: the mean and width of the signal shape, the relative normalisation and relative width of the two Crystal Ball functions, the slope of the exponential function, the offset of the $\Bd\to\Dstarzb\Kp\pim$ shape and the yields of the eight contributions.
The result of the fit is shown in Fig.~\ref{fig:massfit} and the yields are summarised in Table~\ref{tab:massfit}.
The yields are also reported within the signal region used in the Dalitz plot fit, corresponding to $\pm2.5$ widths of the \Bd signal shape ($5248.55$--$5309.05\mev$).
The distribution of candidates in the signal region over the Dalitz plot is shown in Fig.~\ref{fig:sigcands}(a).
Note that a $\Bd$ mass constraint is applied to calculate the variables that are used to describe the Dalitz plot~\cite{Dalitz:1953cp}, improving the resolution of those variables and giving a unique kinematic boundary.

\begin{table}[!tb]
  \centering
  \caption{\small
    Yields from the fit to the $\Dzb\Kp\pim$ data sample. The full mass range is $5100$--$5900\mev$ and the signal region is $5248.55$--$5309.05\mev$.
     }
  \label{tab:massfit}
  \vspace{1ex}
  \begin{tabular}{lcc}
    \hline
    Component & Full mass range & Signal region \\
    \hline  \\ [-2.5ex]
    $\Bd \to \Dzb\Kp\pim$             & $      2576 \pm 72\phani$      & $      2344 \pm 66$\\
    $\Bs \to \Dzb\Kp\pim$             & $\phanii 55 \pm 27\phani$      & $\phaniii 1 \pm 1\phani$ \\
    $\text{comb. bkgd.}$              & $      5540 \pm 187$           & $\phani 684 \pm 23$ \\
    $\Bd\to \Dstarzb \Kp\pim$         & $      1750 \pm 99\phani$      & $\phaniii 6 \pm 1\phani$ \\
    $\Bd\to \DorDstarzb\pip\pim$      & $\phani 485 \pm 47\phani$      & $\phanii 51 \pm 5\phani$ \\
    $\Lbbar\to \DorDstarzb\Kp\antiproton$ & $\phanii 95 \pm 26\phani$  & $\phanii 18 \pm 5\phani$ \\
    $\Bd\to \DorDstarzb\Kp\Km$        & $\phani 127 \pm 27\phani$      & $\phanii 10 \pm 2\phani$ \\
    $\Bs\to \DorDstarzb\Kp\Km$        & $\phanii 54 \pm 18\phani$      & $\phanii 14 \pm 5\phani$ \\
    \hline
  \end{tabular}
\end{table}

\section{Dalitz plot analysis formalism}
\label{sec:dalitz-generalities}

In $\Bd \to \Dzb\Kp\pim$ decays, resonances are expected in the $m^2(\Dzb\pim)$ and $m^2(\Kp\pim)$ combinations, so the Dalitz plot shown in Fig.~\ref{fig:sigcands}(a) is defined in terms of these two invariant mass squared terms.
For a fixed $\Bd$ mass, these two invariant mass squared combinations can be used to calculate all other relevant kinematic quantities.

The isobar model~\cite{Fleming:1964zz,Morgan:1968zza,Herndon:1973yn} is used to describe the complex decay amplitude.
The total amplitude is given by the coherent sum of amplitudes from resonant and nonresonant intermediate contributions
and is given by
\begin{equation}\label{eqn:amp}
  {\cal A}\left(m^2(\Dzb\pim), m^2(\Kp\pim)\right) = \sum_{j=1}^{N} c_j F_j\left(m^2(\Dzb\pim), m^2(\Kp\pim)\right) \,,
\end{equation}
where $c_j$ are complex coefficients describing the relative contribution for each intermediate process.
The resonant dynamics are contained in the $F_j\left(m^2(\Dzb\pim),m^2(\Kp\pim)\right)$ terms that are
normalised such that the integral of the squared magnitude over the DP is unity for each term.
For a $\Dzb\pim$ resonance $F_j\left(m^2(\Dzb\pim),m^2(\Kp\pim)\right)$ is given by
\begin{equation}
  \label{eq:ResDynEqn}
  F\left(m^2(\Dzb\pim), m^2(\Kp\pim)\right) =
  R\left(m(\Dzb\pim)\right) \times X(|\vec{p}\,|\,r_{\rm BW}) \times X(|\vec{q}\,|\,r_{\rm BW})
  \times T(\vec{p},\vec{q}\,) \, ,
\end{equation}
where $\vec{p}$ is the bachelor particle momentum and $\vec{q}$ is the momentum of one of the resonance daughters, both evaluated in the $\Dzb\pim$ rest frame.
The functions $R$, $X$ and $T$ are described below.

The $X(z)$ terms are Blatt--Weisskopf barrier factors~\cite{blatt-weisskopf}, where $z=|\vec{q}\,|\,r_{\rm BW}$ or $|\vec{p}\,|\,r_{\rm BW}$ and $r_{\rm BW}$ is the barrier radius which is set to $4.0\gev^{-1}\approx 0.8\fm$~\cite{LHCb-PAPER-2014-036} for all resonances.
The barrier factors are angular momentum dependent and are given by
\begin{equation}\begin{array}{rcl}
L = 0 \ : \ X(z) & = & 1\,, \\
L = 1 \ : \ X(z) & = & \sqrt{\frac{1 + z_0^2}{1 + z^2}}\,, \\
L = 2 \ : \ X(z) & = & \sqrt{\frac{z_0^4 + 3z_0^2 + 9}{z^4 + 3z^2 + 9}}\,,\\
L = 3 \ : \ X(z) & = & \sqrt{\frac{z_0^6 + 6z_0^4 + 45z_0^2 + 225}{z^6 + 6z^4 + 45z^2 + 225}}\,,
\end{array}\label{eq:BWFormFactors}\end{equation}
where $z_0$ is the value of $z$ at the pole mass of the resonance and $L$ is the orbital angular momentum between the resonance and the bachelor particle.
Since the parent and daughter particles all have zero spin, $L$ is also the spin of the resonance.

The $T(\vec{p},\vec{q})$ terms describe the angular distributions in the Zemach tensor formalism~\cite{Zemach:1963bc,Zemach:1968zz} and are given by
\begin{equation}\begin{array}{rcl}
L = 0 \ : \ T(\vec{p},\vec{q}) & = & 1\,,\\
L = 1 \ : \ T(\vec{p},\vec{q}) & = & -\,2\,\vec{p}\cdot\vec{q}\,,\\
L = 2 \ : \ T(\vec{p},\vec{q}) & = & \frac{4}{3} \left[3(\vec{p}\cdot\vec{q}\,)^2 - (|\vec{p}\,||\vec{q}\,|)^2\right]\,,\\
L = 3 \ : \ T(\vec{p},\vec{q}) & = & -\,\frac{24}{15} \left[5(\vec{p}\cdot\vec{q}\,)^3 - 3(\vec{p}\cdot\vec{q}\,)(|\vec{p}\,||\vec{q}\,|)^2\right]\,.
\end{array}\label{eq:ZTFactors}\end{equation}
These expressions are proportional to the Legendre polynomials, $P_L(x)$, where $x$ is the cosine of the angle between $\vec{p}$ and $\vec{q}$ (referred to as the helicity angle).

The $R\left(m(\Dzb\pim)\right)$ functions are the mass lineshapes.
Resonant contributions are typically described by the relativistic Breit--Wigner (RBW) function
\begin{equation}
\label{eq:RelBWEqn}
R(m) = \frac{1}{(m_0^2 - m^2) - i\, m_0 \Gamma(m)} \,,
\end{equation}
where the mass-dependent decay width is
\begin{equation}
\label{eq:GammaEqn}
\Gamma(m) = \Gamma_0 \left(\frac{q}{q_0}\right)^{2L+1}
\left(\frac{m_0}{m}\right) X^2(q\,r_{\rm BW}) \,,
\end{equation}
where $q_0$ is the value of $q = |\vec{q}\,|$ when the invariant mass is equal to the pole mass of the resonance, $m_0$.

The large phase space available in \B decays allows for the presence of nonresonant amplitudes (\ie\ contributions that do not proceed via a known resonance) that vary across the Dalitz plot.
An exponential form factor (EFF) has been found to describe nonresonant contributions well in several DP analyses of \B decays~\cite{Garmash:2004wa},
\begin{equation}
  R(m) = \exp\left[-\alpha m^2\right] \, ,
  \label{eq:nonres}
\end{equation}
where $\alpha$ is a shape parameter that must be determined from the data and $m$ is a two-body invariant mass ($m(\Dzb\pim)$ in this example).

The RBW function is a good model for narrow resonances that are well separated from any other resonant or nonresonant
contribution of the same spin.
This approach is known to break down in the $K\pi$ S-wave because the $\Kstarbsubz(1430)$ resonance interferes strongly with a slowly varying nonresonant term, as described in Ref.~\cite{Meadows:2007jm}.
The LASS lineshape~\cite{lass} has been developed to combine these two contributions,
\begin{eqnarray}
 \label{eq:LASSEqn}
  R(m) & = & \frac{m}{q \cot{\delta_B} - iq} + \exp\left[2i \delta_B\right]
  \frac{m_0 \Gamma_0 \frac{m_0}{q_0}}
       {(m_0^2 - m^2) - i m_0 \Gamma_0 \frac{q}{m} \frac{m_0}{q_0}}\, , \\
{\rm where} \ \cot{\delta_B} & = & \frac{1}{aq} + \frac{1}{2} r q \, ,
\end{eqnarray}
and where $m_0$ and $\Gamma_0$ are the pole mass and width of the $\Kstarbsubz(1430)$ state, and $a$ and $r$ are shape parameters.

The $D\pi$ S-wave nonresonant contribution can be described by the ``dabba'' lineshape~\cite{Bugg:2009tu}, defined as
\begin{equation}
  R(m) = \frac{B'(m^2)(m^2-s_A)\rho}{1-\beta(m^2-m^2_{\rm min}) - iB'(m^2)(m^2-s_A)\rho} \, ,
\end{equation}
where
\begin{equation}
B'(m^2)=b\exp\left[-\alpha(m^2-m^2_{\rm min})\right]\,.
\end{equation}
Here $m_{\rm min}$ is the invariant mass at threshold, $s_A = m^2_{D} - 0.5\,m^2_{\pi}$ is the Adler zero, $\rho$ is a phase-space factor and $b$, $\alpha$ and $\beta$ are parameters with values fixed to $24.49 \gev^{-2}$, $0.1 \gev^{-2}$ and $0.1 \gev^{-2}$, respectively, according to Ref.~\cite{Bugg:2009tu}.

Ignoring reconstruction and selection effects, the DP probability density function would be
\begin{equation}
\label{eq:SigDPLike}
{\cal{P}}_{\rm phys}\left(m^2(\Dzb\pim), m^2(\Kp\pim)\right) =
\frac
{|{\cal A}\left(m^2(\Dzb\pim), m^2(\Kp\pim)\right)|^2}
{\int\!\!\int_{\rm DP}~{|{\cal A}|^2}~dm^2(\Dzb\pim)\,dm^2(\Kp\pim)} \, ,
\end{equation}
where the dependence of ${\cal A}$ on the DP position has been suppressed in the denominator for brevity.
The primary results of most Dalitz plot analyses are the complex coefficients, as defined in Eq.~(\ref{eqn:amp}).
These, however, depend on the choice of phase convention, amplitude formalism and normalisation used in each analysis.
The convention-independent quantities of fit fractions and interference fit fractions provide a way to reliably compare results between different analyses.
Fit fractions are defined as the integral of the amplitude for each intermediate component squared, divided by that of the coherent matrix element squared for all intermediate contributions,
\begin{equation}
{\it FF}_j =
\frac
{\int\!\!\int_{\rm DP}\left|c_j F_j\left(m^2(\Dzb\pim), m^2(\Kp\pim)\right)\right|^2~dm^2(\Dzb\pim)\,dm^2(\Kp\pim)}
{\int\!\!\int_{\rm DP}\left|{\cal A}\right|^2~dm^2(\Dzb\pim)\,dm^2(\Kp\pim)} \, .
\label{eq:fitfraction}
\end{equation}
The fit fractions need not sum to unity due to possible net constructive or destructive interference, described by interference fit fractions defined (for $i<j$ only) by
\begin{equation}
  {\it FF}_{ij} =
  \frac
  {\int\!\!\int_{\rm DP} 2 \, \Real\left[c_ic_j^* F_iF_j^*\right]~dm^2(\Dzb\pim)\,dm^2(\Kp\pim)}
  {\int\!\!\int_{\rm DP}\left|{\cal A}\right|^2~dm^2(\Dzb\pim)\,dm^2(\Kp\pim)} \, ,
  \label{eq:intfitfraction}
\end{equation}
where the dependence of $F_i^{(*)}$ and ${\cal A}$ on the DP position has been omitted.

\section{Dalitz plot fit}
\label{sec:dalitz}

\subsection{Signal efficiency}
\label{sec:efficiency}

The variation of the efficiency across the phase space is studied in terms of the square Dalitz plot (SDP).
The SDP is defined by variables \mpr\ and \thpr\ that range between 0 and 1 and are given for the $\Dzb\Kp \pim$ case by
\begin{equation}
\label{eq:sqdp-vars}
\mpr \equiv \frac{1}{\pi} \arccos\left(2\frac{m(\Dzb\pim) - m^{\rm min}_{\Dzb\pim}}{m^{\rm max}_{\Dzb\pim} - m^{\rm min}_{\Dzb\pim}} - 1 \right)
\hspace{10mm}{\rm and}\hspace{10mm}
\thpr \equiv \frac{1}{\pi}\theta(\Dzb\pim)\,,
\end{equation}
where $m^{\rm max}_{\Dzb\pim} = m_{\Bd} - m_{\Kp}$ and $m^{\rm min}_{\Dzb\pim} = m_{\Dzb} + m_{\pim}$ are the kinematic boundaries of $m(\Dzb\pim)$ allowed in the $\Bd \to \Dzb\Kp\pim$ decay and $\theta(\Dzb\pim)$ is the helicity angle of the $\Dzb\pim$ system (the angle between the $\Kp$ and the $\Dzb$ meson in the $\Dzb\pim$ rest frame).
The distribution of selected events across the SDP is shown in Fig.~\ref{fig:sigcands}(b).

\begin{figure}[!tb]
\centering
 \includegraphics[width=0.49\textwidth]{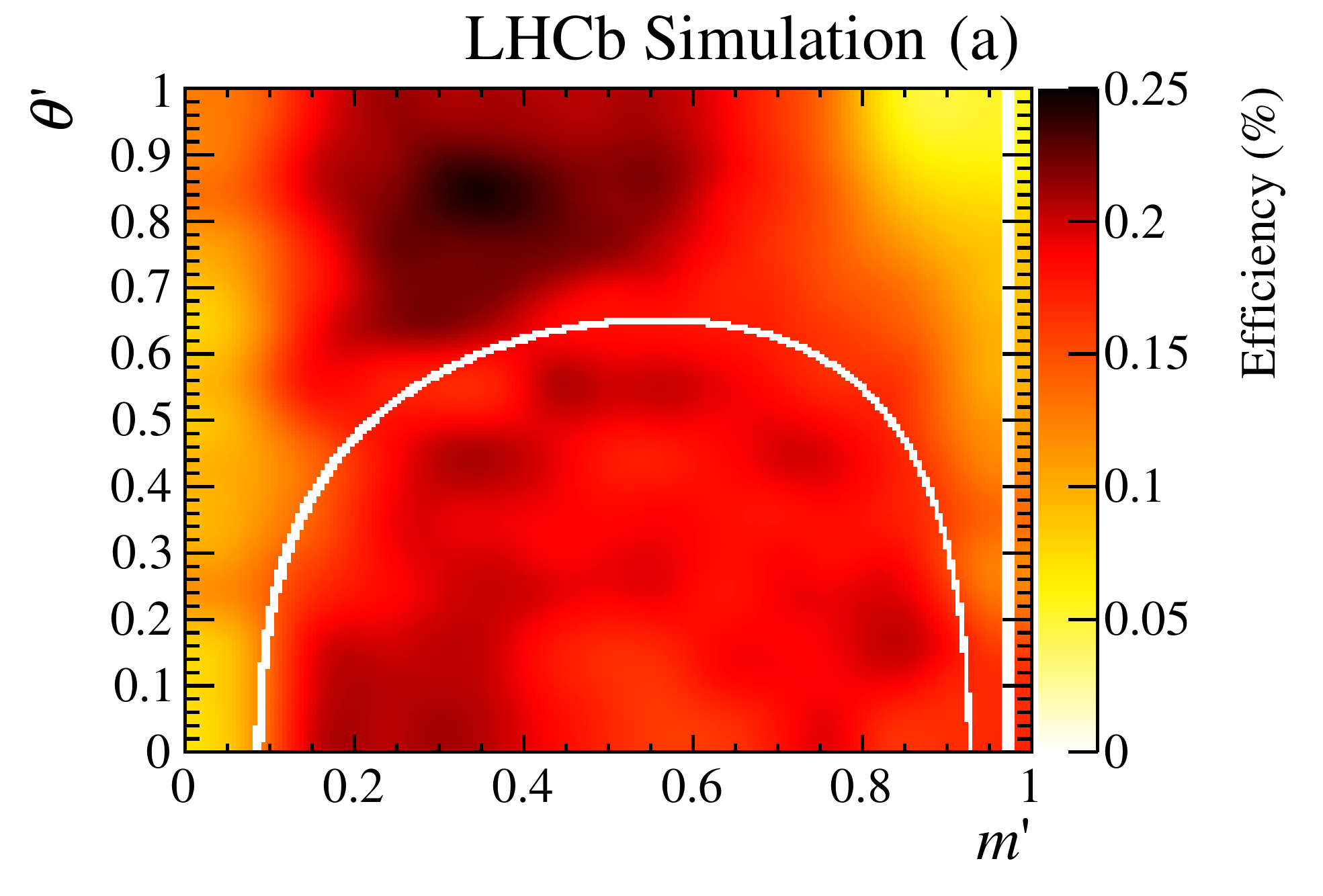}
 \includegraphics[width=0.49\textwidth]{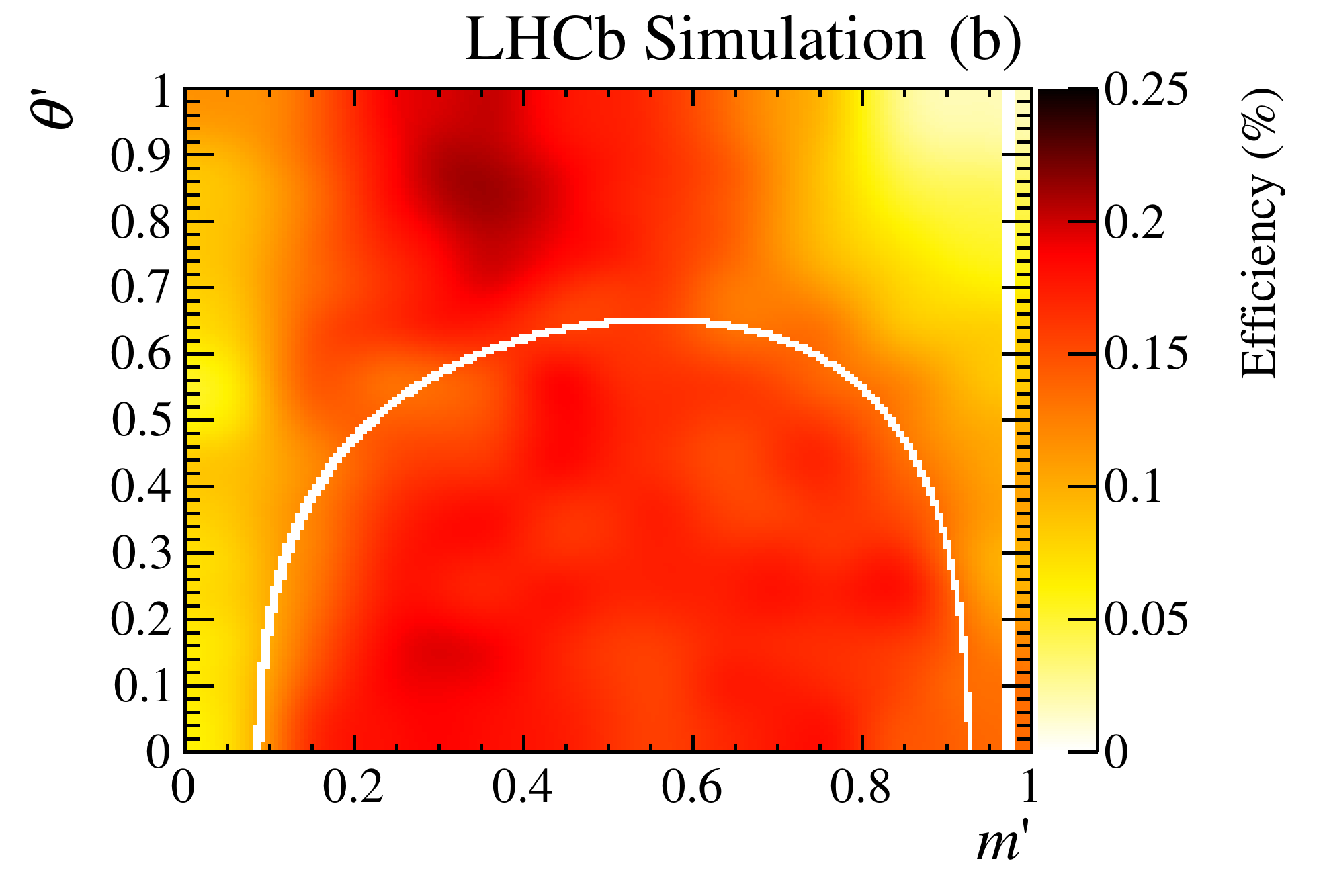}
\caption{Efficiency variation as a function of SDP position for candidates triggered by (a) signal decay products and (b) by the
rest of the event. The vertical white stripe is due to the \Dstar veto and the curved white band is due to the \Dzb veto.}
\label{fig:eff}
\end{figure}

Efficiency variation across the SDP is caused by the detector acceptance and by trigger, selection and PID requirements.
The efficiency is evaluated with simulated samples generated uniformly over the SDP.
Data-driven corrections are applied to correct for known differences between data and simulation in the tracking, trigger and PID efficiencies, using identical methods to those described in Ref.~\cite{LHCb-PAPER-2014-036}.
The efficiency functions are obtained by fitting the corrected simulation with two-dimensional cubic splines.
Figure~\ref{fig:eff} shows the histograms used to model the variation of the efficiency over the SDP for candidates triggered
by (a) signal decays and (b) by the rest of the event.

\subsection{Background studies}
\label{sec:backgrounds}

\begin{figure}[!tb]
\centering
\includegraphics[width=0.49\textwidth]{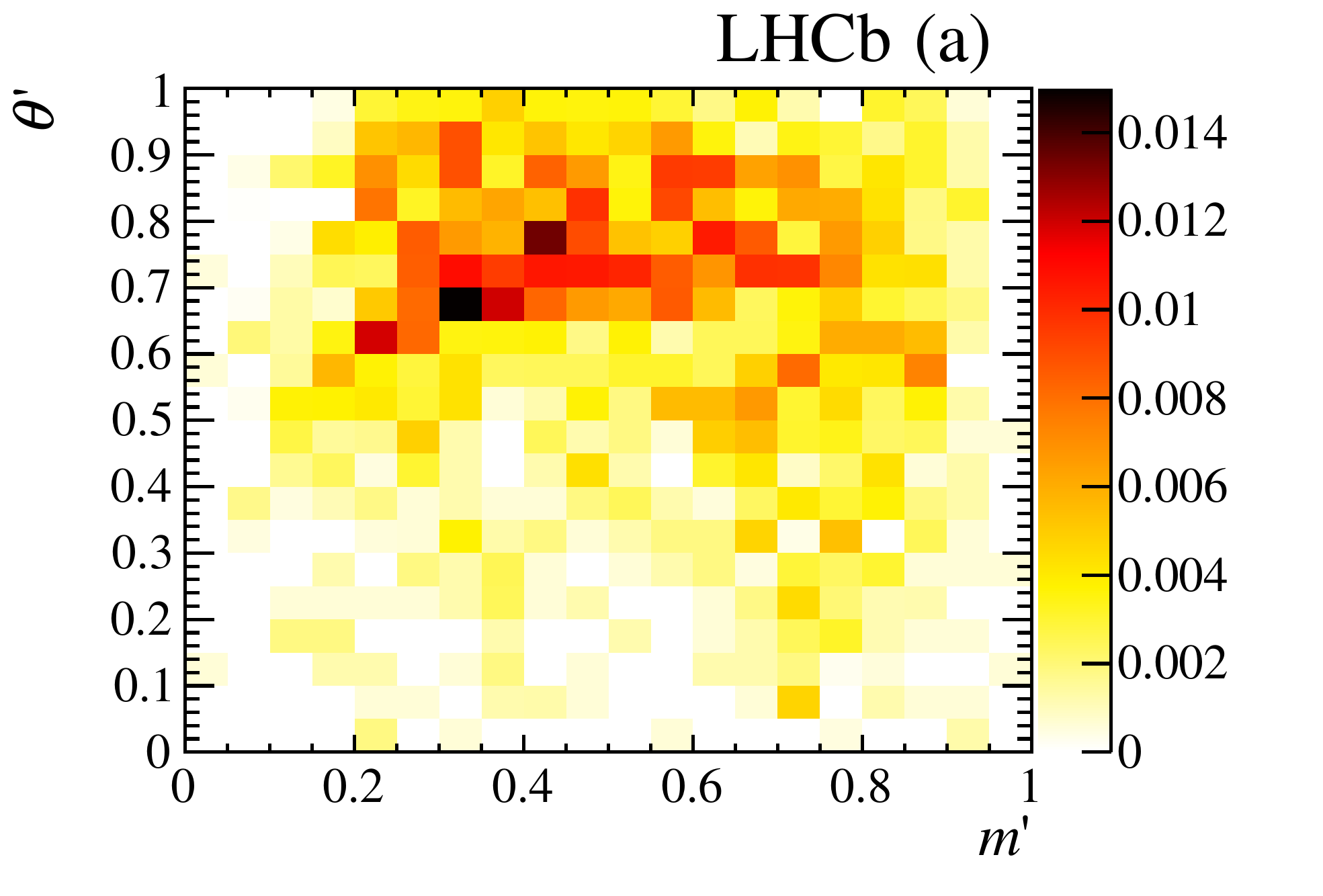}
\includegraphics[width=0.49\textwidth]{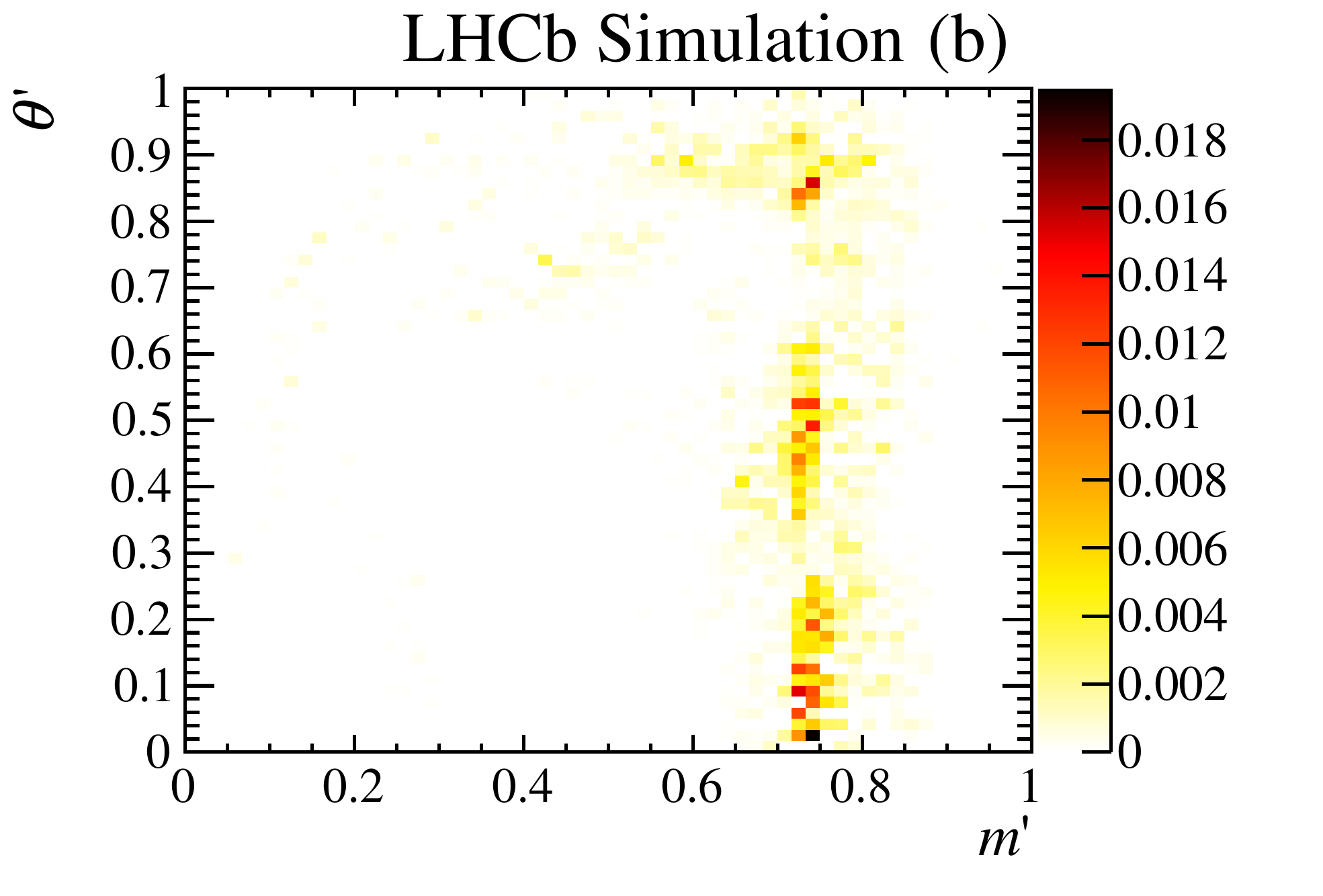}
\caption{SDP distributions of the background contributions from (a) combinatorial background and (b)
$\Bd\to \DorDstarzb\pip\pim$ decays.}
\label{fig:dpbkgs}
\end{figure}

Among the background yields in the signal region, given in Table~\ref{tab:massfit}, the only sizeable components are due to combinatorial background and $\Bd\to \DorDstarzb\pip\pim$ decays.
The SDP distributions of these backgrounds are described by the histograms shown in Fig.~\ref{fig:dpbkgs}.

Combinatorial background contributes $22\,\%$ of candidates in the signal region.
The shape of this contribution over the SDP is obtained from the high $B$ mass sideband ($5400$--$5900\mev$).
As seen in Fig.~\ref{fig:massfit},
this sideband is dominated by combinatorial background, with a small contribution from $\Bd\to \DorDstarzb\pip\pim$ decays.
The SDP distribution of the $\Bd\to \DorDstarzb\pip\pim$ contribution in the sideband region is modelled as described below for the signal region and is subtracted from the sideband sample.
To check that the SDP shape of the combinatorial background does not vary with the $B$ candidate mass, a sample of doubly charged $\Dzb\Kpm\pipm$ candidates is investigated.
This confirms that the SDP distribution from the sideband is a reliable model for combinatorial background in the signal region.

The $\Bd\to \DorDstarzb\pip\pim$ contribution accounts for $1.6\,\%$ of the candidates in the signal region.
Its SDP distribution is obtained from a simulated sample, with the $\Bd\to\Dzb\pip\pim$ and $\Bd\to\Dstarzb\pip\pim$ contributions combined and reweighted as described in Sec.~\ref{sec:mass-fit}.
The $\Bd\to\Dzb\pip\pim$ decay mode is the dominant component of this background in the signal region.

Due to resolution effects, $\Bd\to\Dstar(2010)^-\Kp$, $\Dstar(2010)^- \to \Dzb\pim$ decays are not entirely removed by the \Dstar veto.
Although this component corresponds to real $\Bd\to\Dzb\pim\Kp$ decays, it is treated as incoherent (\ie\ as background) since its interference with the remainder of the signal is negligible.
Its SDP distribution is modelled as a Gaussian peak in $m(\Dzb\pim)$, with mean fixed to the known $\Dstar(2010)^-$ mass and width and yield determined in the fit to data.

\subsection{Amplitude model for $\Bd\to\Dzb\Kp\pim$ decays}
The Dalitz plot fit is performed using the {\sc Laura++}~\cite{Laura++} package.
The likelihood function is given by
\begin{equation}
 {\cal L} =
 \prod_i^{N_c}
 \Bigg[
 \sum_k N_k {\cal P}_k\left(m^2_i(\Dzb\pim),m^2_i(\Kp\pim)\right)
 \Bigg] \,,
\end{equation}
where the index $i$ runs over $N_c$ candidates, $k$ runs over the signal and background components and $N_k$ is the yield in each component.
The probability density function for the signal component, ${\cal P}_{\rm sig}$, is given by Eq.~(\ref{eq:SigDPLike}) with
the $|{\cal A}\left(m^2(\Dzb\pim), m^2(\Kp\pim)\right)|^2$ terms multiplied by the efficiency function described in
Sec.~\ref{sec:efficiency}.
As it is possible for the minimisation procedure to find a local minimum of the negative log likelihood (NLL) profile,
the fit is repeated many times with randomised initial values to ensure the global minimum is found.

The nominal Dalitz plot fit model for $\Bd\to\Dzb\Kp\pim$ decays is composed of several resonant and nonresonant amplitudes.
Only those amplitudes that provide significant contributions or that aid the fit stability are included.
Unnatural spin-parity states are not considered as these do not decay to two pseudoscalars.
The eight amplitudes included in this model are listed in Table~\ref{tab:resonances}.
The width of the narrowest signal contribution to the Dalitz plot ($\sim 50 \mev$) is far larger than the mass resolution ($\sim 2.4\mev$); therefore, resolution effects are neglected.

\begin{table}[!tb]
\centering
\caption{\small
  Signal contributions to the fit model, where parameters and uncertainties are taken from Ref.~\cite{PDG2014}.
  The models are described in Sec.~\ref{sec:dalitz-generalities}.
}
\label{tab:resonances}
\resizebox{\textwidth}{!}{
\begin{tabular}{lcccc}
\hline
Resonance & Spin & DP axis & Model & Parameters \\
\hline \\ [-2.5ex]
$\Kstar(892)^{0}$ &1& $m^2(\Kp\pim)$ & RBW & $m_0 = 895.81 \pm 0.19 \mev$, $\Gamma_0 = 47.4 \pm 0.6 \mev$ \\
$\Kstar(1410)^{0}$ &1& $m^2(\Kp\pim)$ & RBW & $m_0 = 1414 \pm 15 \mev$, $\Gamma_0 = 232 \pm 21 \mev$ \\
$K^*_0(1430)^{0}$ &0& $m^2(\Kp\pim)$ & LASS & Determined from data (see text) \\
$K^*_2(1430)^{0}$ &2& $m^2(\Kp\pim)$ & RBW & $m_0 = 1432.4 \pm 1.3 \mev$, $\Gamma_0 = 109 \pm 5 \mev$  \\
$D^{*}_{0}(2400)^{-}$ &0& $m^{2}(\Dzb\pim)$ & RBW & \multirow{2}{*}{Determined from data (see Table~\ref{tab:masswidth})} \\
$D^{*}_{2}(2460)^{-}$ &2& $m^{2}(\Dzb\pim)$ & RBW \\
\hline \\ [-2.5ex]
Nonresonant &0& $m^{2}(\Dzb\pim)$ & dabba & Fixed (see text) \\
Nonresonant &1& $m^{2}(\Dzb\pim)$ & EFF & Determined from data (see text) \\
\hline \\ [-2.5ex]
\end{tabular}
}
\end{table}
The real and imaginary parts of the complex coefficients $c_j$ defined in Eq.~(\ref{eqn:amp}) are free parameters of the fit except for the coefficient of the $D^{*}_{2}(2460)^{-}$ component, which is fixed to 1 as a reference.
The phases and magnitudes of the complex coefficients, as well as fit fractions and interference fit fractions are derived from these free parameters.
In addition, the masses and widths of the $D^{*}_{0}(2400)^{-}$ and $D^{*}_{2}(2460)^{-}$ resonances are determined from the fit to data and are reported in Table~\ref{tab:masswidth}.
The statistical uncertainties on all parameters of interest are calculated using large samples of simulated pseudoexperiments.
These pseudoexperiments are also used to determine the correlations between the statistical uncertainties on the parameters, which are given in Appendix~\ref{app:correlations}.
The LASS parameters are determined to be $m_0 = 1450\pm80\mev$, $\Gamma_0 = 400\pm230\mev$, $a = 3.2\pm1.8\gev$ and $r = 0.9\pm1.1\gev$, while the parameter of the EFF lineshape of the $D\pi$ P-wave nonresonant amplitude is determined to be $\alpha = 0.88\pm0.10 \gev^{-2}$.

\begin{table}[!tb]
\centering
\caption{\small Masses and widths $(\mevnsp)$ determined in the fit to data, with statistical uncertainties only.}
\label{tab:masswidth}
\begin{tabular}{lcc}
\hline
Resonance & Mass & Width \\
\hline \\ [-2.5ex]
$D^{*}_{0}(2400)^{-}$ & $ 2360 \pm 15 $ & $ 255 \pm 26 $\\
$D^{*}_{2}(2460)^{-}$ & $ 2465.6 \pm 1.8\phantom{1} $ & $ 46.0 \pm 3.4 $\\
\hline
\end{tabular}
\end{table}

The values of the fit fractions and complex coefficients obtained from the fit are shown in Table~\ref{tab:ffstat},
while the values of the interference fit fractions are given in Appendix~\ref{app:IFF-results}.
The sum of the fit fractions is seen to exceed unity, mostly due to interference within the $D\pi$ and $K\pi$ S-waves.
Note that in Table~\ref{tab:ffstat}, and all results for fit fractions and derived quantities, values are reported separately for both the $\kaon^*_0(1430)^{0}$ and nonresonant components of the LASS lineshape, as well as their coherent sum.

\begin{table}[!tb]
\centering
\caption{\small Complex coefficients and fit fractions determined from the
  Dalitz plot fit. Uncertainties are statistical only. Note that the fit fractions, magnitudes and phases are derived quantities.
}
\label{tab:ffstat}
\resizebox{\textwidth}{!}{
\begin{tabular}{lccccc}
\hline
Resonance & Fit fraction & \multicolumn{4}{c}{Isobar model coefficients} \\
& (\%) & Real part & Imaginary part & Magnitude & Phase \\
\hline \\ [-2.5ex]
$\Kstar(892)^{0}$          & $37.4 \pm 1.5$     & $          -0.00 \pm 0.15$ & $          -1.27 \pm 0.06$ & $1.27 \pm 0.06$ & $          -1.57 \pm 0.11$\\
$\Kstar(1410)^{0}$         & $\phani0.7 \pm 0.3$& $\phantom{-}0.15 \pm 0.06$ & $          -0.09 \pm 0.09$ & $0.18 \pm 0.07$ & $          -0.54 \pm 0.21$\\
$\kaon^*_0(1430)^{0}$      & $\phani5.1 \pm 2.0$& $\phantom{-}0.14 \pm 0.38$ & $\phantom{-}0.45 \pm 0.15$ & $0.47 \pm 0.09$ & $\phantom{-}1.27 \pm 0.95$\\
LASS nonresonant           & $\phani4.8 \pm 3.8$& $          -0.10 \pm 0.24$ & $\phantom{-}0.44 \pm 0.14$ & $0.46 \pm 0.14$ & $\phantom{-}1.79 \pm 0.65$\\
\ \ \ LASS total           & $\phani6.7 \pm 2.7$& & & \\
$\kaon^*_{2}(1430)^{0}$    & $\phani7.4 \pm 1.7$& $          -0.32 \pm 0.09$ & $          -0.47 \pm 0.07$ & $0.57 \pm 0.05$ & $          -2.16 \pm 0.19$\\
$D^{*}_{0}(2400)^{-}$      & $19.3 \pm 2.8$     & $          -0.80 \pm 0.08$ & $          -0.44 \pm 0.14$ & $0.91 \pm 0.07$ & $          -2.64 \pm 0.15$\\
$D^{*}_{2}(2460)^{-}$      & $23.1 \pm 1.2$     & $\phantom{-}1.00$ & $\phantom{-}0.00$ & $1.00$ & $\phantom{-}0.00$\\
\hline
$D\pi$ S-wave (dabba)      & $\phani6.6 \pm 1.4$ &$          -0.39 \pm 0.09$ & $\phantom{-}0.36 \pm 0.17$ & $0.53 \pm 0.07$ & $\phantom{-}2.40 \pm 0.27$\\
$D\pi$ P-wave (EFF)        & $\phani8.9 \pm 1.6$ &$          -0.62 \pm 0.06$ & $          -0.03 \pm 0.06$ & $0.62 \pm 0.06$ & $          -3.09 \pm 0.10$\\
\hline
Total fit fraction         &113.4 & & & \\
\hline
\end{tabular}
}
\end{table}

\begin{figure}[!tb]
\centering
\includegraphics[width=0.47\textwidth]{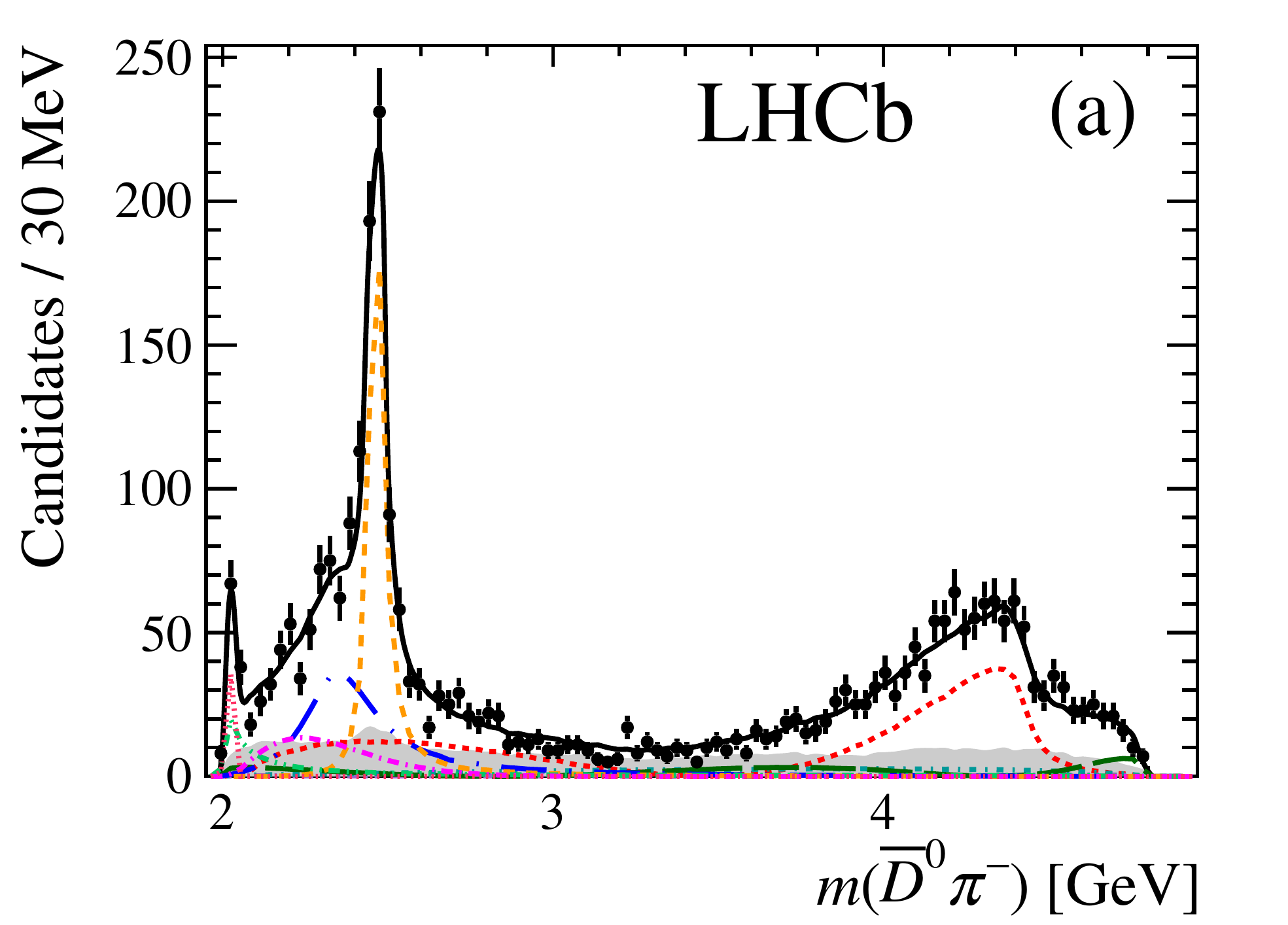}
\includegraphics[width=0.47\textwidth]{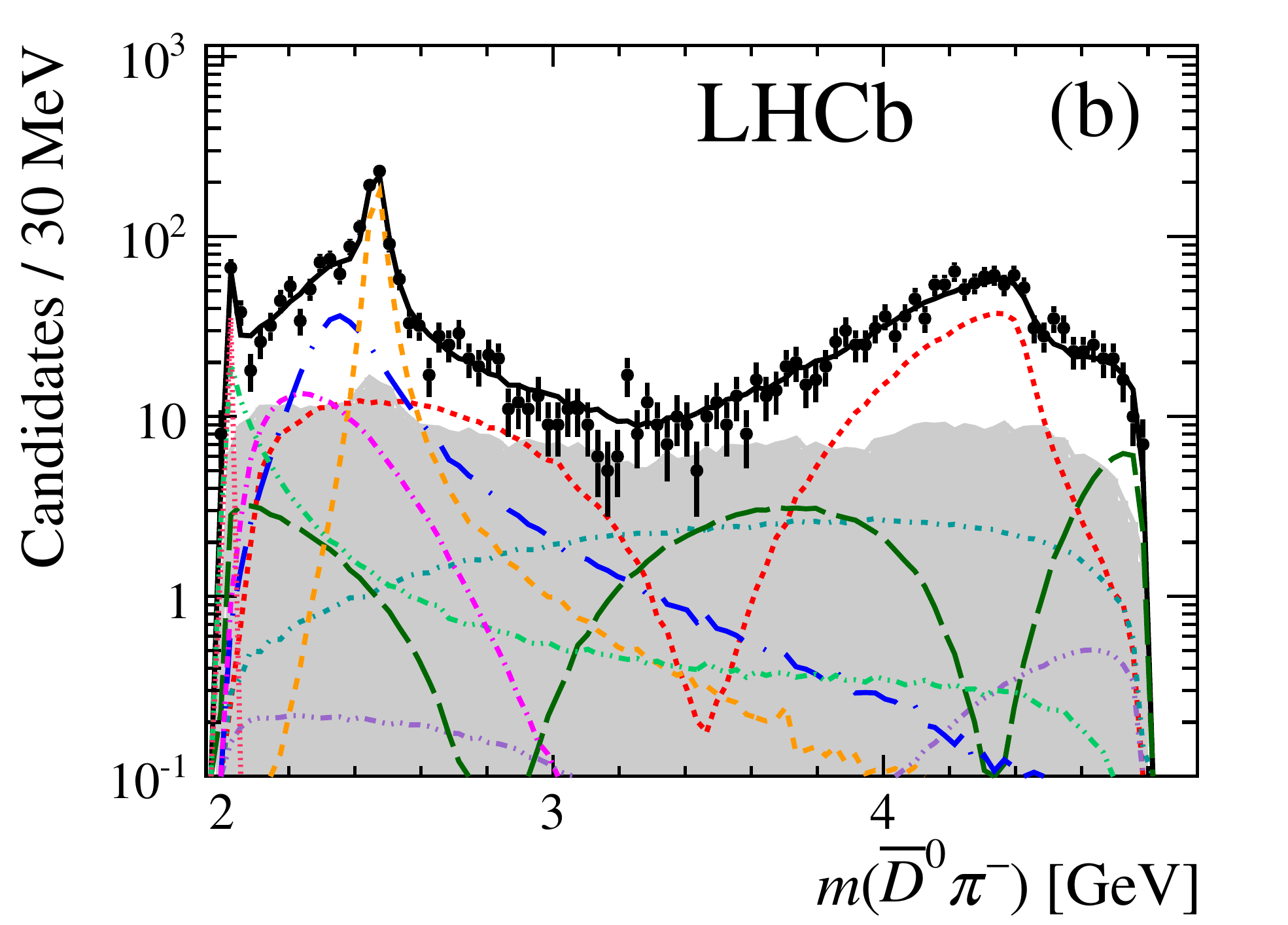}
\includegraphics[width=0.47\textwidth]{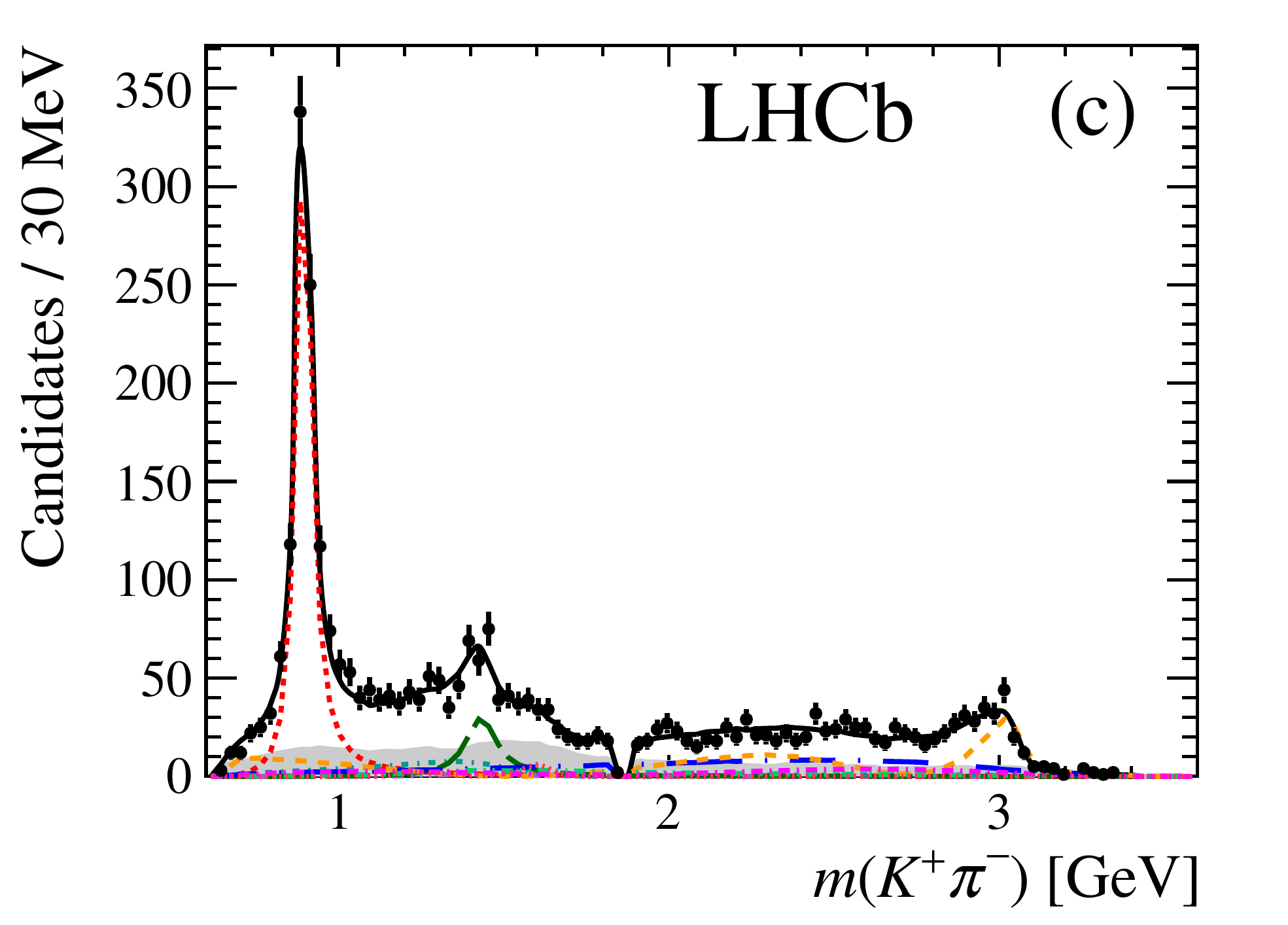}
\includegraphics[width=0.47\textwidth]{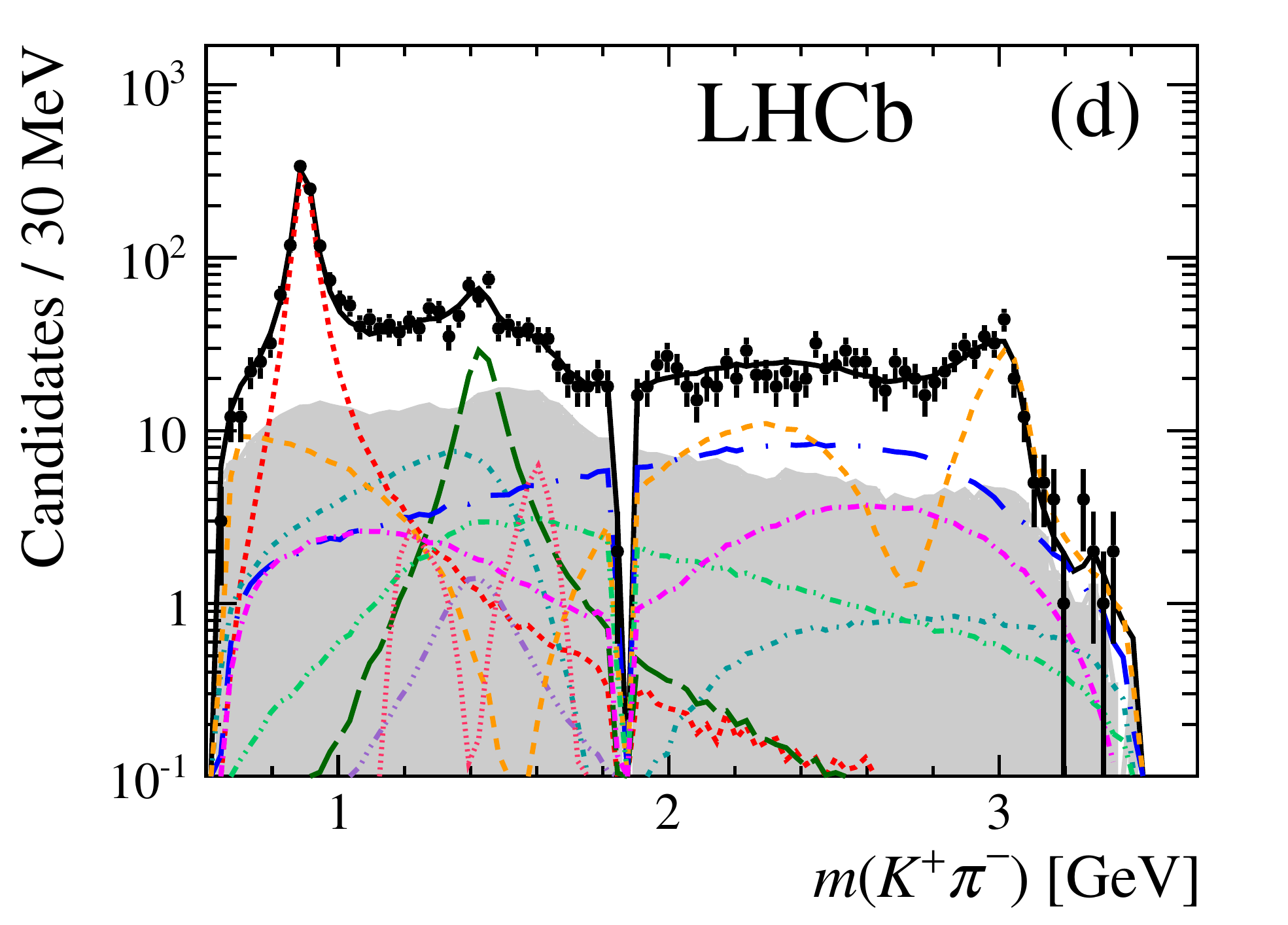}
\includegraphics[width=0.47\textwidth]{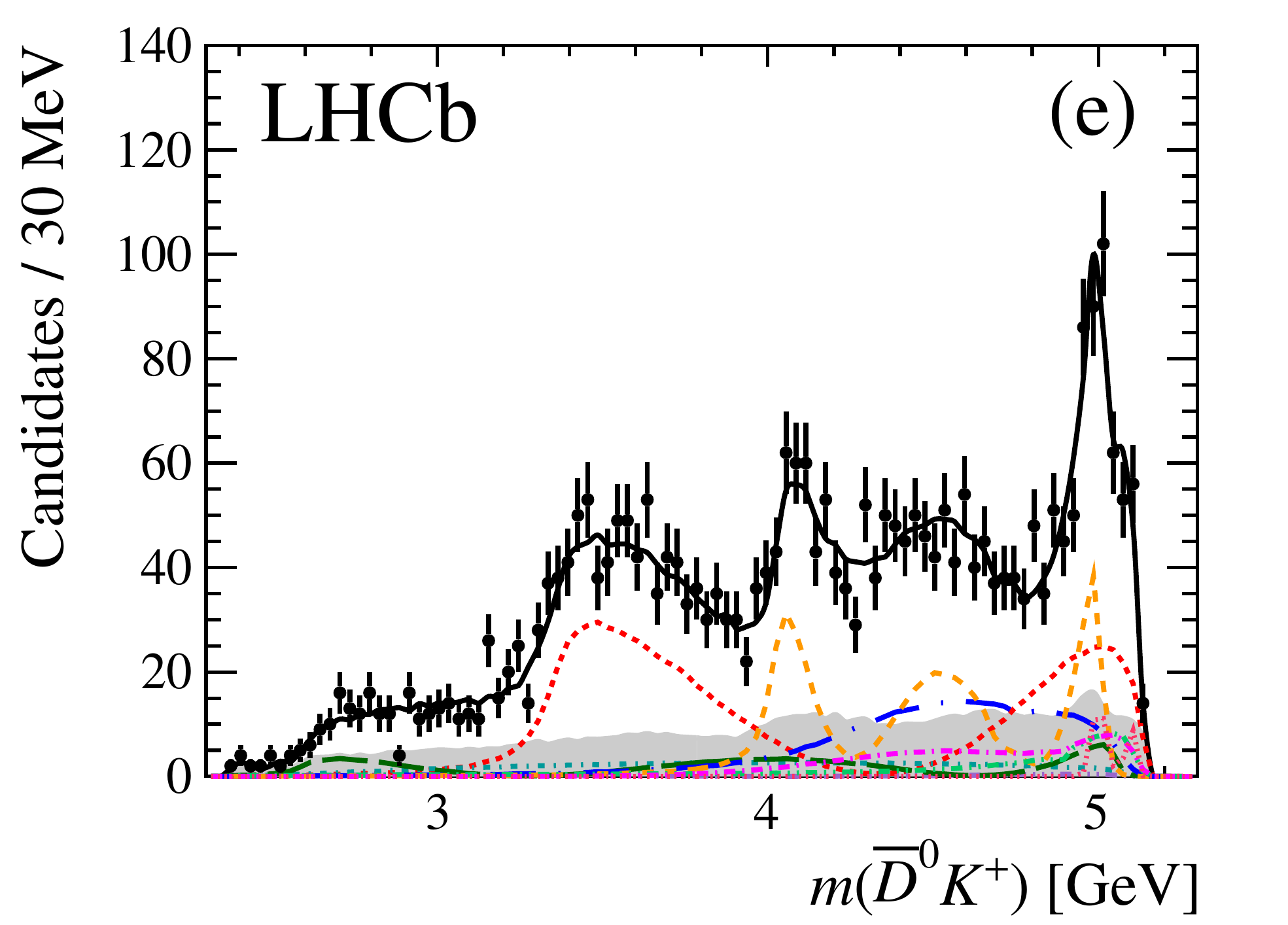}
\includegraphics[width=0.47\textwidth]{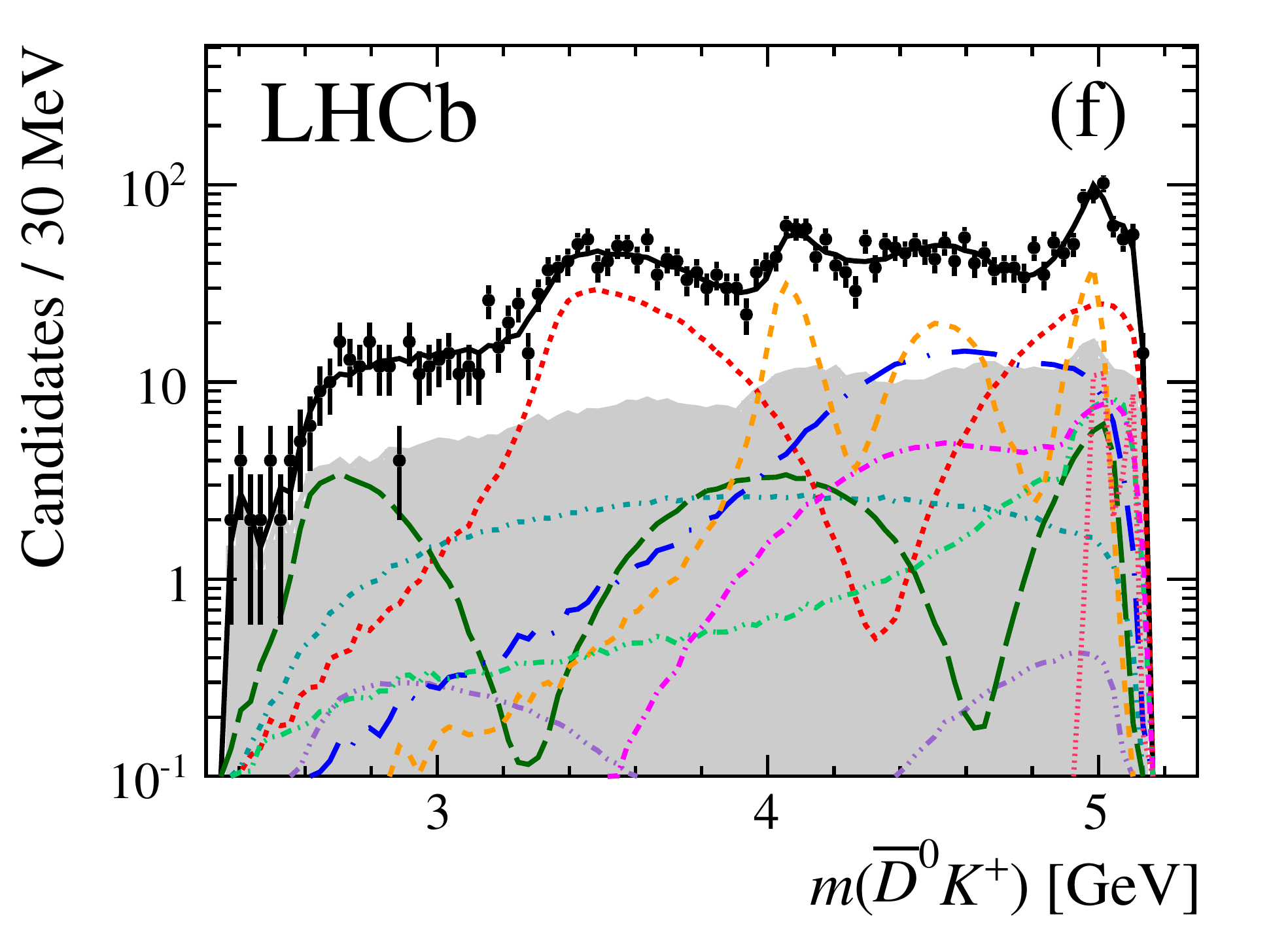}
\includegraphics[width=0.98\textwidth]{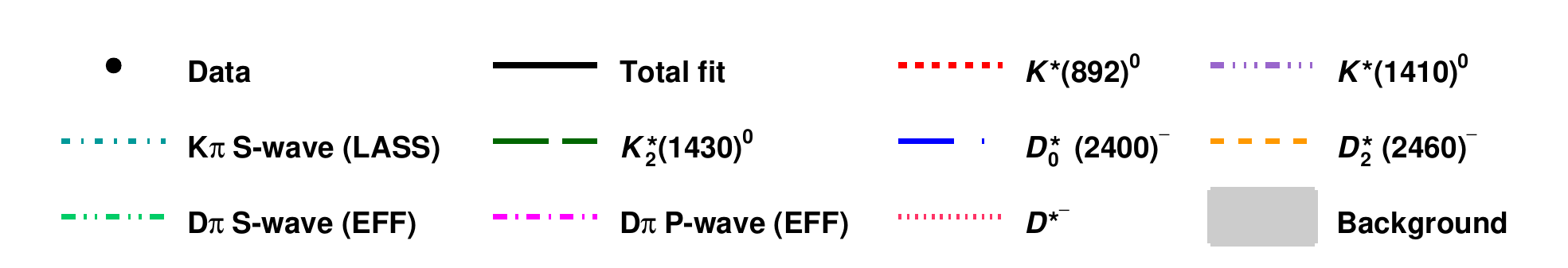}
\caption{\small Projections of the data and amplitude fit results onto (a)~$m(\Dzb\pim)$, (c)~$m(\Kp\pim)$ and (e)~$m(\Dzb\Kp)$, with the same
projections shown in (b), (d) and (f) with a logarithmic $y$-axis scale. Components are described in the legend.
}
\label{fig:fitproj}
\end{figure}

\begin{figure}[!tb]
\centering
\includegraphics[width=0.49\textwidth]{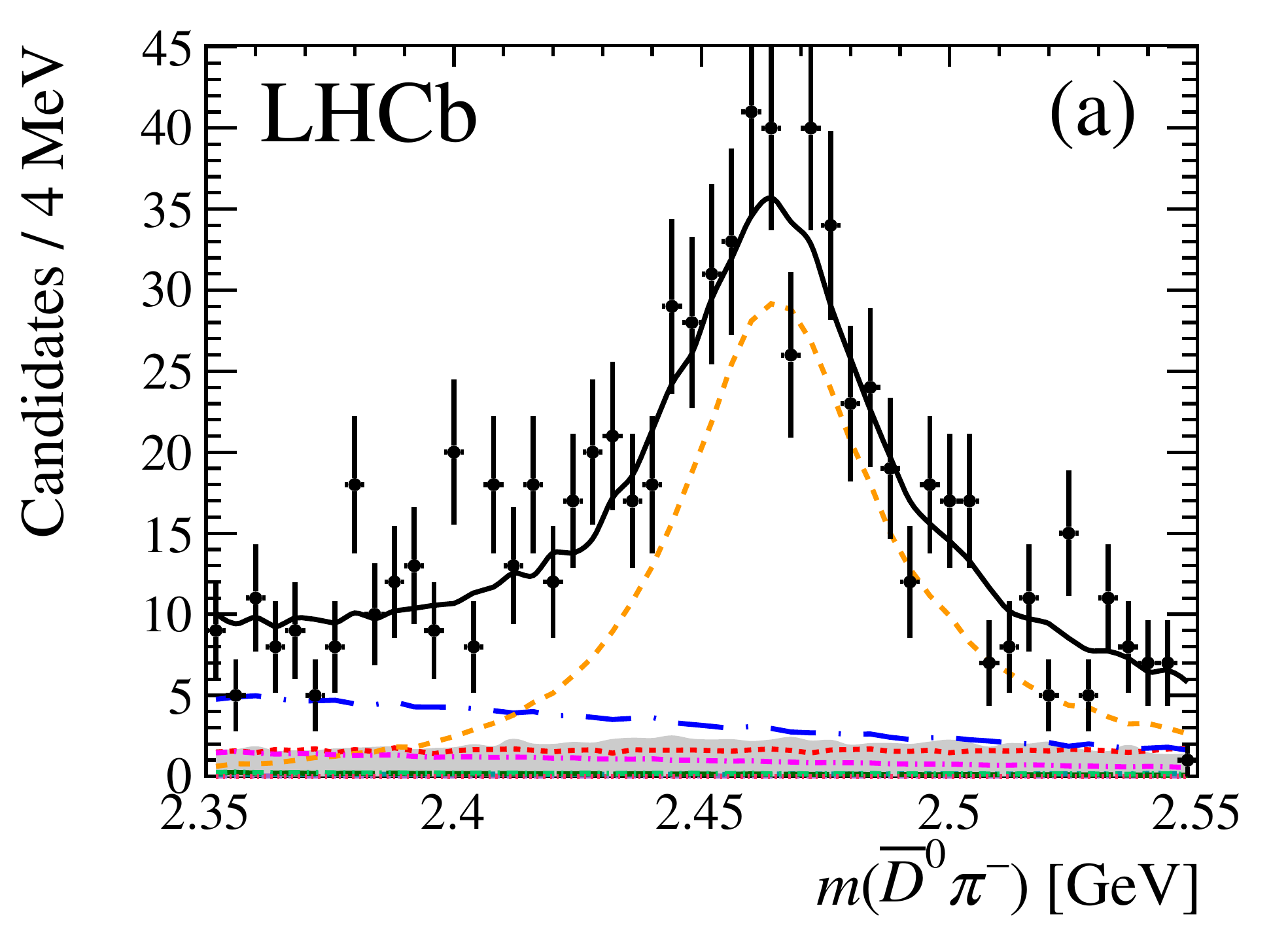}
\includegraphics[width=0.49\textwidth]{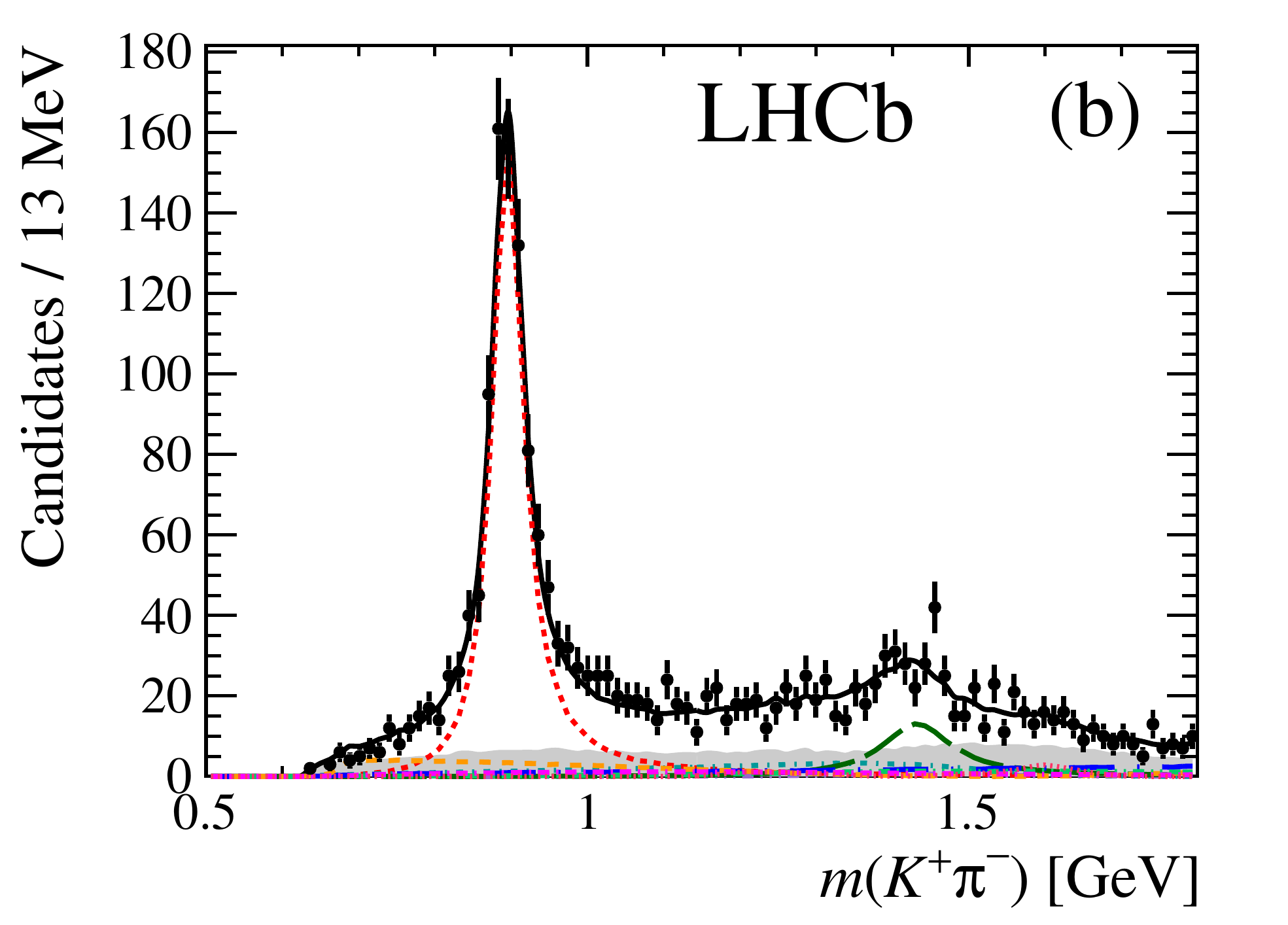}
\caption{\small Projections of the data and amplitude fit results onto (a)~$m(\Dzb\pim)$ in the $D^*_2(2460)^-$ region and (b)~the low $m(\Kp\pim)$ region. Components are as shown in Fig.~\ref{fig:fitproj}.
}
\label{fig:fitprojzoom}
\end{figure}

\begin{figure}[!tb]
\centering
\includegraphics[width=0.49\textwidth]{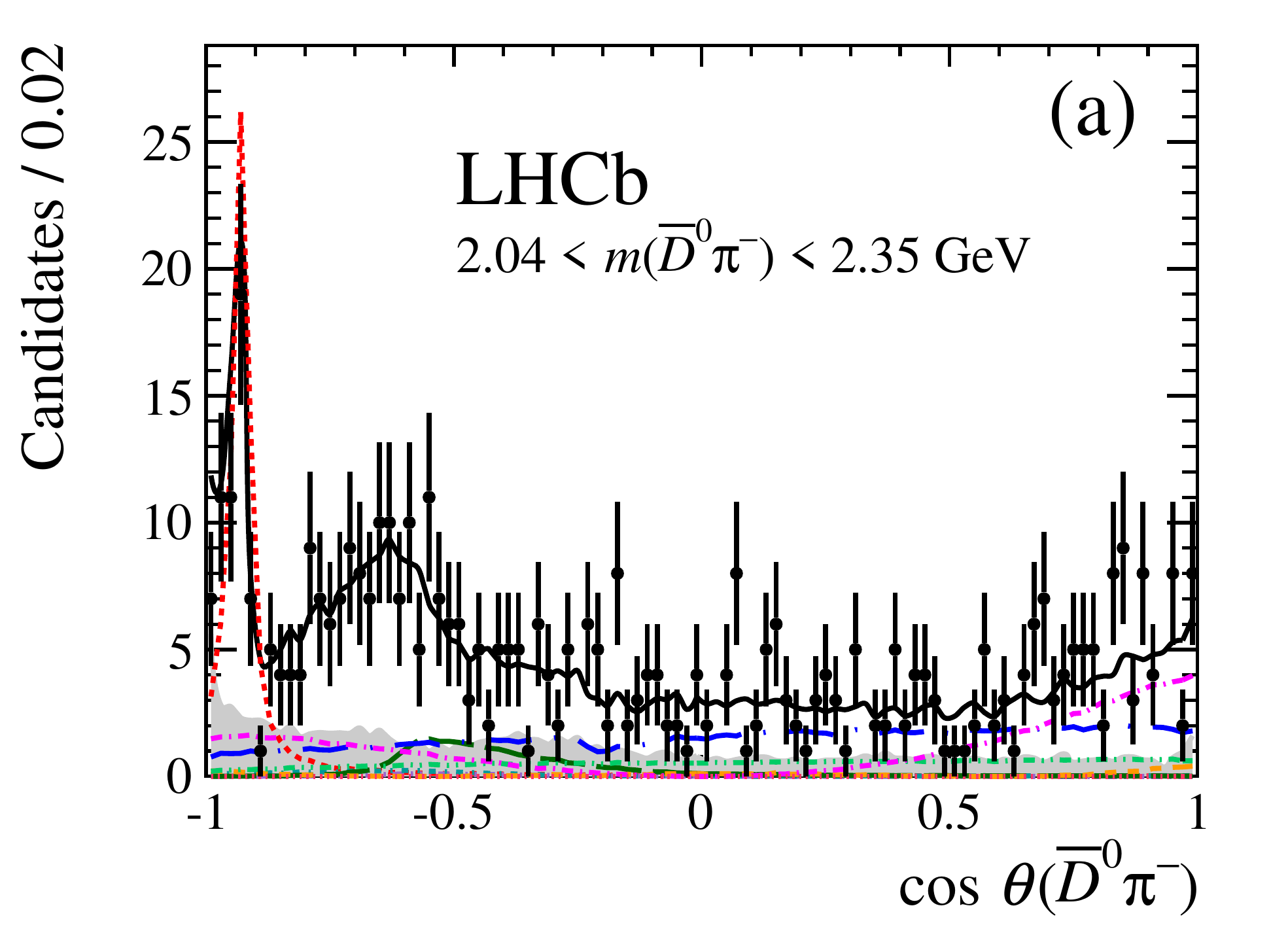}
\includegraphics[width=0.49\textwidth]{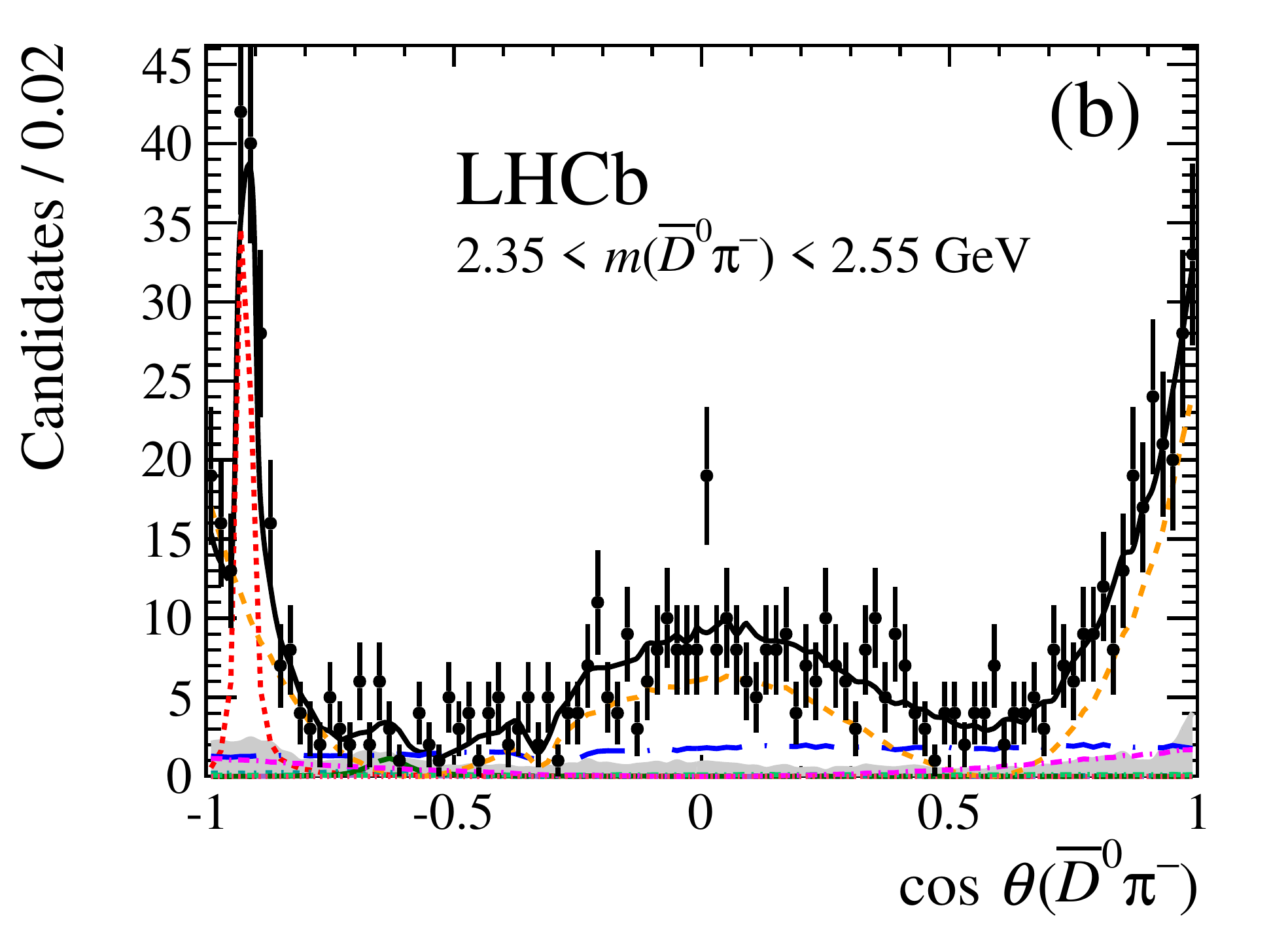}
\caption{\small Projections of the data and amplitude fit results onto $\cos \theta(\Dzb\pim)$ in the mass ranges (a)~$2.04 < m(\Dzb\pim) < 2.35 \gev$ and (b)~$2.35 < m(\Dzb\pim) < 2.55 \gev$.
Components are as shown in Fig.~\ref{fig:fitproj}.
}
\label{fig:fitprojcosdpi}
\end{figure}

\begin{figure}[!tb]
\centering
\includegraphics[width=0.49\textwidth]{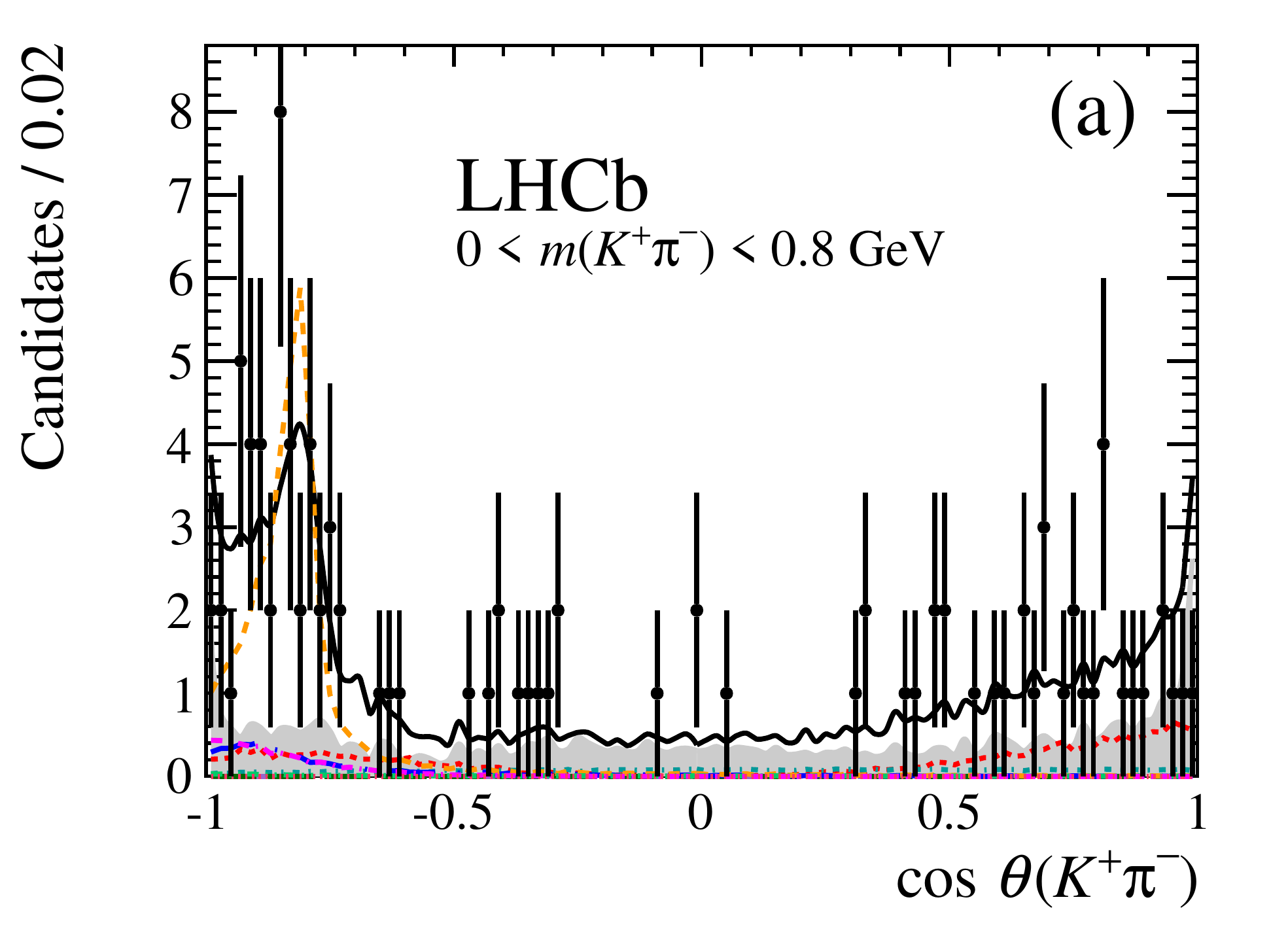}
\includegraphics[width=0.49\textwidth]{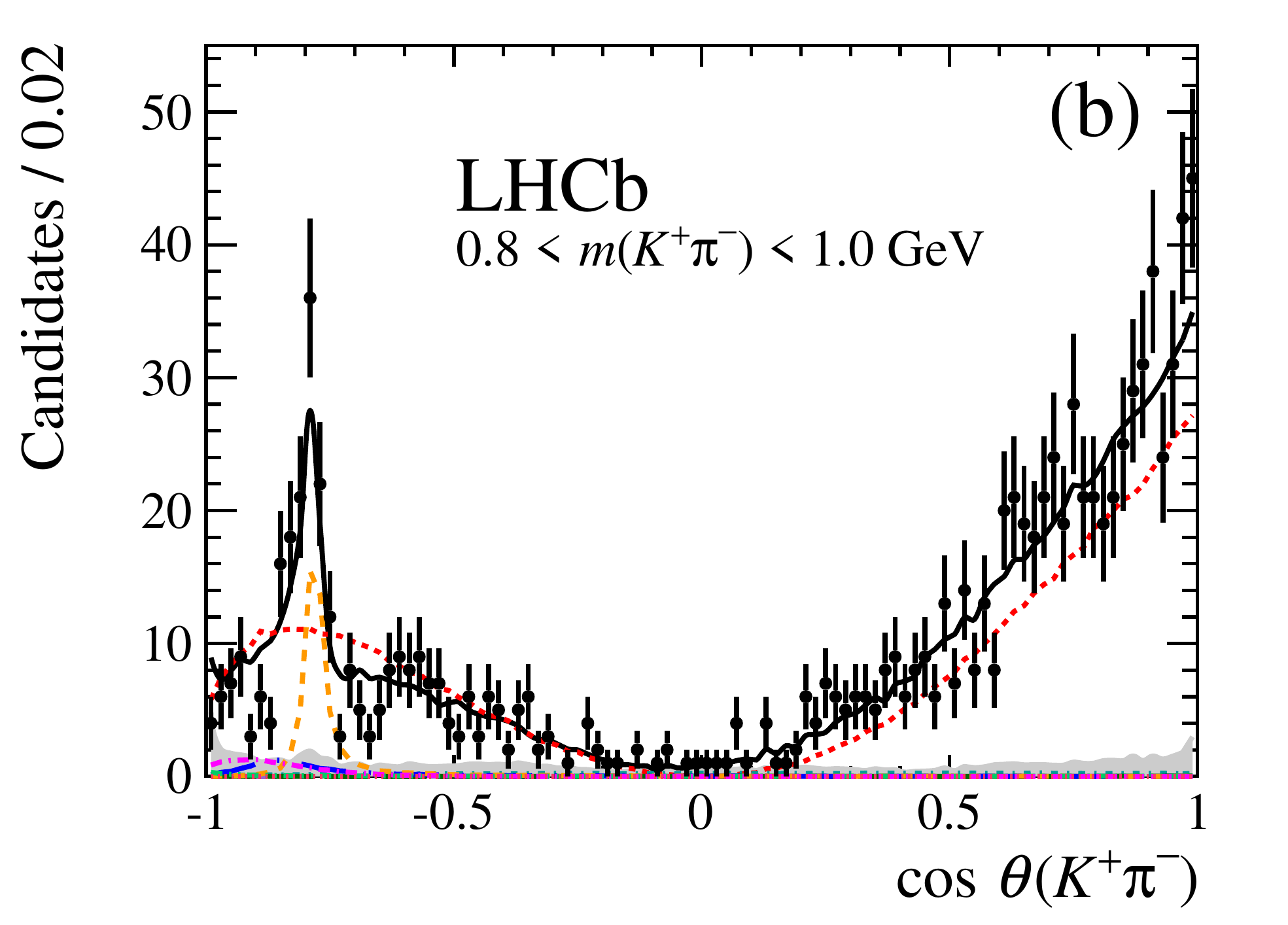}
\includegraphics[width=0.49\textwidth]{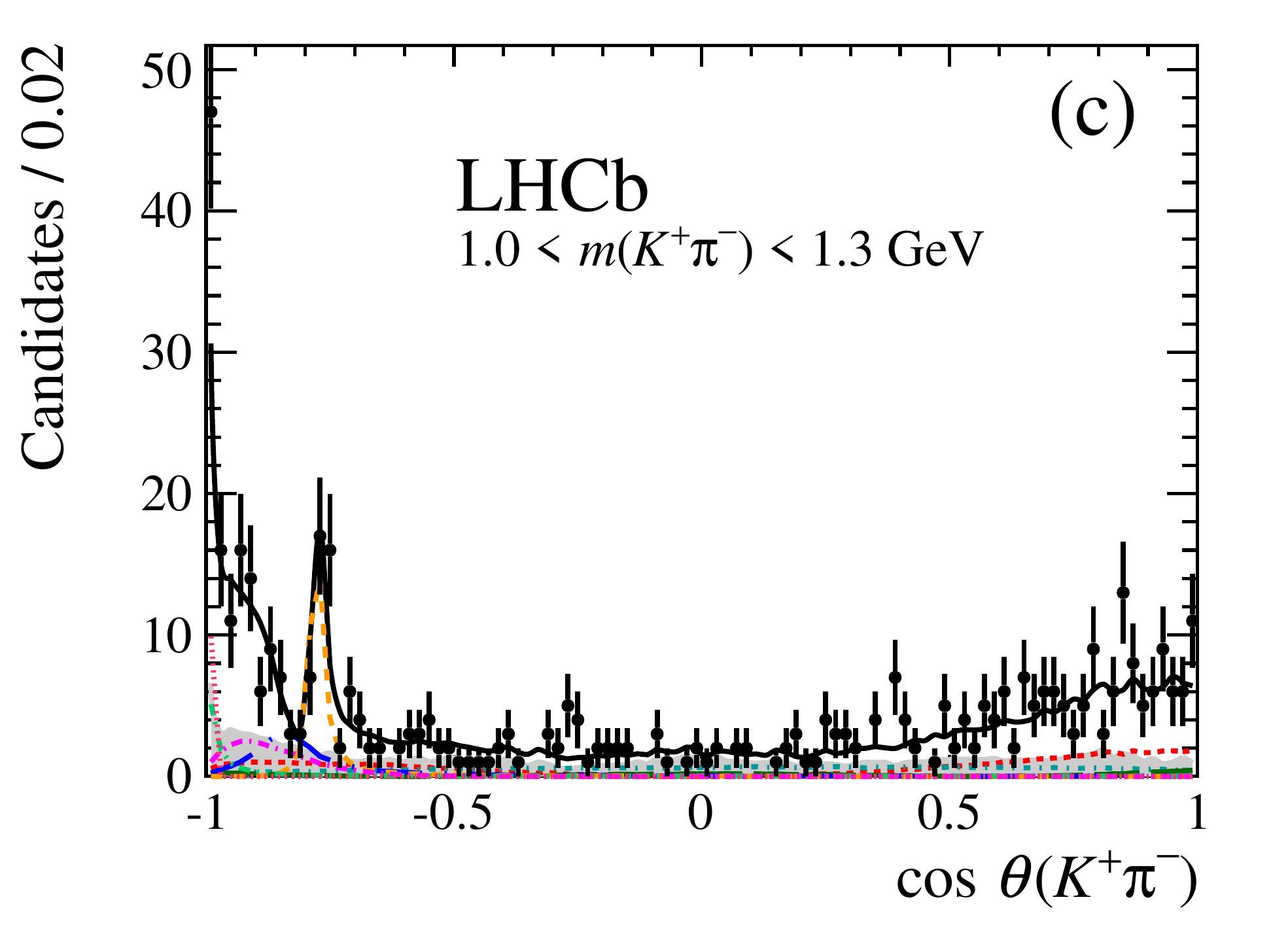}
\includegraphics[width=0.49\textwidth]{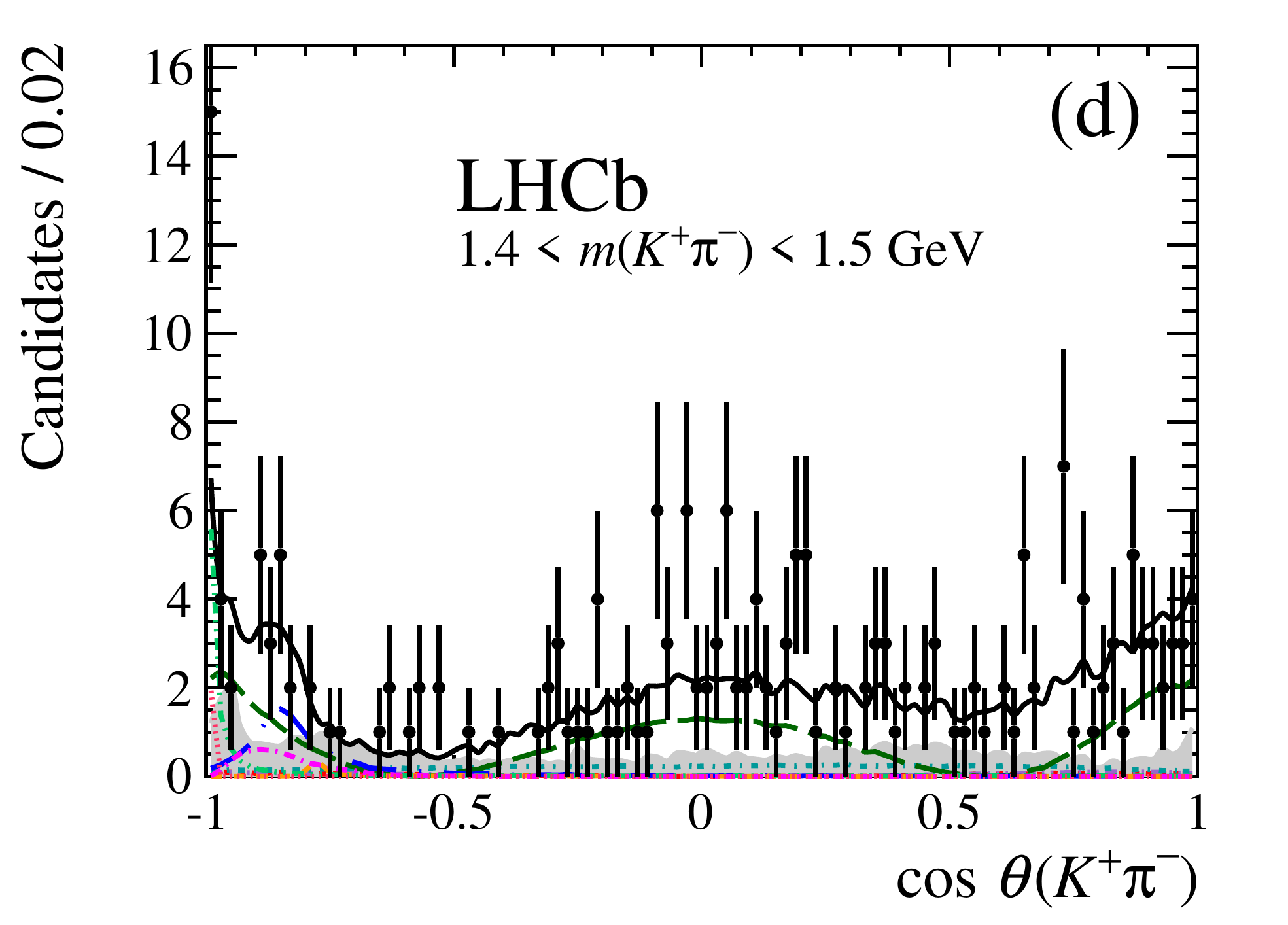}
\caption{\small Projections of the data and amplitude fit results onto $\cos \theta(\Kp\pim)$ in the mass ranges (a)~$m(\Kp\pim) < 0.8 \gev$, (b)~$0.8 < m(\Kp\pim) < 1.0 \gev$, (c)~$1.0 < m(\Kp\pim) < 1.3 \gev$ and (d)~$1.4 < m(\Kp\pim) < 1.5 \gev$.
Components are as shown in Fig.~\ref{fig:fitproj}.
}
\label{fig:fitprojcoskpi}
\end{figure}

Projections of the data and the nominal fit model onto $m(\Kp\pim)$, $m(\Dzb\pim)$ and $m(\Dzb\Kp)$ are shown in Fig.~\ref{fig:fitproj}.
Zooms are provided in the regions of the main resonant contributions in Fig.~\ref{fig:fitprojzoom}.
Projections onto the cosine of the helicity angle of the $\Dzb\pim$ and $\Kp\pim$ systems are shown in Figs.~\ref{fig:fitprojcosdpi} and~\ref{fig:fitprojcoskpi}, respectively.
These projections all show good agreement between data and the fit model.

\begin{figure}[!tb]
\centering
 \includegraphics[width=0.43\textwidth]{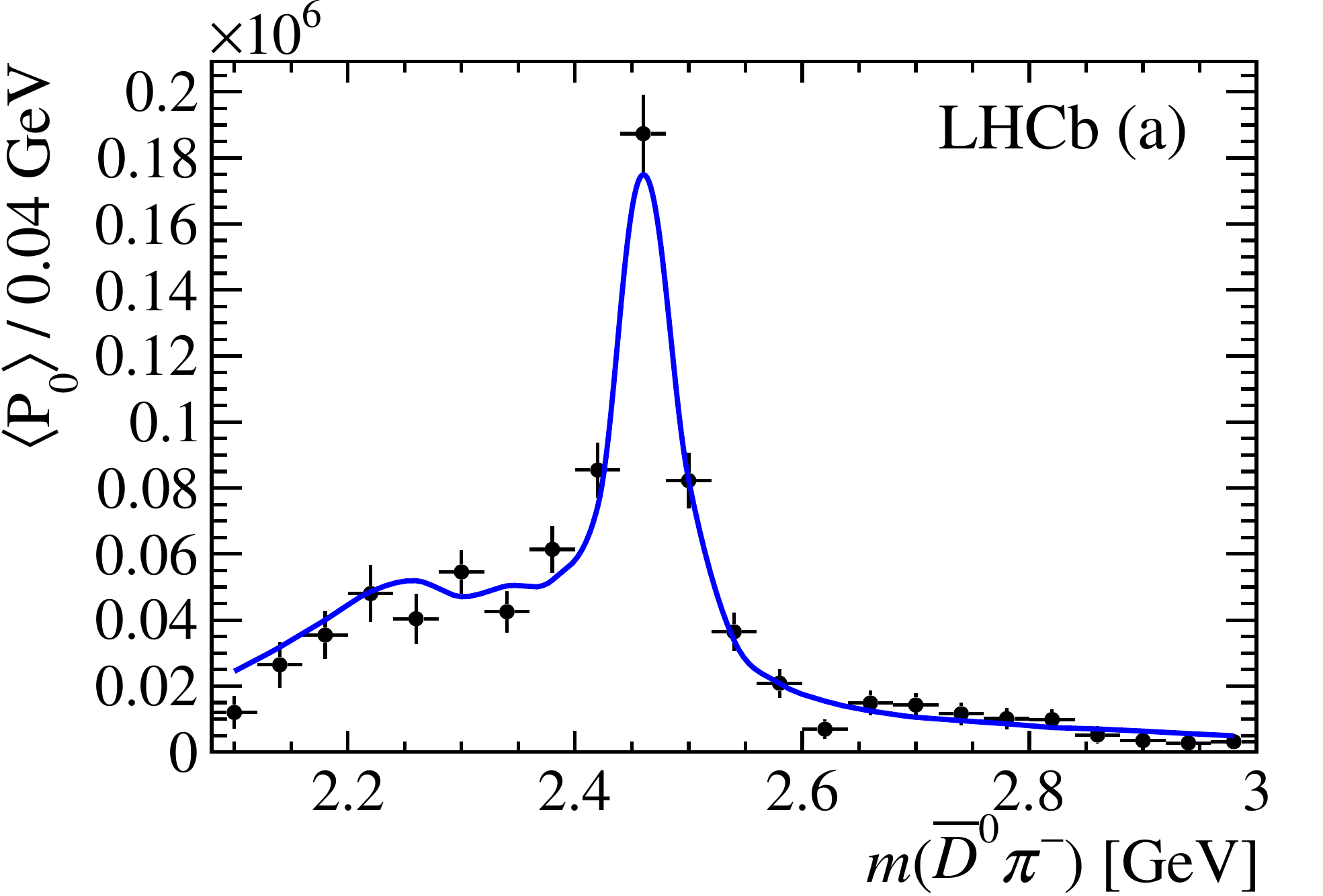}
 \includegraphics[width=0.43\textwidth]{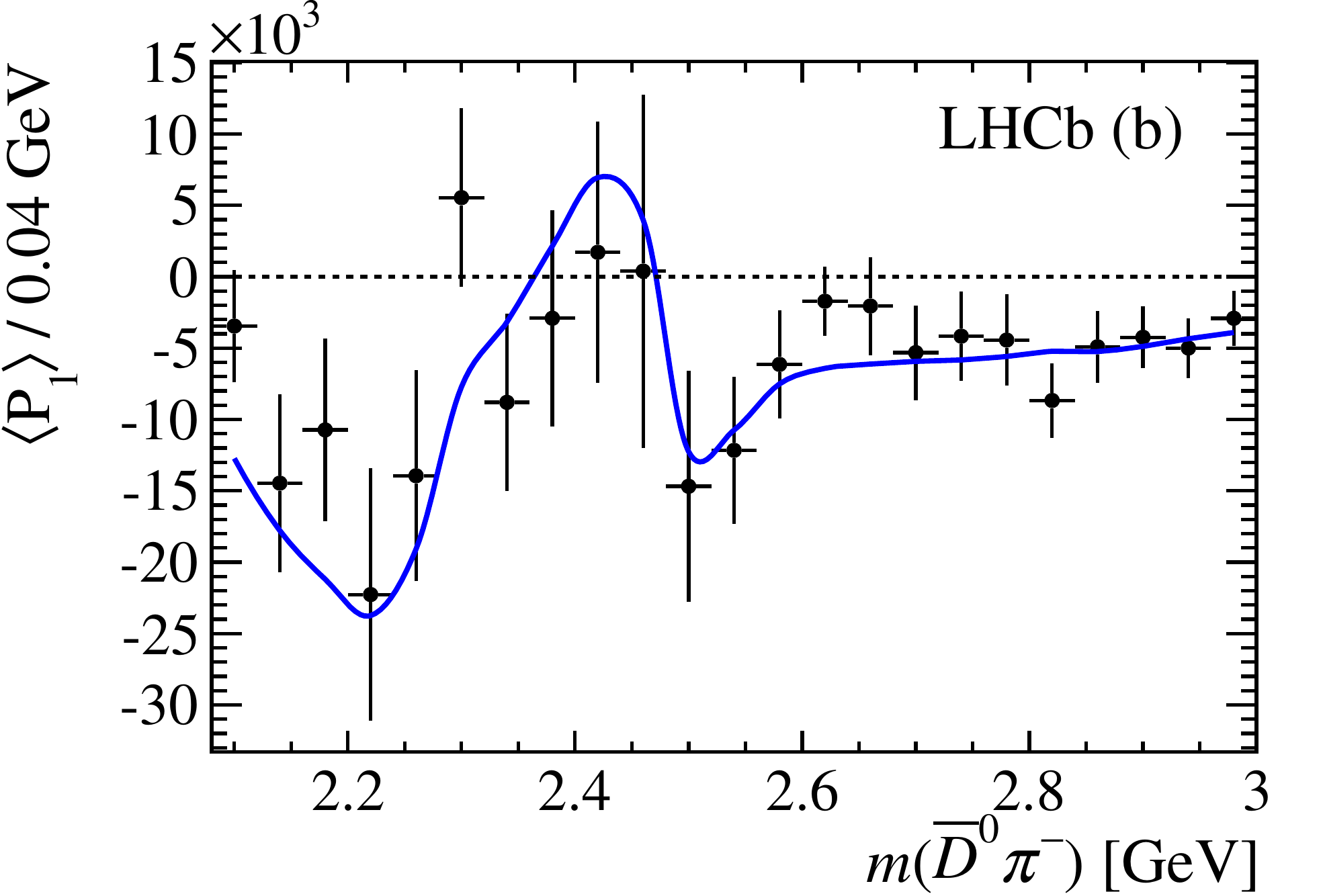}
 \includegraphics[width=0.43\textwidth]{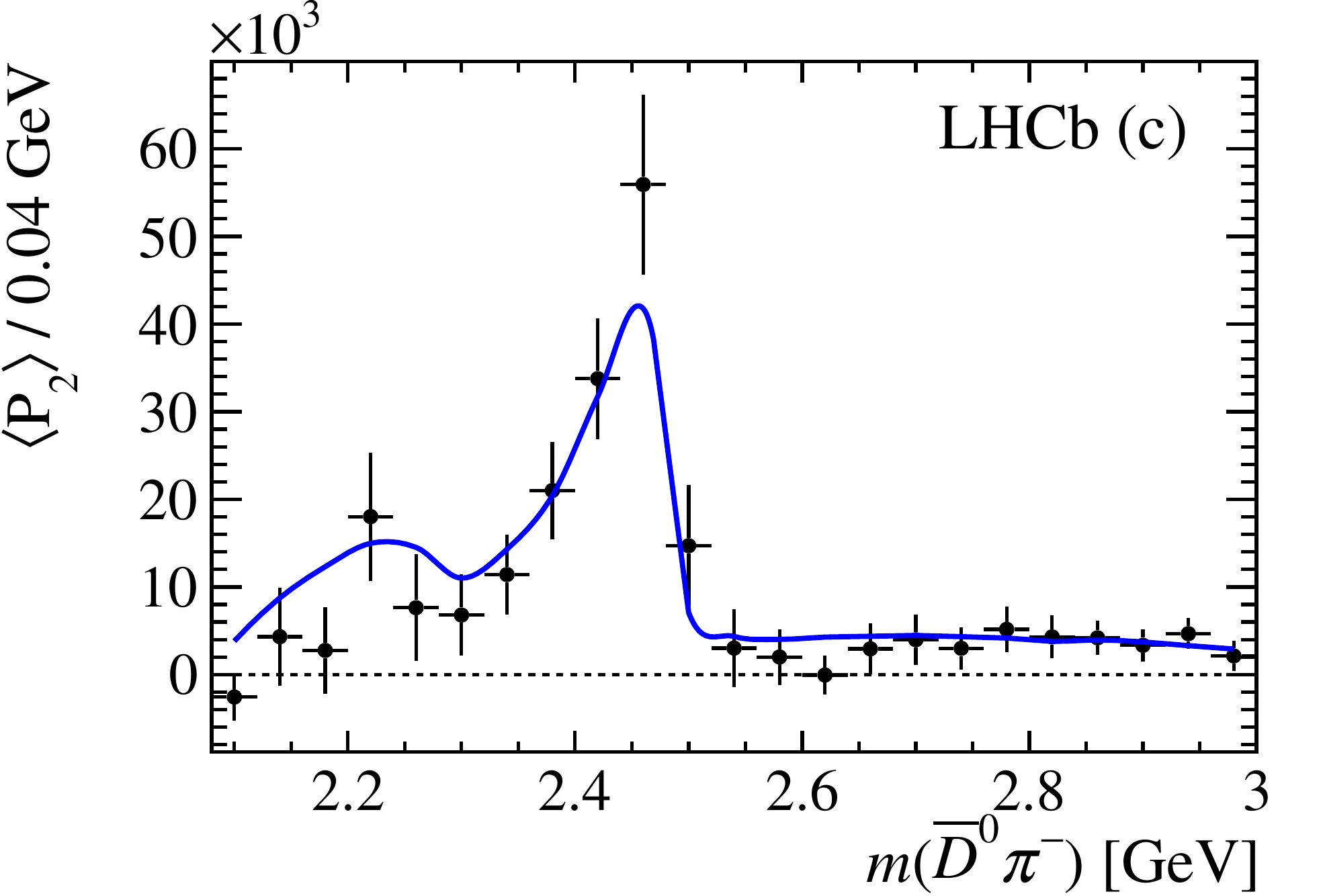}
 \includegraphics[width=0.43\textwidth]{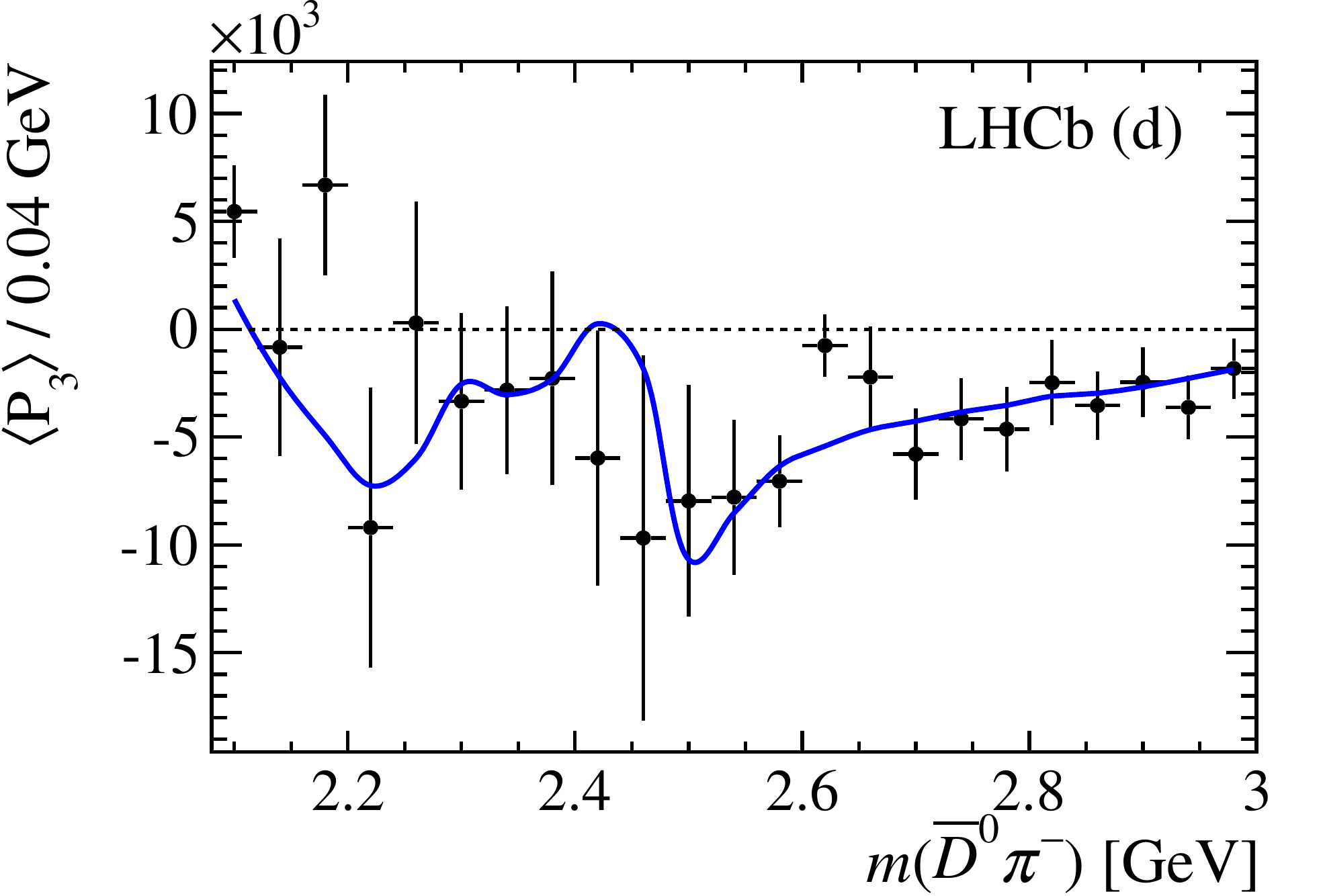}
 \includegraphics[width=0.43\textwidth]{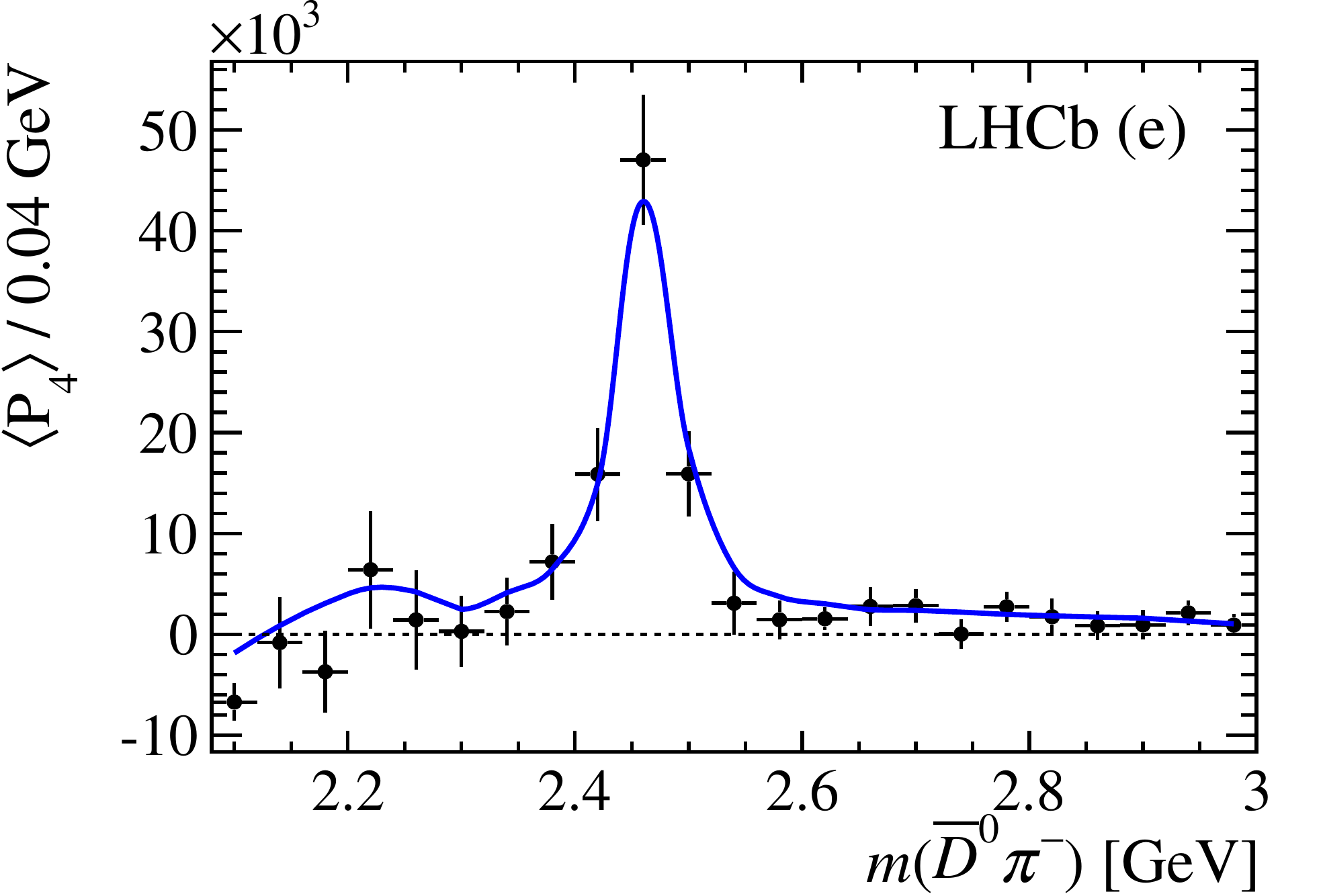}
 \includegraphics[width=0.43\textwidth]{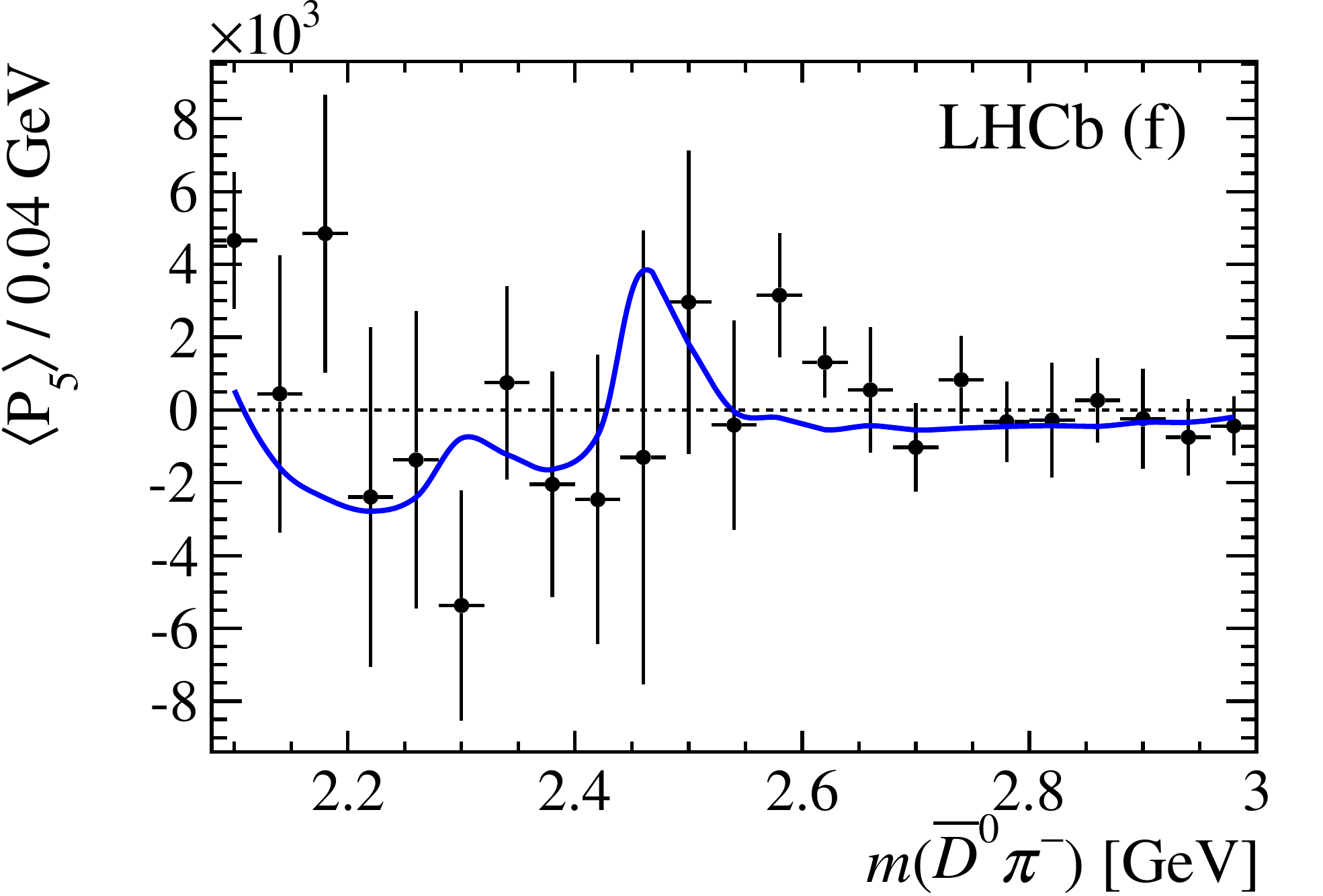}
 \includegraphics[width=0.43\textwidth]{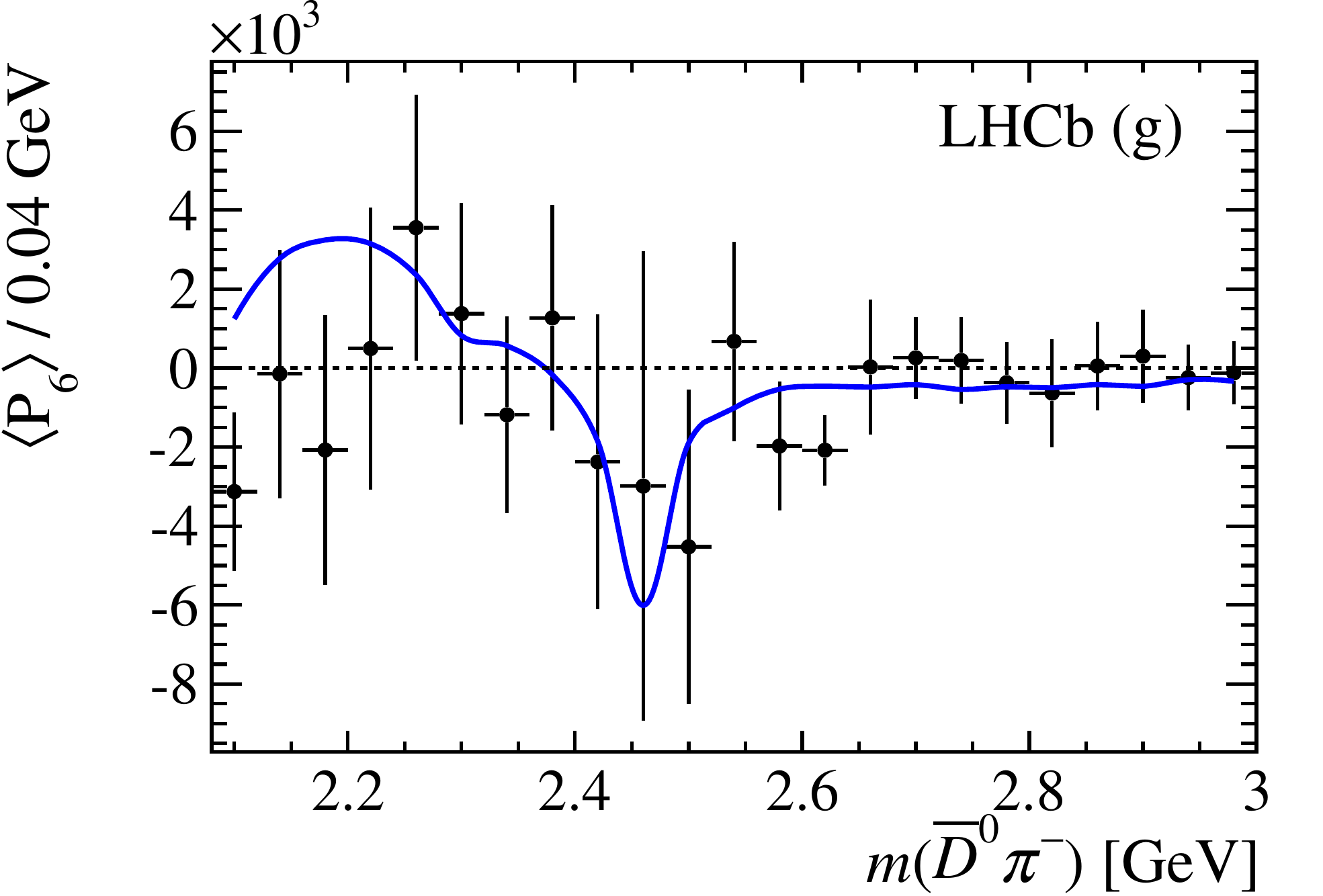}
 \includegraphics[width=0.43\textwidth]{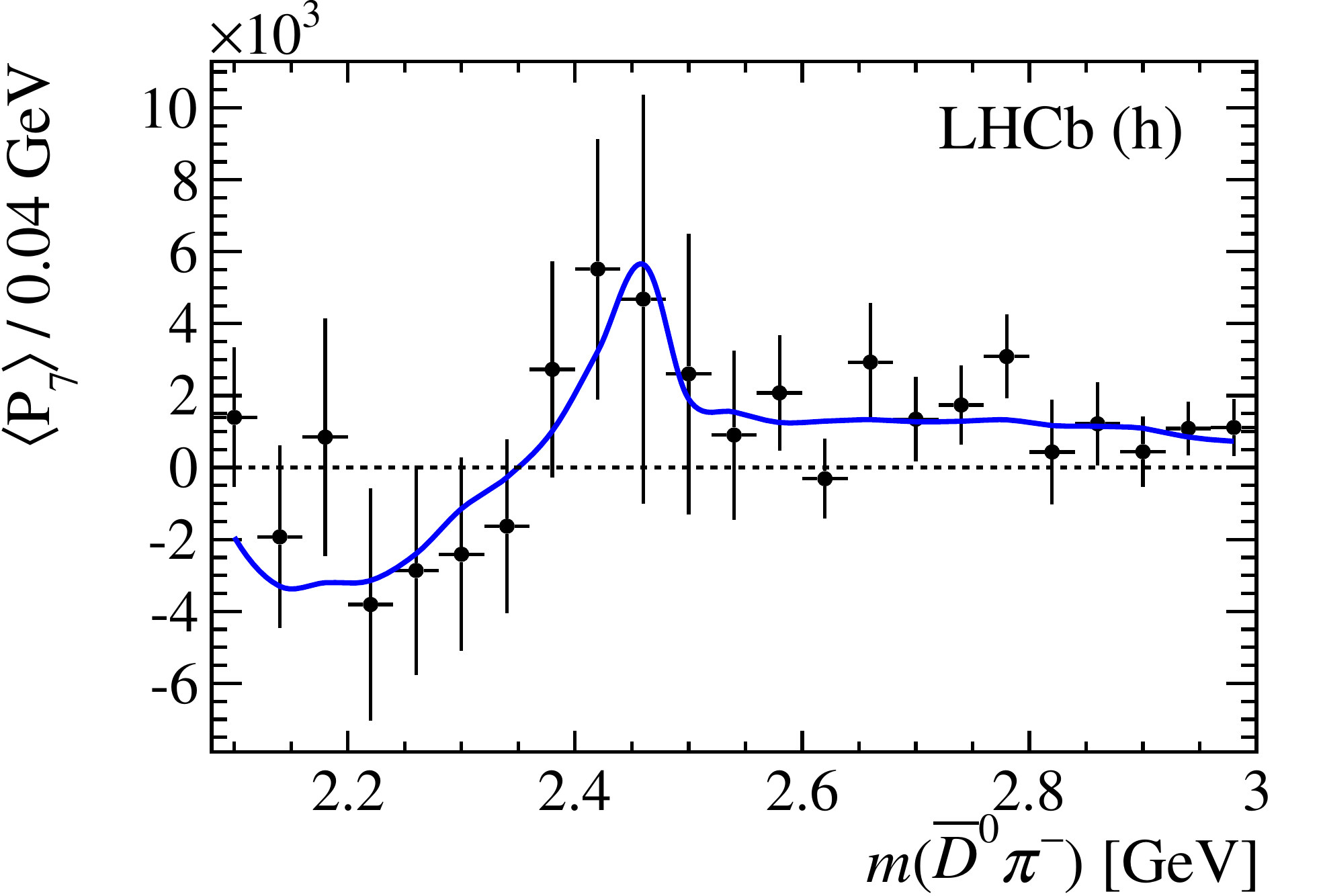}
\caption{\small
  Background-subtracted and efficiency-corrected Legendre moments up to order 7 calculated as a function of $m(\Dzb\pim)$ for data (black data points) and the fit result (solid blue curve).}
\label{fig:momentDpizoom}
\end{figure}

\begin{figure}[!tb]
\centering
 \includegraphics[width=0.43\textwidth]{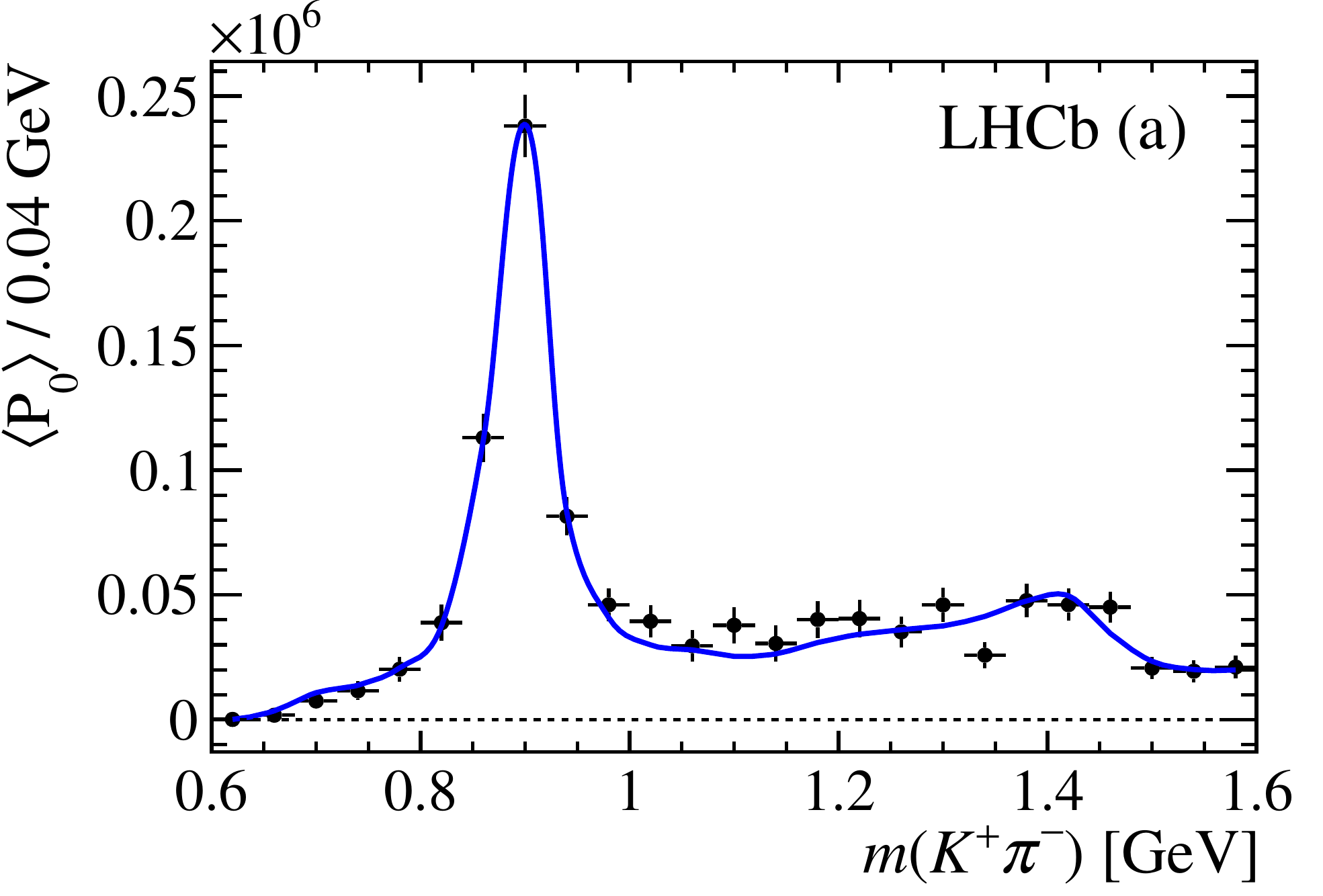}
 \includegraphics[width=0.43\textwidth]{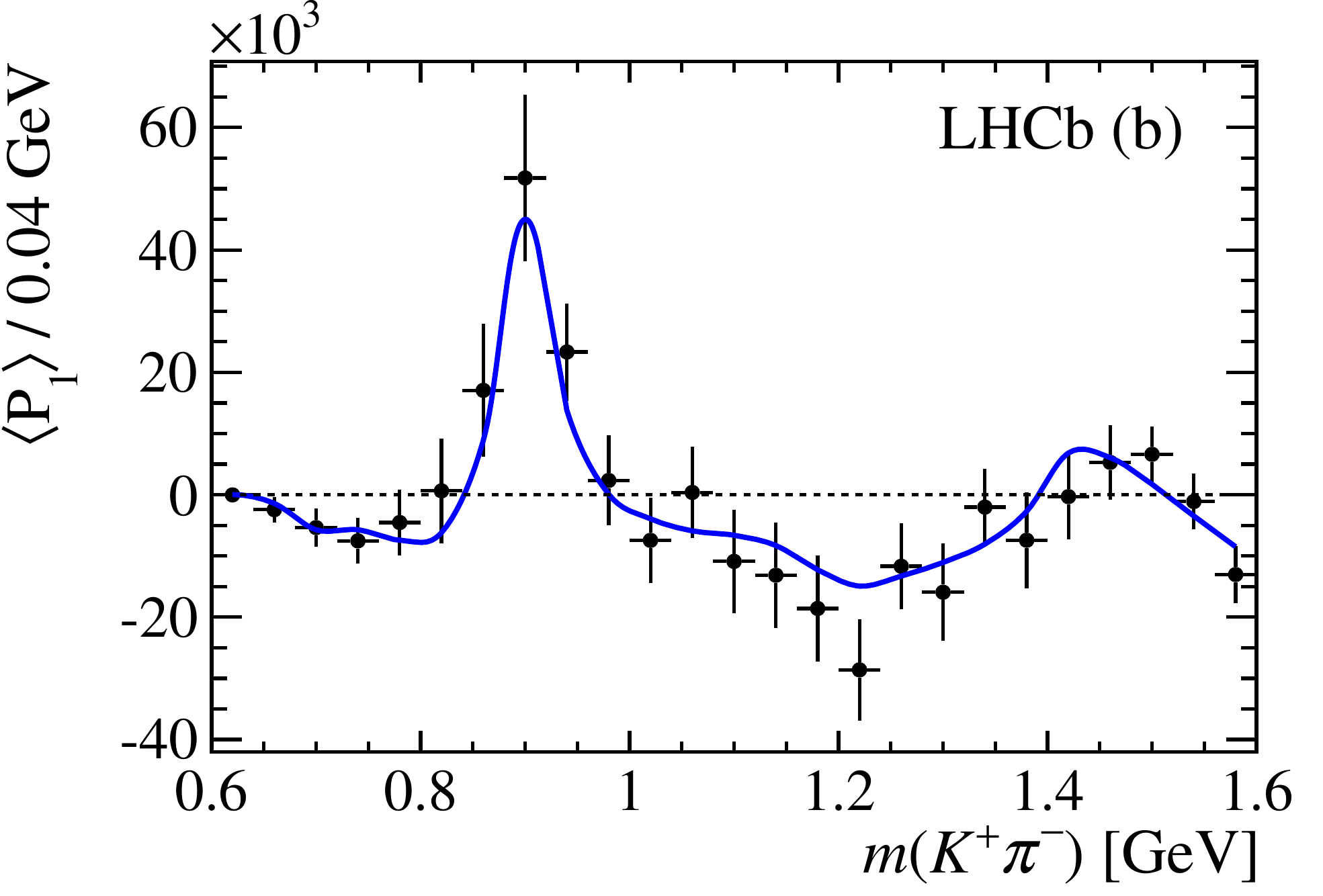}
 \includegraphics[width=0.43\textwidth]{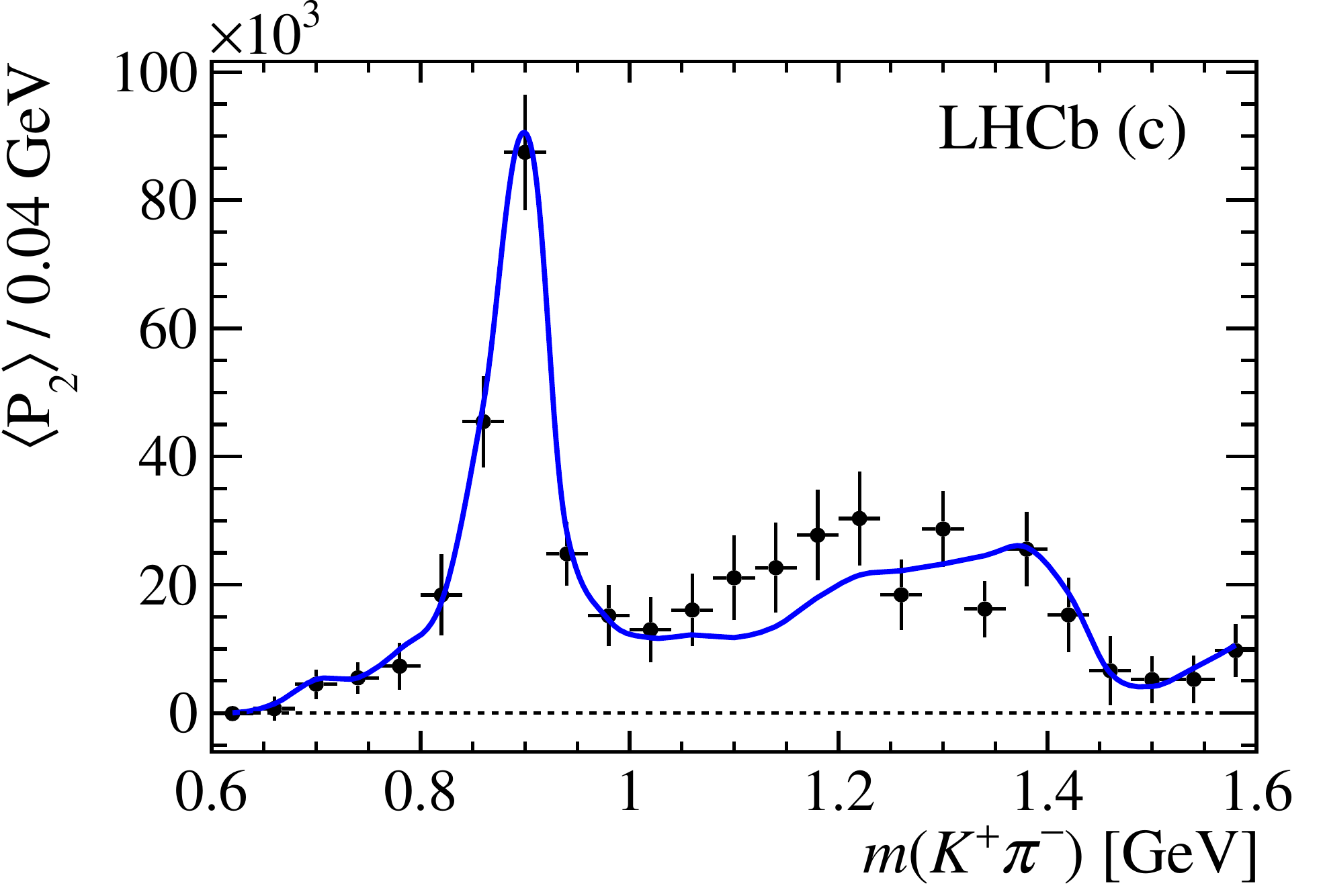}
 \includegraphics[width=0.43\textwidth]{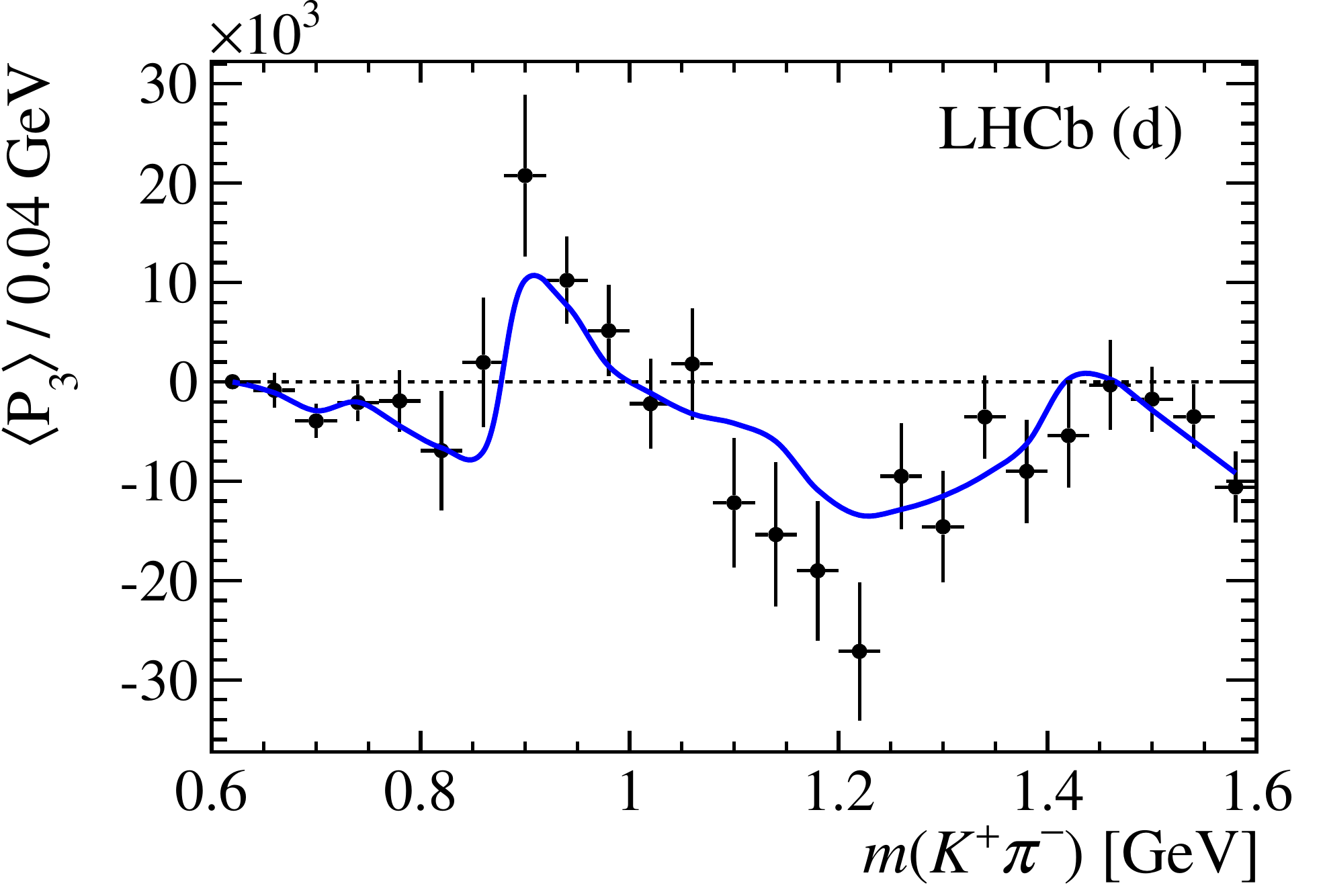}
 \includegraphics[width=0.43\textwidth]{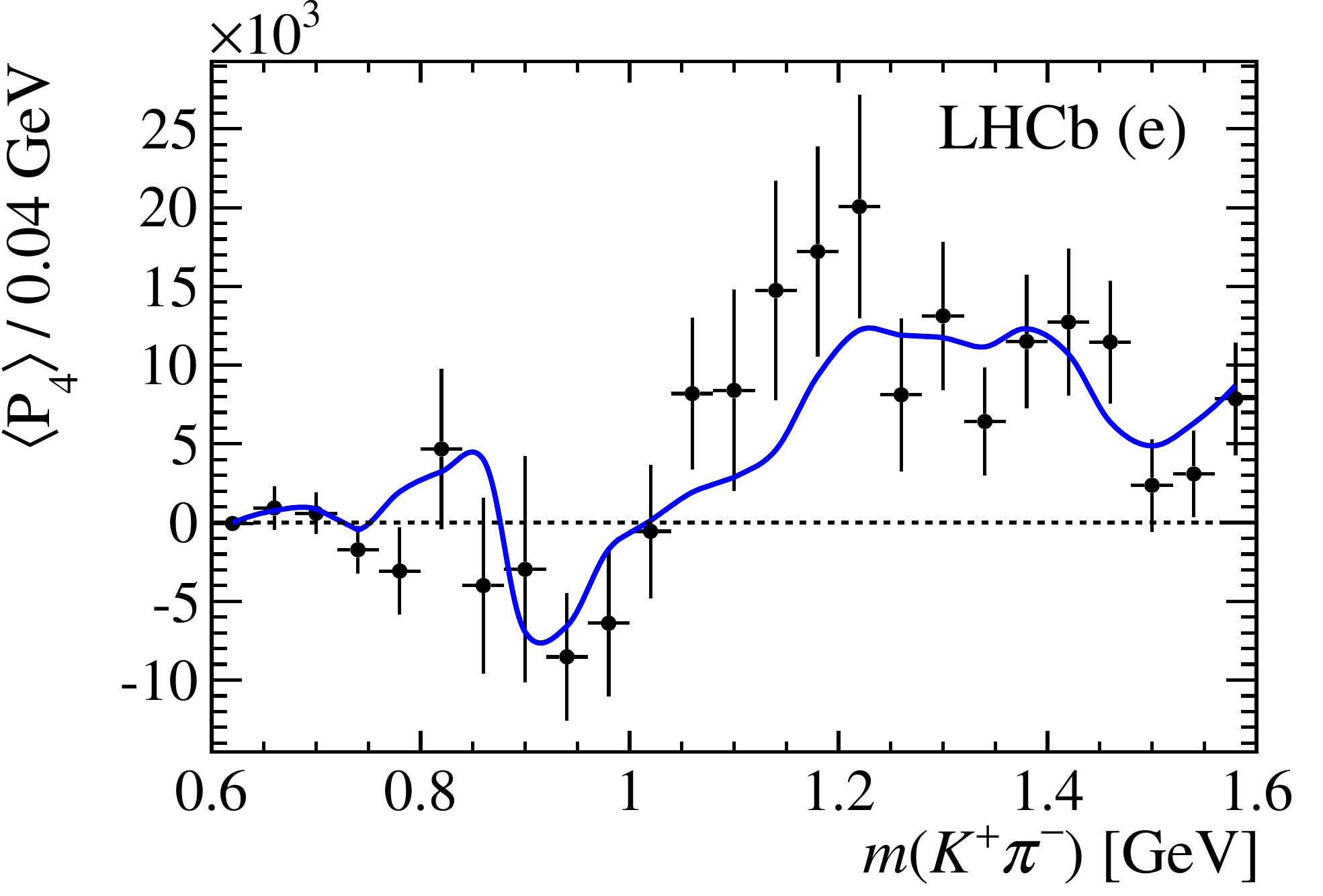}
 \includegraphics[width=0.43\textwidth]{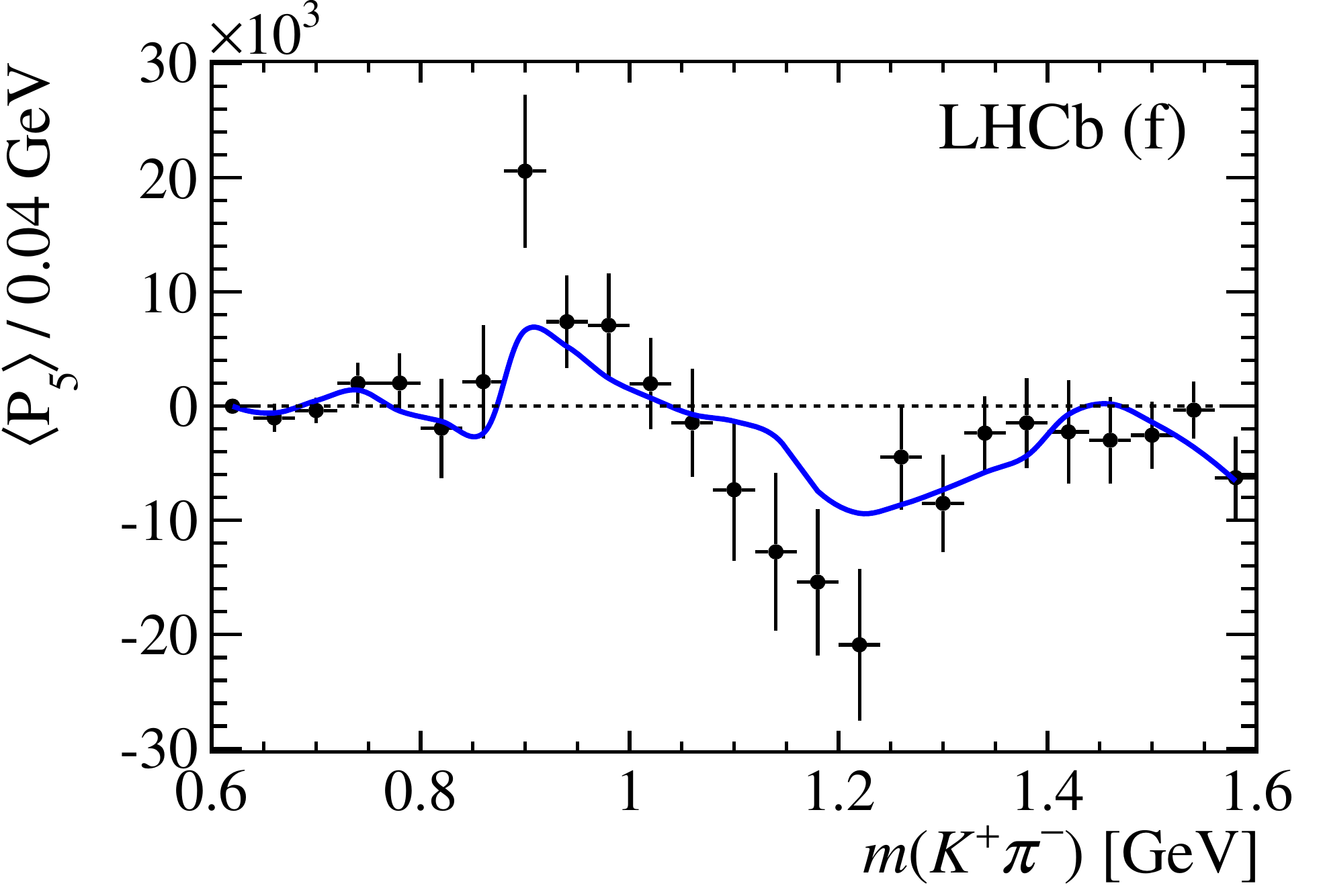}
 \includegraphics[width=0.43\textwidth]{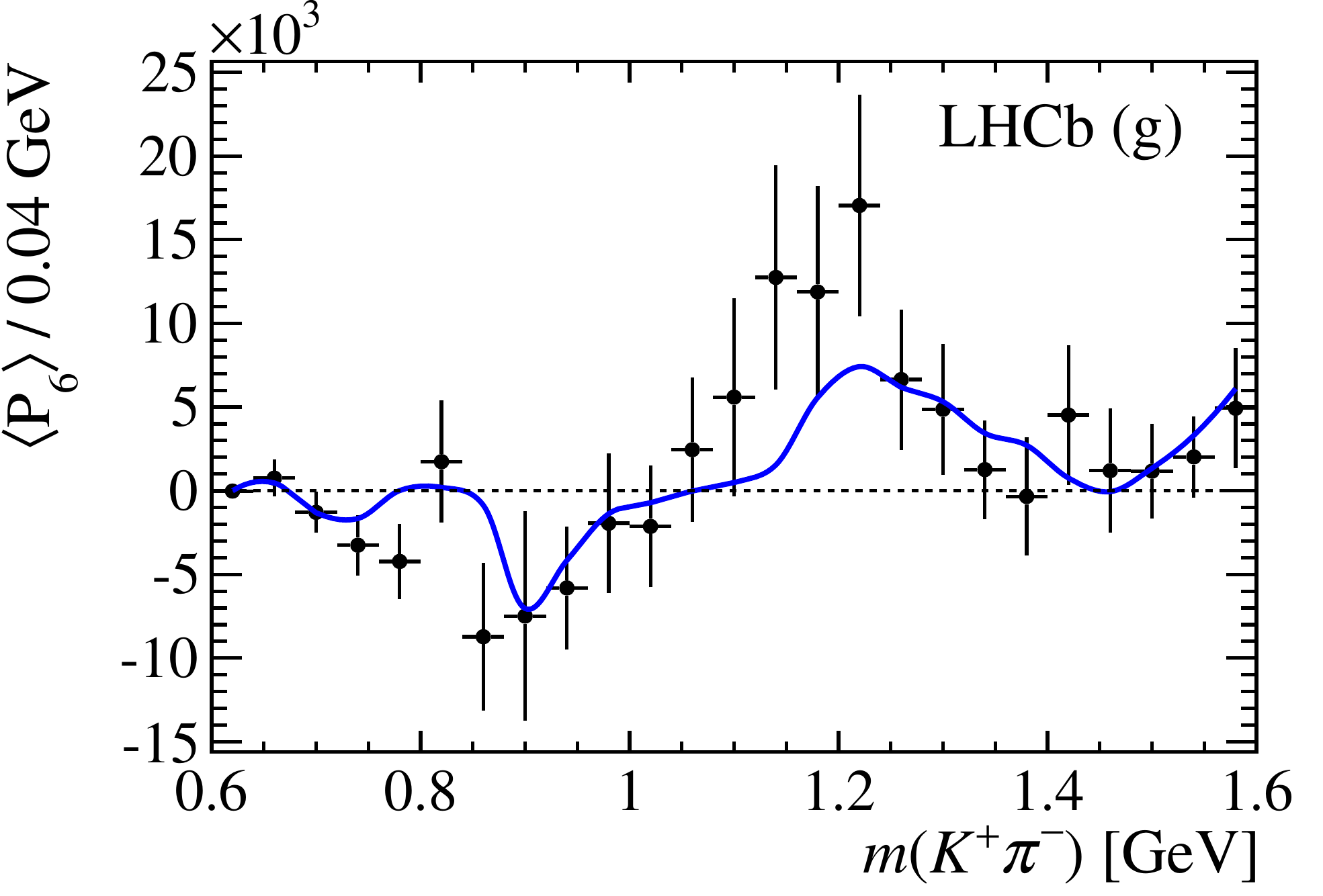}
 \includegraphics[width=0.43\textwidth]{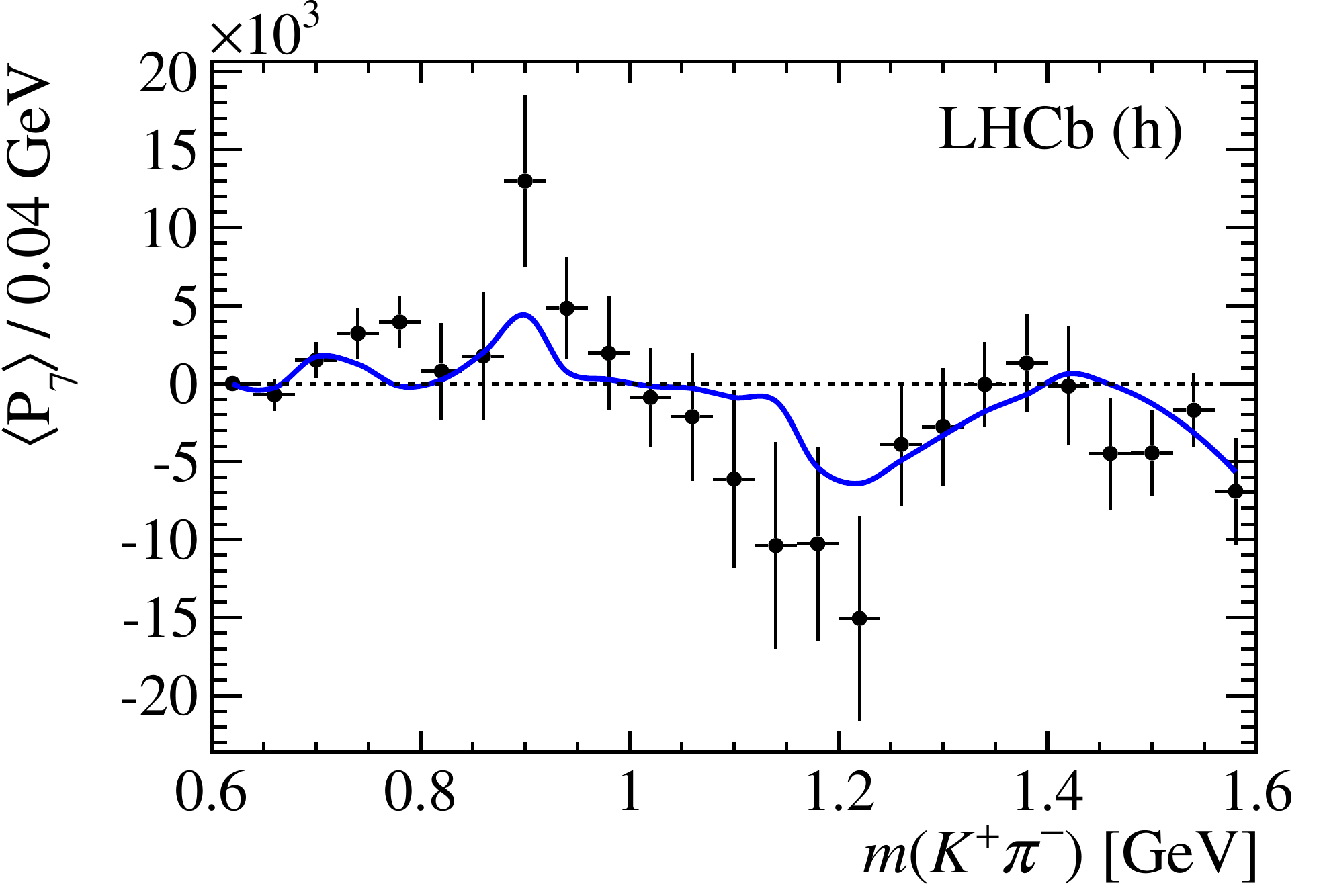}
\caption{\small
  Background-subtracted and efficiency-corrected Legendre moments up to order 7 calculated as a function of $m(\Kp\pim)$ for data (black data points) and the fit result (solid blue curve).}
\label{fig:momentKpizoom}
\end{figure}

Angular moments provide a useful method to investigate the helicity structure of the decays.
The angular moments in $m(\Dzb\pim)$ and $m(\Kp\pim)$, obtained as in Refs.~\cite{LHCb-PAPER-2014-036,LHCb-PAPER-2015-007}, are shown in Figs.~\ref{fig:momentDpizoom} and~\ref{fig:momentKpizoom}, respectively.
The contributions due to the $\Kstar(892)^0$ and $D_2^*(2460)^-$ resonances are seen as peaks in moments up to order 2 and order 4, respectively, as expected for spin-1 and spin-2 resonances.
Reflections make the interpretation of moments at higher masses more difficult.
However, the good agreement seen between data and the fit model provides further support for the fit model being a good description of the data.

\begin{figure}[!tb]
\centering
\includegraphics[width=0.65\textwidth]{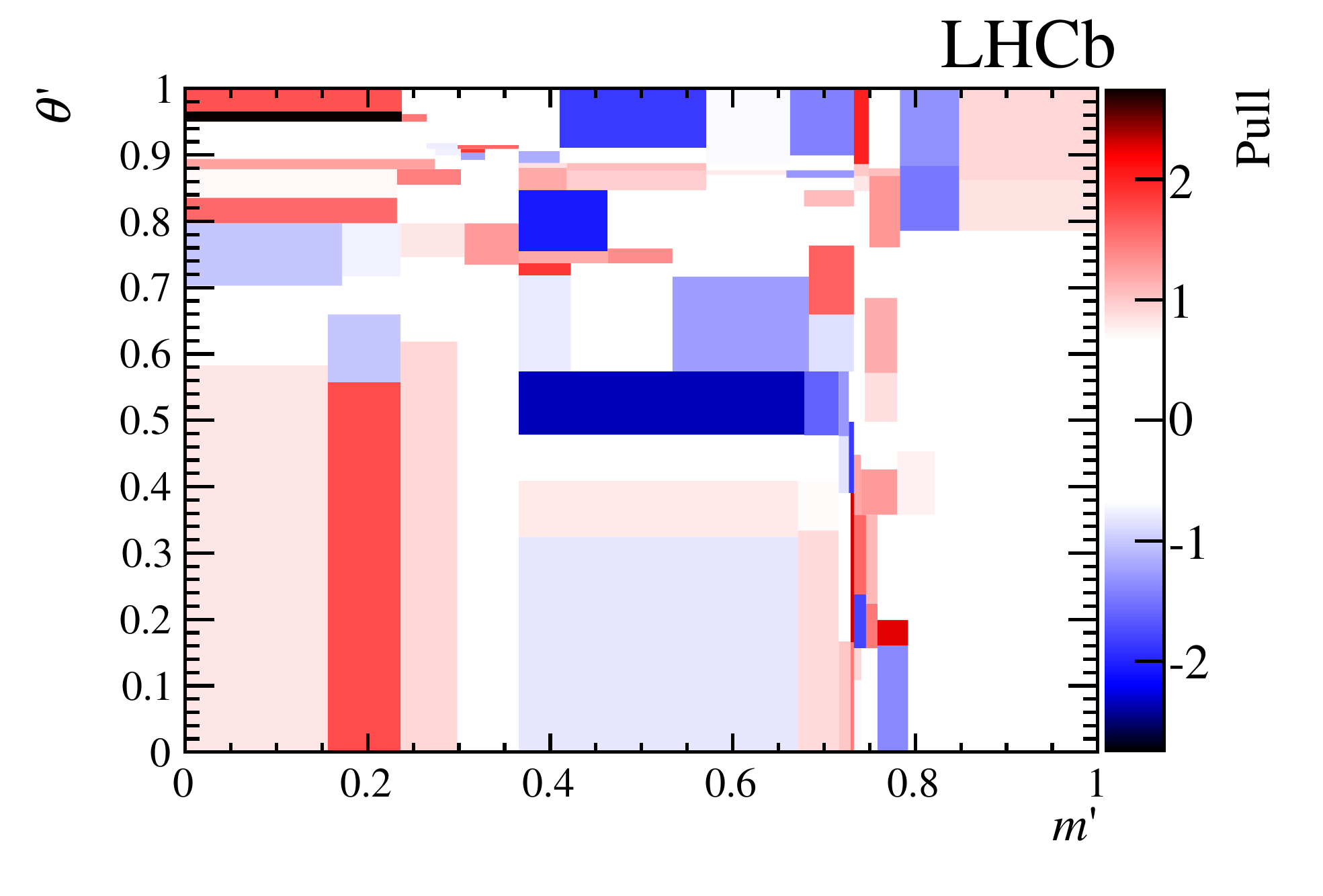}
\caption{\small Differences between the data SDP distribution and the fit model across the SDP, in terms of the per-bin pull.}
\label{fig:sdppull}
\end{figure}

The quality of the fit is evaluated by determining a two-dimensional $\chi^2$ value, comparing the data and fit model in 144 bins across the SDP that are defined adaptively to
ensure they are approximately equally populated.
The pull in each of these bins, defined as the difference between the data and the fit model divided by the uncertainty, is shown in Fig.~\ref{fig:sdppull}.
The effective number of degrees of freedom of the $\chi^2$ is between $N_{\rm bins} - N_{\rm pars} - 1$ and $N_{\rm bins} - 1$,
where $N_{\rm pars}$ is the number of free parameters in the fit, yielding a reduced $\chi^2$ in the range 0.99 -- 1.22.
Additional unbinned tests of fit quality~\cite{Williams:2010vh} also show acceptable agreement between the data and the fit model.

\section{Systematic uncertainties}
\label{sec:systematics}

Systematic uncertainties are considered from sources that can be divided into two categories, experimental and model uncertainties.
Experimental uncertainties arise due to the signal and background yields in the signal region, the distributions of the background components across the SDP,
the variation of efficiency across the SDP, and possible bias due to the fitting procedure.
Model uncertainties are considered due to the fixed parameters in the signal model, the addition or removal of marginal amplitudes, and the choice of models for the $K\pi$ S-wave and the $D\pi$ S- and P-waves.
The uncertainties due to all of these sources are combined in quadrature.

The signal and background yields are determined from the fit to the $B$ candidate invariant mass distribution. Systematic uncertainties are considered on the total yields
due to both statistical and systematic variations of these yields, as evaluated in Ref.~\cite{LHCb-PAPER-2013-022}.
The yields in the signal region are varied appropriately and the effects of these variations on the parameters of interest are assigned as systematic uncertainties.

The histograms describing the background distributions are each varied within statistical uncertainties to establish the uncertainty due to the imperfect knowledge of
the distribution of these backgrounds across the SDP. The histograms describing the efficiency variation across the SDP are also varied within their uncertainties prior to the spline fit.
The efficiency histograms are also varied with local correlations to account for the effects of any unknown correlations between neighbouring bins.
For each parameter the larger uncertainty is considered from the correlated and uncorrelated variations.
In addition, the binning scheme of the control sample used to evaluate PID performance is varied.

An ensemble of pseudoexperiments is used to establish systematic uncertainties related to any potential fit bias.
Pseudoexperiments are generated using the parameters determined in the fit to data, and mean fitted values of the parameters are determined from Gaussian fits to the distributions of results obtained in the ensemble.
The differences between the generated and mean fitted values are found to be small.
Uncertainties are determined as the sum in quadrature of the difference between the generated and mean fitted values, and the uncertainty on the mean fitted value from the Gaussian fit.

Systematic uncertainties due to fixed parameters in the fit model are determined by repeating the fit with these parameters varied.
The fixed masses and widths are varied within their uncertainties as shown in Table~\ref{tab:resonances}, while the \mbox{Blatt--Weisskopf} barrier radii, $r_{\rm BW}$, are varied independently for $\Dzb\pim$ and $\Kp\pim$ resonances between 3 and 5\,$\gev^{-1}$~\cite{LHCb-PAPER-2014-036}.

The $\Kstar(1410)^0$ amplitude does not contribute significantly to the nominal fit model so the fit is repeated with this component removed and the change in results is assigned as the associated systematic uncertainty.
In addition, uncertainties are assigned from the changes in the results when an amplitude due to the $\Kstar(1680)^0$ resonance or from a virtual $B^*(5325)^+$ component, as described in Ref.~\cite{LHCb-PAPER-2014-036}, is included in the model.

A Dalitz plot analysis of $\Bs\to\Dzb\Km\pip$ observed both spin-1 and spin-3 resonances at $m(\Dzb\Km) \sim 2.86 \gev$~\cite{LHCb-PAPER-2014-035,LHCb-PAPER-2014-036}.
A spin-3 resonance has also been seen at $m(\Dzb\pim) \sim 2.76 \gev$ in $\Bd\to\Dzb\pip\pim$ decays~\cite{LHCb-PAPER-2014-070},
while an analysis of $\Bp\to\Dm\Kp\pip$ shows a spin-1 resonance at similar $m(\Dp\pim)$ values~\cite{LHCb-PAPER-2015-007}.
Uncertainties are assigned corresponding to the changes in results when either a spin-1 or spin-3 state at $m(\Dzb\pim) \sim 2.76 \gev$ is added to the nominal model.

The LASS model used to describe the $K\pi$ S-wave is replaced with a Flatt\'e shape~\cite{Flatte:1976xu}, which accounts for the $\kaon\etapr$ threshold near the $\kaon^*_0(1430)^0$ mass, and a resonant term with a modified mass dependent width~\cite{Bugg:2009tu} for the $\kappa$ (also known as $\kaon^*_0(800)$) state. This model gives an NLL that is worse by 4.4 units.
A model-independent description of the $K\pi$ S-wave is also used to fit the data, with the uncertainty on each parameter taken as the larger of the two differences.
This alternative improves the NLL by 8.8 units but introduces an additional four free parameters into the fit.
The dabba lineshape used to describe the $\Dzb\pim$ S-wave is replaced with an EFF lineshape, while a power-law model is introduced as an alternative to the EFF lineshape that describes the $\Dzb\pim$ P-wave. These alterations worsen the NLL by 3.0 units and improve it by 0.8 units respectively.

The total systematic uncertainties due to experimental and model effects for fit fractions and complex coefficients are given together with the results in Sec.~\ref{sec:results}.
The contributions from different sources to the systematic uncertainties on the fit fractions, masses and widths are given in Tables~\ref{tab:exptsystbreak} and~\ref{tab:modsystbreak}.
The dominant experimental systematic uncertainty on most parameters is due to either the efficiency variation or background distributions across the SDP, while the model uncertainties are generally dominated by the effects of alternative descriptions of the $K\pi$ S-wave and the $D\pi$ S- and P-waves.
For the parameters of broad components, the model uncertainties dominate; however for other parameters the statistical, experimental and model uncertainties are all of a similar magnitude.

\begin{table}[!tb]
\centering
\caption{\small Experimental systematic uncertainties on the fit fractions (\%) and masses and widths $(\mevnsp)$. Uncertainties given on the central values are statistical only.}
\label{tab:exptsystbreak}
\resizebox{\textwidth}{!}{
\begin{tabular}{lc|cccc|c}
\hline
Resonance & Central value & S/B fraction & Efficiency & Background SDP & Fit bias & Total \\
\hline &&&&&&\\ [-2.5ex]
$\Kstar(892)^{0}$        & $\phani37.4 \pm 1.5$          & 0.60 & 0.83 & 0.50 & 0.31 & 1.17  \\
$\Kstar(1410)^{0}$       & $\phanii0.7 \pm 0.3$     & 0.06 & 0.39 & 0.69 & 0.05 & 0.80  \\
$\kaon^*_{0}(1430)^{0}$  & $\phanii5.1 \pm 2.0$     & 0.28 & 1.48 & 1.85 & 0.33 & 2.41  \\
LASS nonresonant         & $\phanii4.8 \pm 3.8$     & 0.51 & 2.25 & 2.86 & 0.86 & 3.77  \\
\ \ \ LASS total         & $\phanii6.7 \pm 2.7$     & 0.26 & 1.86 & 1.60 & 1.02 & 2.67  \\
$\kaon^*_{2}(1430)^{0}$  & $\phanii7.4 \pm 1.7$     & 0.23 & 0.72 & 0.53 & 0.54 & 1.07  \\
$D^{*}_{0}(2400)^{-}$    & $\phani19.3 \pm 2.8$          & 0.21 & 1.39 & 1.43 & 0.40 & 2.04  \\
$D^{*}_{2}(2460)^{-}$    & $\phani23.1 \pm 1.2$          & 0.70 & 0.70 & 0.49 & 0.15 & 1.11  \\
\hline
$D\pi$ S-wave (dabba)    & $\phanii6.6 \pm 1.4$     & 0.03 & 0.81 & 0.59 & 0.57 & 1.15  \\
$D\pi$ P-wave (EFF)      & $\phanii8.9 \pm 1.6$     & 0.86 & 1.91 & 0.52 & 0.38 & 2.19  \\
\hline
$m\left(D^{*}_{0}(2400)^{-}\right)$
& $2360 \pm 15$           & 4.6 & 8.1 & 7.0 & 3.7 & 12.2  \\
$m\left(D^{*}_{2}(2460)^{-}\right)$
& $2465.6 \pm 1.8\phani$  & 0.01& 0.37& 0.22& 0.29& 0.51 \\
\hline
$\Gamma\left(D^{*}_{0}(2400)^{-}\right)$
& $\phani255 \pm 26$      & 2.8 & 13.1 & 13.9 & 4.8 & 19.9  \\
$\Gamma\left(D^{*}_{2}(2460)^{-}\right)$
& \phani$46.0 \pm 3.4$    & 0.5 & 0.9 & 0.9 & 0.5 & 1.4  \\
\hline
\end{tabular}
}
\end{table}

\begin{table}[!tb]
\centering
\caption{\small Model uncertainties on the fit fractions (\%) and masses and widths $(\mevnsp)$. Uncertainties given on the central values are statistical only.}
\label{tab:modsystbreak}
\resizebox{\textwidth}{!}{
\begin{tabular}{lc|ccc|c}
\hline
Resonance & Central value & Fixed parameters & Add/remove & Alternative models & Total \\
\hline &&&&&\\ [-2.5ex]
$\Kstar(892)^{0}$        & $\phani37.4 \pm 1.5$          & 0.75 & 1.14 & 1.09 & 1.74  \\
$\Kstar(1410)^{0}$       & $\phanii0.7 \pm 0.3$     & 0.18 & 0.70 & 0.22 & 0.76  \\
$\kaon^*_{0}(1430)^{0}$  & $\phanii5.1 \pm 2.0$     & 0.79 & 3.30 & 0.23 & 3.40  \\
LASS nonresonant         & $\phanii4.8 \pm 3.8$     & 1.10 & 3.99 & 5.20 & 6.65  \\
\ \ \ LASS total         & $\phanii6.7 \pm 2.7$     & 0.53 & 1.42 & 5.21 & 5.43  \\
$\kaon^*_{2}(1430)^{0}$  & $\phanii7.4 \pm 1.7$     & 0.36 & 1.87 & 0.56 & 1.98  \\
$D^{*}_{0}(2400)^{-}$    & $\phani19.3 \pm 2.8$          & 0.55 & 1.95 & 7.11 & 7.40  \\
$D^{*}_{2}(2460)^{-}$    & $\phani23.1 \pm 1.2$          & 0.18 & 0.73 & 0.99 & 1.24  \\
\hline
$D\pi$ S-wave (dabba)    & $\phanii6.6 \pm 1.4$     & 0.27 & 1.40 & 3.46 & 3.74  \\
$D\pi$ P-wave (EFF)      & $\phanii8.9 \pm 1.6$     & 0.31 & 1.99 & 2.15 & 2.95  \\
\hline
$m\left( D^{*}_{0}(2400)^{-}\right)$
& $2360 \pm 15$           & $6.1$ & $9.3$ & $25.6$ & $27.9$  \\
$m\left( D^{*}_{2}(2460)^{-}\right)$
& $2465.6 \pm 1.8\phani$  & $0.09$ & $1.05$ & $0.48$ & $1.15$  \\
\hline
$\Gamma\left( D^{*}_{0}(2400)^{-}\right)$
& $\phani255 \pm 26$      & $4.0$ & $18.0$ & $43.5$ & $47.2$  \\
$\Gamma\left( D^{*}_{2}(2460)^{-}\right)$
& $\phani46.0 \pm 3.4$    & $1.4$ & $0.5$ & $2.4$ & $2.9$  \\
\hline
\end{tabular}
}
\end{table}

The stability of the fit results is confirmed by several cross-checks. The data sample is divided into subsamples according to
the flavour of the $B$ candidate, the trigger decision, the polarity of the magnet and the year of data taking.
Each subsample is fitted separately and all are seen to be consistent with the nominal fit results.
The fit is also repeated with additional resonance components to test the fit model.
Resonances up to spin-3 were considered for all three pairs of daughters with the mass and width allowed to vary in the fit.
No additional significant contributions were observed.

\section{Results and summary}
\label{sec:results}

The results for the complex coefficients are reported in Table~\ref{tab:amplitude-results}, with results for the fit fractions given in Table~\ref{tab:cfitfrac-results}.
The results for the interference fit fractions are given in Appendix~\ref{app:IFF-results}.
Upper limits are determined on the fit fractions of the $K^*(1410)^0$ and $D_3^*(2760)^-$ components, which do not give significant contributions, using the method described in Ref.~\cite{LHCb-PAPER-2014-036}.

The fit fractions for resonant contributions are converted into quasi-two-body product branching fractions by multiplying by $\mathcal{B}(\Bd\to\Dzb\Kp\pim) = (9.2 \pm 0.6 \pm 0.7 \pm 0.6) \times 10^{-5}$,
which is obtained from the ratio of branching fractions $\mathcal{B}(\Bd\to\Dzb\Kp\pim)/\mathcal{B}(\Bd\to\Dzb\pip\pim)$~\cite{LHCb-PAPER-2013-022} multiplied by the latest result for $\mathcal{B}(\Bd\to\Dzb\pip\pim)$~\cite{LHCb-PAPER-2014-070}, accounting for the different $\Dstar(2010)^-$ veto windows used in the analyses and for the 3.7\,\% residual background due to $\Bd\to\Dstar(2010)^-\Kp$ decays.
The statistical correlation between the results of Ref.~\cite{LHCb-PAPER-2013-022} and this analysis is found to be negligible.
The results for the product branching fractions are shown in Table~\ref{tab:BFresults}.
For the $\Kp\pim$ resonances, where the branching fractions to $K\pi$ are known, the product branching fractions are converted to the \B decay branching fractions, shown in Table~\ref{tab:BFresults2}.
The results for $\mathcal{B}(\Bd\to\Dzb\Kstar(892)^0)$ and $\mathcal{B}(\Bd\to D^{*}_{2}(2460)^{-}\Kp)\times \mathcal{B}(D^{*}_{2}(2460)^{-}\to \Dzb\pim)$ are consistent with, and more precise than, previous measurements~\cite{Aubert:2005yt,Krokovny:2002ua,Aubert:2006qn}.

\begin{table}[!tb]
\centering
\caption{\small
  Results for the complex amplitudes and their uncertainties presented (top) in terms of real and imaginary parts and (bottom) in terms and magnitudes and phases.
  The three quoted errors are statistical, experimental systematic and model uncertainties, respectively.
}
\label{tab:amplitude-results}
\begin{tabular}{lc@{$\:\pm\:$}c@{$\:\pm\:$}c@{$\:\pm\:$}cc@{$\:\pm\:$}c@{$\:\pm\:$}c@{$\:\pm\:$}c}
\hline
Resonance & \multicolumn{4}{c}{Real part} & \multicolumn{4}{c}{Imaginary part} \\
\hline \\ [-2.5ex]
$\Kstar(892)^{0}$         & $          -0.00$ & $0.15$ &0.24 &0.34 & $          -1.27$ & $0.06$ &0.03 &0.06 \\
$\Kstar(1410)^{0}$        & $\phantom{-}0.15$ & $0.06$ &0.04 &0.09 & $          -0.09$ & $0.09$ &0.18 &0.18 \\
$\kaon^*_{0}(1430)^{0}$   & $\phantom{-}0.14$ & $0.38$ &0.48 &0.38 & $\phantom{-}0.45$ & $0.15$ &0.37 &0.17 \\
LASS nonresonant          & $          -0.10$ & $0.24$ &0.16 &0.42 & $\phantom{-}0.44$ & $0.14$ &0.17 &0.23 \\
$\kaon^*_{2}(1430)^{0}$   & $          -0.32$ & $0.09$ &0.15 &0.23 & $          -0.47$ & $0.07$ &0.14 &0.15 \\
$D^{*}_{0}(2400)^{-}$     & $          -0.80$ & $0.08$ &0.07 &0.22 & $          -0.44$ & $0.14$ &0.12 &0.18 \\
$D^{*}_{2}(2460)^{-}$     & \multicolumn{4}{c}{1.00}               & \multicolumn{4}{c}{0.00}               \\
\hline
$D\pi$ S-wave (dabba)     & $          -0.39$ & $0.09$ &0.09 &0.14 & $\phantom{-}0.36$ & $0.17$ &0.14 &0.23 \\
$D\pi$ P-wave (EFF)       & $          -0.62$ & $0.06$ &0.03 &0.11 & $          -0.03$ & $0.06$ &0.05 &0.10 \\
\hline
\end{tabular}
\vspace{2mm} \\
\begin{tabular}{lc@{$\:\pm\:$}c@{$\:\pm\:$}c@{$\:\pm\:$}cc@{$\:\pm\:$}c@{$\:\pm\:$}c@{$\:\pm\:$}c}
\hline
Resonance & \multicolumn{4}{c}{Magnitude} & \multicolumn{4}{c}{Phase} \\
\hline \\ [-2.5ex]
$\Kstar(892)^{0}$         & $\phantom{-}1.27$ & 0.06 &0.03 &0.05 & $          -1.57$ & 0.11 &0.16 &0.27 \\
$\Kstar(1410)^{0}$        & $\phantom{-}0.18$ & 0.07 &0.10 &0.11 & $          -0.54$ & 0.21 &0.55 &1.04 \\
$\kaon^*_{0}(1430)^{0}$   & $\phantom{-}0.47$ & 0.09 &0.10 &0.14 & $\phantom{-}1.27$ & 0.95 &1.04 &0.81 \\
LASS nonresonant          & $\phantom{-}0.46$ & 0.14 &0.16 &0.29 & $\phantom{-}1.79$ & 0.65 &0.35 &0.69 \\
$\kaon^*_{2}(1430)^{0}$   & $\phantom{-}0.57$ & 0.05 &0.04 &0.08 & $          -2.16$ & 0.19 &0.43 &0.43 \\
$D^{*}_{0}(2400)^{-}$     & $\phantom{-}0.91$ & 0.07 &0.06 &0.17 & $          -2.64$ & 0.15 &0.14 &0.23 \\
$D^{*}_{2}(2460)^{-}$     & \multicolumn{4}{c}{1.00}  & \multicolumn{4}{c}{0.00}             \\
\hline
$D\pi$ S-wave (dabba)     & $\phantom{-}0.53$ & 0.07 &0.04 &0.14 & $\phantom{-}2.40$ & 0.27 &0.24 &0.44 \\
$D\pi$ P-wave (EFF)       & $\phantom{-}0.62$ & 0.06 &0.04 &0.11 & $          -3.09$ & 0.10 &0.07 &0.17 \\
\hline
\end{tabular}
\end{table}

\begin{table}[!tb]
\centering
\caption{\small
  Results for the fit fractions and their uncertainties (\%).
  The three quoted errors are statistical, experimental systematic and
  model uncertainties, respectively. Upper limits are given at 90\,\% (95\,\%) confidence level.
}
\label{tab:cfitfrac-results}
\begin{tabular}{lr@{$\:\pm\:$}c@{$\:\pm\:$}c@{$\:\pm\:$}cc}
\hline
Resonance & \multicolumn{4}{c}{Fit fraction} & Upper limit \\
\hline
$\Kstar(892)^{0}$         & 37.4 & 1.5 & 1.2 & 1.7 \\
$\Kstar(1410)^{0}$        &  0.7 & 0.3 & 0.8 & 0.8 & $<$ 3.2 (3.7)\\
$\kaon^*_{0}(1430)^{0}$   &  5.1 & 2.0 & 2.4 & 3.4 \\
LASS nonresonant          &  4.8 & 3.8 & 3.8 & 6.7 \\
\ \ \ LASS total          &  6.7 & 2.7 & 2.7 & 5.4 \\
$\kaon^*_{2}(1430)^{0}$   &  7.4 & 1.7 & 1.1 & 2.0 \\
$D^{*}_{0}(2400)^{-}$     & 19.3 & 2.8 & 2.0 & 7.4 \\
$D^{*}_{2}(2460)^{-}$     & 23.1 & 1.2 & 1.1 & 1.2 \\
$D_3^*(2760)^-$ & \multicolumn{4}{c}{} & $<$ 1.0 (1.1) \\
\hline
$D\pi$ S-wave (dabba)     &  6.6 & 1.4 & 1.2 & 3.7 \\
$D\pi$ P-wave (EFF)       &  8.9 & 1.6 & 2.2 & 3.0 \\
\hline
\end{tabular}
\end{table}

\begin{table}[!tb]
\centering
\caption{\small
  Results for the product branching fractions.
  The four quoted errors are statistical, experimental systematic, model and PDG uncertainties, respectively.
  Upper limits are given at 90\,\% (95\,\%) confidence level.
}
\label{tab:BFresults}
\begin{tabular}{lc@{$\:\pm\:$}c@{$\:\pm\:$}c@{$\:\pm\:$}c@{$\:\pm\:$}cc}
\hline
Resonance & \multicolumn{5}{c}{Product branching fraction ($10^{-5}$)} & Upper limit ($10^{-5}$) \\
\hline
$\Kstar(892)^{0}$         & 3.42 & 0.13 & 0.10 & 0.16 & 0.40 \\
$\Kstar(1410)^{0}$        & 0.07 & 0.03 & 0.08 & 0.07 & 0.01 & $<$ 0.29 (0.34)\\
$\kaon^*_{0}(1430)^{0}$   & 0.47 & 0.18 & 0.22 & 0.31 & 0.05 \\
LASS nonresonant          & 0.44 & 0.34 & 0.34 & 0.61 & 0.05 \\
\ \ \ LASS total          & 0.61 & 0.25 & 0.25 & 0.49 & 0.07 \\
$\kaon^*_{2}(1430)^{0}$   & 0.68 & 0.15 & 0.10 & 0.18 & 0.08 \\
$D^{*}_{0}(2400)^{-}$     & 1.77 & 0.26 & 0.19 & 0.67 & 0.20 \\
$D^{*}_{2}(2460)^{-}$     & 2.12 & 0.10 & 0.11 & 0.11 & 0.25 \\
$D_3^*(2760)^-$ & \multicolumn{5}{c}{} & $<$ 0.10 (0.11) \\
\hline
$D\pi$ S-wave (dabba)     & 0.60 & 0.13 & 0.11 & 0.34 & 0.07 \\
$D\pi$ P-wave (EFF)       & 0.81 & 0.15 & 0.20 & 0.27 & 0.09 \\
\hline
\end{tabular}
\end{table}

\begin{table}[!tb]
\centering
\caption{\small
  Results for the branching fractions.
  The four quoted errors are statistical, experimental systematic, model and PDG uncertainties, respectively.
  Upper limits are given at 90\,\% (95\,\%) confidence level.
}
\label{tab:BFresults2}
\begin{tabular}{lc@{$\:\pm\:$}c@{$\:\pm\:$}c@{$\:\pm\:$}c@{$\:\pm\:$}cc}
\hline
Resonance & \multicolumn{5}{c}{Branching fraction ($10^{-5}$)} & Upper limit ($10^{-5}$)\\
\hline
$\Kstar(892)^{0}$         & 5.13 & 0.20 & 0.15 & 0.24 & 0.60 \\
$\Kstar(1410)^{0}$        & 1.59 & 0.68 & 1.81 & 1.59 & 0.36 & $<$ 6.7 (7.8)\\
$\kaon^*_{0}(1430)^{0}$    & 0.71 & 0.27 & 0.33 & 0.47 & 0.08 \\
LASS nonresonant          & 0.66 & 0.51 & 0.51 & 0.92 & 0.08 \\
\ \ \ LASS total          & 0.92 & 0.38 & 0.38 & 0.74 & 0.11 \\
$\kaon^*_{2}(1430)^{0}$    & 2.04 & 0.45 & 0.30 & 0.54 & 0.25 \\
\hline
\end{tabular}
\end{table}

The masses and widths of the $D^{*}_{0}(2400)^{-}$ and $D^{*}_{2}(2460)^{-}$ states are found to be
\begin{eqnarray}
\nonumber  m(D^{*}_{0}(2400)^{-})      & = & ( 2360   \pm 15   \pm 12   \pm 28  ) \mev \\
\nonumber  \Gamma(D^{*}_{0}(2400)^{-}) & = & ( 255    \pm 26   \pm 20   \pm 47  ) \mev \\
\nonumber  m(D^{*}_{2}(2460)^{-})      & = & (2465.6  \pm 1.8  \pm 0.5  \pm 1.2 ) \mev \\
\nonumber  \Gamma(D^{*}_{2}(2460)^{-}) & = & (46.0    \pm 3.4  \pm 1.4  \pm 2.9 ) \mev\,,
\end{eqnarray}
where the three uncertainties are statistical, experimental systematic and model systematic, respectively.
These are consistent with, though less precise than, recent results from a DP analysis of $\Bd\to\Dzb\pip\pim$ decays~\cite{LHCb-PAPER-2014-070}.
They also show good agreement with, and have similar precision to, earlier measurements of these quantities~\cite{Kuzmin:2006mw,PDG2014}.

In summary, the first amplitude analysis of $\Bd\to\Dzb\Kp\pim$ decays has been presented, using a data sample corresponding to 3.0\invfb of $pp$ collision data collected by the LHCb experiment.
A good description of the data is obtained with a model containing contributions from intermediate $\Kstar(892)^{0}$, $\Kstar(1410)^{0}$, $K^*_2(1430)^{0}$ and $D^{*}_{2}(2460)^{-}$ resonances, with additional components to describe broad structures in the $K\pi$ S-wave and the $D\pi$ S- and P-waves.
The masses and widths of the $D^{*}_{0}(2400)^{-}$ and $D^{*}_{2}(2460)^{-}$ resonances are measured, as are the complex amplitudes and fit fractions for all components included in the amplitude model.
The results can be used in conjunction with an analysis of $\Bd\to\D\Kp\pim$ decays, where the neutral \D meson is reconstructed in final states such as $\Kp\Km$, to measure the CKM unitarity triangle parameter $\gamma$~\cite{Gershon:2008pe,Gershon:2009qc}.

\section*{Acknowledgements}

\noindent We express our gratitude to our colleagues in the CERN
accelerator departments for the excellent performance of the LHC. We
thank the technical and administrative staff at the LHCb
institutes. We acknowledge support from CERN and from the national
agencies: CAPES, CNPq, FAPERJ and FINEP (Brazil); NSFC (China);
CNRS/IN2P3 (France); BMBF, DFG, HGF and MPG (Germany); INFN (Italy);
FOM and NWO (The Netherlands); MNiSW and NCN (Poland); MEN/IFA (Romania);
MinES and FANO (Russia); MinECo (Spain); SNSF and SER (Switzerland);
NASU (Ukraine); STFC (United Kingdom); NSF (USA).
The Tier1 computing centres are supported by IN2P3 (France), KIT and BMBF
(Germany), INFN (Italy), NWO and SURF (The Netherlands), PIC (Spain), GridPP
(United Kingdom).
We are indebted to the communities behind the multiple open
source software packages on which we depend. We are also thankful for the
computing resources and the access to software R\&D tools provided by Yandex LLC (Russia).
Individual groups or members have received support from
EPLANET, Marie Sk\l{}odowska-Curie Actions and ERC (European Union),
Conseil g\'{e}n\'{e}ral de Haute-Savoie, Labex ENIGMASS and OCEVU,
R\'{e}gion Auvergne (France), RFBR (Russia), XuntaGal and GENCAT (Spain), Royal Society and Royal
Commission for the Exhibition of 1851 (United Kingdom).
\clearpage

{\noindent\bf\Large Appendices}
\appendix
\section{Correlation matrices}
\label{app:correlations}

The statistical correlations between the complex coefficients, masses and widths determined from the fit to data are given in terms of real and imaginary parts, and in terms of magnitudes and phases in Tables~\ref{tab:corrs-realimag}, and~\ref{tab:corrs-magphase}.
The correlations are determined using the same sample of simulated pseudoexperiments used to calculate the statistical uncertainties on fit parameters.

\begin{sidewaystable}[!tb]
\centering
\caption{\small
  Statistical correlations between the real ($x$) and imaginary ($y$) parts of the complex coefficients that are free parameters of the fit.
  Correlations with the masses ($m$) and widths ($\Gamma$) that are determined from the fit are also included.
  The correlations are determined from the same sample of simulated pseudoexperiments used to evaluate systematic uncertainties.
  The labels correspond to:
 (0) $\Kstar(892)^{0}$,
 (1) $\Kstar(1410)^{0}$,
 (2) $\kaon^*_{0}(1430)^{0}$,
 (3) LASS nonresonant,
 (4) $\kaon^*_{2}(1430)^{0}$,
 (5) $D^{*}_{0}(2400)^{-}$,
 (6) $D^{*}_{2}(2460)^{-}$,
 (7) $D\pi$ S-wave (dabba),
 (8) $D\pi$ P-wave (EFF).
}
\label{tab:corrs-realimag}
\resizebox{\textwidth}{!}{
\begin{tabular}{lcccccccccccccccccccc}
\hline
 &            $x_{0}$ & $y_{0}$ & $x_{1}$ & $y_{1}$ & $x_{2}$ & $y_{2}$ & $x_{3}$ & $y_{3}$ & $x_{4}$ & $y_{4}$ & $x_{5}$ & $y_{5}$ & $x_{7}$ & $y_{7}$ & $x_{8}$ & $y_{8}$ & $m_{6}$ & $\Gamma_{6}$ & $m_{5}$ & $\Gamma_{5}$ \\
\hline
$x_{0}$      & $\phanm$1.00 & $\phanm$0.14 & $\phanm$0.02 & $\phanm$0.17 &      $-0.18$ & $\phanm$0.15 &      $-0.17$ & $\phanm$0.03 & $\phanm$0.39 &      $-0.37$ & $\phanm$0.16 &      $-0.09$ &      $-0.18$ &      $-0.29$ & $\phanm$0.07 &      $-0.37$ &      $-0.19$ & $\phanm$0.02 & $\phanm$0.04 & $\phanm$0.12   \\
$y_{0}$      &              & $\phanm$1.00 &      $-0.06$ &      $-0.08$ & $\phanm$0.09 &      $-0.11$ &      $-0.00$ & $\phanm$0.14 & $\phanm$0.03 & $\phanm$0.03 & $\phanm$0.12 & $\phanm$0.13 &      $-0.12$ &      $-0.24$ & $\phanm$0.35 &      $-0.07$ & $\phanm$0.09 & $\phanm$0.12 & $\phanm$0.01 &      $-0.04$   \\
$x_{1}$      &              &              & $\phanm$1.00 &      $-0.09$ &      $-0.08$ &      $-0.14$ &      $-0.04$ & $\phanm$0.29 &      $-0.01$ & $\phanm$0.03 &      $-0.20$ & $\phanm$0.18 &      $-0.21$ &      $-0.31$ &      $-0.17$ &      $-0.36$ &      $-0.08$ &      $-0.17$ &      $-0.23$ & $\phanm$0.02   \\
$y_{1}$      &              &              &              & $\phanm$1.00 & $\phanm$0.05 & $\phanm$0.34 & $\phanm$0.09 &      $-0.01$ & $\phanm$0.60 &      $-0.39$ & $\phanm$0.02 &      $-0.03$ &      $-0.03$ & $\phanm$0.07 &      $-0.02$ &      $-0.04$ &      $-0.16$ &      $-0.02$ & $\phanm$0.07 & $\phanm$0.02   \\
$x_{2}$      &              &              &              &              & $\phanm$1.00 &      $-0.13$ & $\phanm$0.81 & $\phanm$0.09 & $\phanm$0.14 &      $-0.16$ &      $-0.15$ & $\phanm$0.21 & $\phanm$0.10 & $\phanm$0.13 & $\phanm$0.04 &      $-0.15$ &      $-0.07$ & $\phanm$0.01 &      $-0.25$ & $\phanm$0.01   \\
$y_{2}$      &              &              &              &              &              & $\phanm$1.00 & $\phanm$0.08 & $\phanm$0.28 & $\phanm$0.24 &      $-0.36$ &      $-0.13$ & $\phanm$0.08 &      $-0.04$ &      $-0.01$ &      $-0.04$ & $\phanm$0.04 &      $-0.02$ &      $-0.01$ &      $-0.09$ &      $-0.10$   \\
$x_{3}$      &              &              &              &              &              &              & $\phanm$1.00 & $\phanm$0.20 & $\phanm$0.13 &      $-0.34$ &      $-0.10$ & $\phanm$0.33 &      $-0.08$ & $\phanm$0.08 &      $-0.04$ &      $-0.05$ &      $-0.07$ &      $-0.03$ &      $-0.33$ &      $-0.09$   \\
$y_{3}$      &              &              &              &              &              &              &              & $\phanm$1.00 &      $-0.12$ &      $-0.12$ &      $-0.51$ & $\phanm$0.29 &      $-0.34$ &      $-0.53$ & $\phanm$0.09 &      $-0.12$ & $\phanm$0.09 &      $-0.14$ &      $-0.36$ & $\phanm$0.01   \\
$x_{4}$      &              &              &              &              &              &              &              &              & $\phanm$1.00 &      $-0.53$ & $\phanm$0.32 &      $-0.10$ &      $-0.20$ & $\phanm$0.03 & $\phanm$0.02 &      $-0.27$ &      $-0.17$ &      $-0.06$ & $\phanm$0.10 &      $-0.00$   \\
$y_{4}$      &              &              &              &              &              &              &              &              &              & $\phanm$1.00 & $\phanm$0.06 &      $-0.09$ & $\phanm$0.40 & $\phanm$0.10 & $\phanm$0.14 & $\phanm$0.16 & $\phanm$0.11 &      $-0.05$ & $\phanm$0.07 &      $-0.03$   \\
$x_{5}$      &              &              &              &              &              &              &              &              &              &              & $\phanm$1.00 &      $-0.29$ &      $-0.03$ & $\phanm$0.22 & $\phanm$0.02 & $\phanm$0.04 &      $-0.08$ & $\phanm$0.02 & $\phanm$0.45 &      $-0.06$   \\
$y_{5}$      &              &              &              &              &              &              &              &              &              &              &              & $\phanm$1.00 &      $-0.31$ &      $-0.34$ & $\phanm$0.06 & $\phanm$0.03 & $\phanm$0.24 &      $-0.14$ &      $-0.72$ &      $-0.33$   \\
$x_{7}$      &              &              &              &              &              &              &              &              &              &              &              &              & $\phanm$1.00 & $\phanm$0.46 & $\phanm$0.05 & $\phanm$0.10 &      $-0.11$ &      $-0.01$ & $\phanm$0.27 & $\phanm$0.17   \\
$y_{7}$      &              &              &              &              &              &              &              &              &              &              &              &              &              & $\phanm$1.00 &      $-0.16$ & $\phanm$0.36 &      $-0.01$ & $\phanm$0.17 & $\phanm$0.32 & $\phanm$0.08   \\
$x_{8}$      &              &              &              &              &              &              &              &              &              &              &              &              &              &              & $\phanm$1.00 &      $-0.03$ & $\phanm$0.01 &      $-0.06$ &      $-0.09$ &      $-0.14$   \\
$y_{8}$      &              &              &              &              &              &              &              &              &              &              &              &              &              &              &              & $\phanm$1.00 & $\phanm$0.29 & $\phanm$0.02 & $\phanm$0.15 &      $-0.02$   \\
$m_{6}$      &              &              &              &              &              &              &              &              &              &              &              &              &              &              &              &              & $\phanm$1.00 & $\phanm$0.09 & $\phanm$0.01 &      $-0.04$   \\
$\Gamma_{6}$ &              &              &              &              &              &              &              &              &              &              &              &              &              &              &              &              &              & $\phanm$1.00 &      $-0.03$ & $\phanm$0.00   \\
$m_{5}$      &              &              &              &              &              &              &              &              &              &              &              &              &              &              &              &              &              &              & $\phanm$1.00 & $\phanm$0.27   \\
$\Gamma_{5}$ &              &              &              &              &              &              &              &              &              &              &              &              &              &              &              &              &              &              &              & $\phanm$1.00   \\
\hline
\end{tabular}
}
\end{sidewaystable}
\begin{sidewaystable}[!tb]
\centering
\caption{\small
  Statistical correlations between the magnitudes ($a$) and phases ($\Delta$) of the complex coefficients that are free parameters of the fit.
  Correlations with the masses ($m$) and widths ($\Gamma$) that are determined from the fit are also included.
  The correlations are determined from the same sample of simulated pseudoexperiments used to evaluate systematic uncertainties.
  The labels correspond to:
 (0) $\Kstar(892)^{0}$,
 (1) $\Kstar(1410)^{0}$,
 (2) $\kaon^*_{0}(1430)^{0}$,
 (3) LASS nonresonant,
 (4) $\kaon^*_{2}(1430)^{0}$,
 (5) $D^{*}_{0}(2400)^{-}$,
 (6) $D^{*}_{2}(2460)^{-}$,
 (7) $D\pi$ S-wave (dabba),
 (8) $D\pi$ P-wave (EFF).
}
\label{tab:corrs-magphase}
\resizebox{\textwidth}{!}{
\begin{tabular}{lcccccccccccccccccccc}
\hline
 &            $a_{0}$ & $\Delta_{0}$ & $a_{1}$ & $\Delta_{1}$ & $a_{2}$ & $\Delta_{2}$ & $a_{3}$ & $\Delta_{3}$ & $a_{4}$ & $\Delta_{4}$ & $a_{5}$ & $\Delta_{5}$ & $a_{7}$ & $\Delta_{7}$ & $a_{8}$ & $\Delta_{8}$ & $m_{6}$ & $\Gamma_{6}$ & $m_{5}$ & $\Gamma_{5}$ \\
\hline
$a_{0}$      & $\phanm$1.00 &      $-0.11$ &      $-0.08$ & $\phanm$0.05 & $\phanm$0.24 & $\phanm$0.03 &      $-0.14$ & $\phanm$0.04 & $\phanm$0.04 & $\phanm$0.01 & $\phanm$0.21 & $\phanm$0.02 & $\phanm$0.22 &      $-0.22$ & $\phanm$0.38 &      $-0.00$ &      $-0.04$ &      $-0.09$ &      $-0.01$ & $\phanm$0.04   \\
$\Delta_{0}$ &              & $\phanm$1.00 &      $-0.10$ & $\phanm$0.12 & $\phanm$0.03 & $\phanm$0.20 & $\phanm$0.14 & $\phanm$0.14 &      $-0.01$ & $\phanm$0.45 &      $-0.05$ & $\phanm$0.10 &      $-0.08$ & $\phanm$0.35 &      $-0.07$ & $\phanm$0.05 &      $-0.22$ & $\phanm$0.06 & $\phanm$0.07 & $\phanm$0.09   \\
$a_{1}$      &              &              & $\phanm$1.00 &      $-0.35$ &      $-0.25$ & $\phanm$0.06 & $\phanm$0.32 & $\phanm$0.01 &      $-0.01$ &      $-0.38$ & $\phanm$0.04 &      $-0.22$ &      $-0.06$ & $\phanm$0.31 & $\phanm$0.02 & $\phanm$0.24 & $\phanm$0.07 &      $-0.18$ &      $-0.20$ & $\phanm$0.01   \\
$\Delta_{1}$ &              &              &              & $\phanm$1.00 & $\phanm$0.24 & $\phanm$0.15 & $\phanm$0.16 & $\phanm$0.00 & $\phanm$0.14 & $\phanm$0.38 &      $-0.18$ &      $-0.13$ & $\phanm$0.16 & $\phanm$0.26 & $\phanm$0.22 &      $-0.06$ &      $-0.09$ &      $-0.04$ &      $-0.12$ & $\phanm$0.05   \\
$a_{2}$      &              &              &              &              & $\phanm$1.00 & $\phanm$0.04 & $\phanm$0.11 &      $-0.28$ &      $-0.07$ & $\phanm$0.39 &      $-0.09$ & $\phanm$0.07 & $\phanm$0.27 & $\phanm$0.10 & $\phanm$0.01 & $\phanm$0.01 &      $-0.15$ & $\phanm$0.04 &      $-0.03$ &      $-0.13$   \\
$\Delta_{2}$ &              &              &              &              &              & $\phanm$1.00 & $\phanm$0.15 & $\phanm$0.77 &      $-0.20$ &      $-0.10$ &      $-0.01$ & $\phanm$0.24 & $\phanm$0.09 & $\phanm$0.13 & $\phanm$0.03 &      $-0.09$ & $\phanm$0.05 & $\phanm$0.01 & $\phanm$0.23 &      $-0.10$   \\
$a_{3}$      &              &              &              &              &              &              & $\phanm$1.00 &      $-0.06$ & $\phanm$0.00 &      $-0.11$ & $\phanm$0.16 &      $-0.30$ &      $-0.12$ & $\phanm$0.56 & $\phanm$0.01 & $\phanm$0.12 & $\phanm$0.02 &      $-0.07$ &      $-0.27$ &      $-0.04$   \\
$\Delta_{3}$ &              &              &              &              &              &              &              & $\phanm$1.00 &      $-0.28$ &      $-0.19$ & $\phanm$0.15 & $\phanm$0.41 &      $-0.24$ &      $-0.08$ &      $-0.00$ &      $-0.13$ & $\phanm$0.04 & $\phanm$0.02 & $\phanm$0.41 & $\phanm$0.08   \\
$a_{4}$      &              &              &              &              &              &              &              &              & $\phanm$1.00 & $\phanm$0.13 & $\phanm$0.05 &      $-0.17$ & $\phanm$0.14 & $\phanm$0.18 &      $-0.00$ & $\phanm$0.06 & $\phanm$0.00 & $\phanm$0.14 &      $-0.19$ &      $-0.09$   \\
$\Delta_{4}$ &              &              &              &              &              &              &              &              &              & $\phanm$1.00 &      $-0.16$ & $\phanm$0.16 & $\phanm$0.17 & $\phanm$0.12 & $\phanm$0.00 &      $-0.02$ &      $-0.19$ &      $-0.03$ & $\phanm$0.03 & $\phanm$0.01   \\
$a_{5}$      &              &              &              &              &              &              &              &              &              &              & $\phanm$1.00 & $\phanm$0.24 &      $-0.26$ &      $-0.06$ & $\phanm$0.11 & $\phanm$0.05 &      $-0.07$ & $\phanm$0.01 & $\phanm$0.19 & $\phanm$0.39   \\
$\Delta_{5}$ &              &              &              &              &              &              &              &              &              &              &              & $\phanm$1.00 &      $-0.00$ &      $-0.28$ &      $-0.01$ &      $-0.20$ &      $-0.20$ & $\phanm$0.15 & $\phanm$0.76 & $\phanm$0.27   \\
$a_{7}$      &              &              &              &              &              &              &              &              &              &              &              &              & $\phanm$1.00 &      $-0.01$ & $\phanm$0.17 &      $-0.06$ & $\phanm$0.04 & $\phanm$0.13 &      $-0.05$ &      $-0.24$   \\
$\Delta_{7}$ &              &              &              &              &              &              &              &              &              &              &              &              &              & $\phanm$1.00 &      $-0.12$ & $\phanm$0.37 & $\phanm$0.03 &      $-0.00$ &      $-0.27$ &      $-0.12$   \\
$a_{8}$      &              &              &              &              &              &              &              &              &              &              &              &              &              &              & $\phanm$1.00 &      $-0.04$ &      $-0.01$ & $\phanm$0.04 & $\phanm$0.06 & $\phanm$0.15   \\
$\Delta_{8}$ &              &              &              &              &              &              &              &              &              &              &              &              &              &              &              & $\phanm$1.00 &      $-0.37$ &      $-0.02$ &      $-0.33$ &      $-0.01$   \\
$m_{6}$      &              &              &              &              &              &              &              &              &              &              &              &              &              &              &              &              & $\phanm$1.00 & $\phanm$0.09 & $\phanm$0.01 &      $-0.04$   \\
$\Gamma_{6}$ &              &              &              &              &              &              &              &              &              &              &              &              &              &              &              &              &              & $\phanm$1.00 &      $-0.03$ & $\phanm$0.00   \\
$m_{5}$      &              &              &              &              &              &              &              &              &              &              &              &              &              &              &              &              &              &              & $\phanm$1.00 & $\phanm$0.27   \\
$\Gamma_{5}$ &              &              &              &              &              &              &              &              &              &              &              &              &              &              &              &              &              &              &              & $\phanm$1.00   \\
\hline
\end{tabular}
}
\end{sidewaystable}

\section{Results for interference fit fractions}
\label{app:IFF-results}

The central values of the interference fit fractions are given in Table~\ref{tab:interferencefrac}.
The statistical, experimental systematic and model uncertainties on these quantities are given in Tables~\ref{tab:IFF-statErrs},~\ref{tab:expt-systIFF} and~\ref{tab:model-systIFF}.

\begin{table}[!hb]
\centering
\caption{\small
  Interference fit fractions (\%) from the nominal DP fit.
  The amplitudes are all pairwise products involving:
 ($A_{0}$) $\Kstar(892)^{0}$,
 ($A_{1}$) $\Kstar(1410)^{0}$,
 ($A_{2}$) $\kaon^*_{0}(1430)^{0}$,
 ($A_{3}$) LASS nonresonant,
 ($A_{4}$) $\kaon^*_{2}(1430)^{0}$,
 ($A_{5}$) $D^{*}_{0}(2400)^{-}$,
 ($A_{6}$) $D^{*}_{2}(2460)^{-}$,
 ($A_{7}$) $D\pi$ S-wave (dabba),
 ($A_{8}$) $D\pi$ P-wave (EFF).
  The diagonal elements are the same as the conventional fit fractions.
}
\label{tab:interferencefrac}
\begin{tabular}{lccccccccc}
\hline
          &     $\phanm A_{0}$  &     $\phanm A_{1}$  &     $\phanm A_{2}$  &     $\phanm A_{3}$  &     $\phanm A_{4}$  &     $\phanm A_{5}$  &     $\phanm A_{6}$  &     $\phanm A_{7}$  &     $\phanm A_{8}$  \\
\hline
$A_{0}$   & $\phanm37.4$  & $ \phanm1.8$  & $      -0.0$  & $ \phanm0.0$  & $ \phanm0.0$  & $      -2.4$  & $      -0.2$  & $      -2.6$  & $      -5.0$  \\
$A_{1}$   &               & $ \phanm0.7$  & $      -0.0$  & $ \phanm0.0$  & $ \phanm0.0$  & $      -1.1$  & $      -0.3$  & $      -0.6$  & $      -1.1$  \\
$A_{2}$   &               &               & $ \phanm5.1$  & $      -3.2$  & $      -0.0$  & $ \phanm2.9$  & $ \phanm1.5$  & $ \phanm1.6$  & $ \phanm1.9$  \\
$A_{3}$   &               &               &               & $ \phanm4.8$  & $ \phanm0.0$  & $      -8.4$  & $ \phanm0.1$  & $ \phanm2.1$  & $ \phanm0.2$  \\
$A_{4}$   &               &               &               &               & $ \phanm7.4$  & $      -0.3$  & $      -1.0$  & $      -1.5$  & $      -0.4$  \\
$A_{5}$   &               &               &               &               &               & $\phanm19.3$  & $      -0.0$  & $ \phanm2.8$  & $ \phanm0.0$  \\
$A_{6}$   &               &               &               &               &               &               & $\phanm23.1$  & $      -0.0$  & $ \phanm0.0$  \\
$A_{7}$   &               &               &               &               &               &               &               & $ \phanm6.6$  & $      -0.0$  \\
$A_{8}$   &               &               &               &               &               &               &               &               & $ \phanm8.9$  \\
\hline
\end{tabular}
\end{table}

\begin{table}[!tb]
\centering
\caption{\small
  Statistical uncertainties on the interference fit fractions (\%).
  The amplitudes are all pairwise products involving:
 ($A_{0}$) $\Kstar(892)^{0}$,
 ($A_{1}$) $\Kstar(1410)^{0}$,
 ($A_{2}$) $\kaon^*_{0}(1430)^{0}$,
 ($A_{3}$) LASS nonresonant,
 ($A_{4}$) $\kaon^*_{2}(1430)^{0}$,
 ($A_{5}$) $D^{*}_{0}(2400)^{-}$,
 ($A_{6}$) $D^{*}_{2}(2460)^{-}$,
 ($A_{7}$) $D\pi$ S-wave (dabba),
 ($A_{8}$) $D\pi$ P-wave (EFF).
  The diagonal elements are the same as the conventional fit fractions.
}
\label{tab:IFF-statErrs}
\begin{tabular}{lccccccccc}
\hline
          &           $A_{0}$  &           $A_{1}$  &           $A_{2}$  &           $A_{3}$  &           $A_{4}$  &           $A_{5}$  &           $A_{6}$  &           $A_{7}$  &           $A_{8}$ \\
\hline
$A_{0}$   & 1.5 & 0.7 & 0.0 & 0.0 & 0.0 & 0.9 & 0.4 & 0.4 & 0.5  \\
$A_{1}$   &     & 0.3 & 0.0 & 0.0 & 0.0 & 0.7 & 0.3 & 0.4 & 0.5  \\
$A_{2}$   &     &     & 2.0 & 2.6 & 0.0 & 1.1 & 0.5 & 1.5 & 0.5  \\
$A_{3}$   &     &     &     & 3.8 & 0.0 & 3.2 & 0.4 & 3.4 & 0.8  \\
$A_{4}$   &     &     &     &     & 1.7 & 0.6 & 0.1 & 0.6 & 0.5  \\
$A_{5}$   &     &     &     &     &     & 2.8 & 0.0 & 3.1 & 0.0  \\
$A_{6}$   &     &     &     &     &     &     & 1.2 & 0.0 & 0.0  \\
$A_{7}$   &     &     &     &     &     &     &     & 1.4 & 0.0  \\
$A_{8}$   &     &     &     &     &     &     &     &     & 1.6  \\
\hline
\end{tabular}
\end{table}

\begin{table}[!tb]
\centering
\caption{\small
  Experimental systematic uncertainties on the interference fit fractions (\%).
  The amplitudes are all pairwise products involving:
 ($A_{0}$) $\Kstar(892)^{0}$,
 ($A_{1}$) $\Kstar(1410)^{0}$,
 ($A_{2}$) $\kaon^*_{0}(1430)^{0}$,
 ($A_{3}$) LASS nonresonant,
 ($A_{4}$) $\kaon^*_{2}(1430)^{0}$,
 ($A_{5}$) $D^{*}_{0}(2400)^{-}$,
 ($A_{6}$) $D^{*}_{2}(2460)^{-}$,
 ($A_{7}$) $D\pi$ S-wave (dabba),
 ($A_{8}$) $D\pi$ P-wave (EFF).
  The diagonal elements are the same as the conventional fit fractions.
}
\label{tab:expt-systIFF}
\begin{tabular}{lccccccccc}
\hline
          &           $A_{0}$  &           $A_{1}$  &           $A_{2}$  &           $A_{3}$  &           $A_{4}$  &           $A_{5}$  &           $A_{6}$  &           $A_{7}$  &           $A_{8}$ \\
\hline
$A_{0}$   & 1.2 & 0.5 & 0.0 & 0.0 & 0.0 & 2.3 & 0.2 & 0.4 & 0.5  \\
$A_{1}$   &     & 0.8 & 0.0 & 0.0 & 0.0 & 0.4 & 0.8 & 0.7 & 0.9  \\
$A_{2}$   &     &     & 2.4 & 3.1 & 0.0 & 1.4 & 2.0 & 2.6 & 1.1  \\
$A_{3}$   &     &     &     & 3.8 & 0.0 & 3.6 & 0.2 & 2.6 & 0.3  \\
$A_{4}$   &     &     &     &     & 1.1 & 2.2 & 0.2 & 0.5 & 1.0  \\
$A_{5}$   &     &     &     &     &     & 2.0 & 0.0 & 4.0 & 0.0  \\
$A_{6}$   &     &     &     &     &     &     & 1.1 & 0.0 & 0.0  \\
$A_{7}$   &     &     &     &     &     &     &     & 1.2 & 0.0  \\
$A_{8}$   &     &     &     &     &     &     &     &     & 2.2  \\
\hline
\end{tabular}
\end{table}

\begin{table}[!tb]
\centering
\caption{\small
  Model uncertainties on the interference fit fractions (\%).
  The amplitudes are all pairwise products involving:
 ($A_{0}$) $\Kstar(892)^{0}$,
 ($A_{1}$) $\Kstar(1410)^{0}$,
 ($A_{2}$) $\kaon^*_{0}(1430)^{0}$,
 ($A_{3}$) LASS nonresonant,
 ($A_{4}$) $\kaon^*_{2}(1430)^{0}$,
 ($A_{5}$) $D^{*}_{0}(2400)^{-}$,
 ($A_{6}$) $D^{*}_{2}(2460)^{-}$,
 ($A_{7}$) $D\pi$ S-wave (dabba),
 ($A_{8}$) $D\pi$ P-wave (EFF).
  The diagonal elements are the same as the conventional fit fractions.
}
\label{tab:model-systIFF}
\begin{tabular}{lccccccccc}
\hline
          &           $A_{0}$  &           $A_{1}$  &           $A_{2}$  &           $A_{3}$  &           $A_{4}$  &           $A_{5}$  &           $A_{6}$  &           $A_{7}$  &           $A_{8}$ \\
\hline
$A_{0}$   & 1.7 & 0.8 & 0.0 & 0.0 & 0.0 & 2.6 & 0.7 & 0.8 & 1.4  \\
$A_{1}$   &     & 0.8 & 0.0 & 0.0 & 0.0 & 0.8 & 0.6 & 0.8 & 1.1  \\
$A_{2}$   &     &     & 3.4 & 2.9 & 0.0 & 5.8 & 1.7 & 3.6 & 2.7  \\
$A_{3}$   &     &     &     & 6.6 & 0.0 & 8.3 & 0.3 & 3.5 & 0.7  \\
$A_{4}$   &     &     &     &     & 2.0 & 1.7 & 0.2 & 1.2 & 1.0  \\
$A_{5}$   &     &     &     &     &     & 7.4 & 0.0 & 4.5 & 0.0  \\
$A_{6}$   &     &     &     &     &     &     & 1.2 & 0.0 & 0.0  \\
$A_{7}$   &     &     &     &     &     &     &     & 3.7 & 0.0  \\
$A_{8}$   &     &     &     &     &     &     &     &     & 2.9  \\
\hline
\end{tabular}
\end{table}

\clearpage

\ifx\mcitethebibliography\mciteundefinedmacro
\PackageError{LHCb.bst}{mciteplus.sty has not been loaded}
{This bibstyle requires the use of the mciteplus package.}\fi
\providecommand{\href}[2]{#2}

\clearpage
\centerline{\large\bf LHCb collaboration}
\begin{flushleft}
\small
R.~Aaij$^{38}$,
B.~Adeva$^{37}$,
M.~Adinolfi$^{46}$,
A.~Affolder$^{52}$,
Z.~Ajaltouni$^{5}$,
S.~Akar$^{6}$,
J.~Albrecht$^{9}$,
F.~Alessio$^{38}$,
M.~Alexander$^{51}$,
S.~Ali$^{41}$
% ...

{\footnotesize \it
% ...
$ ^{5}$Clermont Universit\'{e}, Universit\'{e} Blaise Pascal, CNRS/IN2P3, LPC, Clermont-Ferrand, France\\
$ ^{6}$CPPM, Aix-Marseille Universit\'{e}, CNRS/IN2P3, Marseille, France\\
% ...
$ ^{9}$Fakult\"{a}t Physik, Technische Universit\"{a}t Dortmund, Dortmund, Germany\\
% ...
$ ^{37}$Universidad de Santiago de Compostela, Santiago de Compostela, Spain\\
$ ^{38}$European Organization for Nuclear Research (CERN), Geneva, Switzerland\\
% ...
$ ^{41}$Nikhef National Institute for Subatomic Physics, Amsterdam, The Netherlands\\
% ...
$ ^{46}$H.H. Wills Physics Laboratory, University of Bristol, Bristol, United Kingdom\\
% ...
$ ^{51}$School of Physics and Astronomy, University of Glasgow, Glasgow, United Kingdom\\
$ ^{52}$Oliver Lodge Laboratory, University of Liverpool, Liverpool, United Kingdom\\
% ...
}
\end{flushleft}

\end{document}